\documentclass{aa}  

\usepackage{graphicx}
\usepackage{txfonts}
\usepackage{verbatim}
\begin{document}

   \title{The eROSITA Final Equatorial Depth Survey (eFEDS):}

   \subtitle{The hard X-ray selected sample}

   \author{K. Nandra\inst{1},
    S. G. H. Waddell\inst{1},
    T. Liu\inst{1},
    J. Buchner\inst{1},
    T. Dwelly\inst{1},
    M. Salvato\inst{1},
    Y. Shen\inst{2},
    Q. Wu\inst{2},
    R. Arcodia\inst{1},
    Th. Boller\inst{1},  
    H. Brunner\inst{1},
    M. Brusa\inst{3,4}, 
    W. Collmar\inst{1},
    J. Comparat\inst{1},
    A. Georgakakis\inst{5},
    M. Grau\inst{1}, 
    S. Hämmerich\inst{6},
    H. Ibarra-Medel\inst{8},
    Z. Igo\inst{1},  
    M. Krumpe\inst{7},
    G. Lamer\inst{7},
    A. Merloni\inst{1},
    B. Musiimenta\inst{3,4},
    J. Wolf\inst{1},
    R.J. Assef\inst{9},
    F.E. Bauer\inst{10,11,12,13},
    W.N. Brandt\inst{14,15,16}
    \and
    H.-W. Rix\inst{17}
    }

   \institute{Max Planck Institute for Extraterrestrial Physics (MPE),
              Giessenbachstrasse 1, 85748 Garching bei München, Germany\\
              \email{knandra@mpe.mpg.de}
        \and
             Department of Astronomy, University of Illinois at Urbana-Champaign, Urbana, IL 61801, USA
        \and
            Dipartimento di Fisica e Astronomia "Augusto Righi", Alma Mater Studiorum Università di Bologna, via Gobetti 93/2, 40129 Bologna, Italy
        \and
            INAF-Osservatorio di Astrofisica e Scienza dello Spazio di Bologna, via Gobetti 93/3, 40129 Bologna, Italy
        \and
            Institute for Astronomy and Astrophysics, National Observatory of Athens, V. Paulou and I. Metaxa, 11532, Greece
        \and
            Dr. Karl Remeis-Sternwarte and Erlangen Centre for Astroparticle Physics, Friedrich-Alexander Universität Erlangen-Nürnberg, Sternwartstra{\ss}e 7, 96049 Bamberg, Germany
        \and
            Leibniz-Institut für Astrophysik Potsdam (AIP), An der Sternwarte 16, 14482 Potsdam, Germany
        \and
            Instituto de Astronomía y Ciencias Planetarias, Universidad de Atacama, Copayapu 485, Copiapó, Chile
        \and
            Instituto de Estudios Astrof\'isicos, Facultad de Ingenier\'ia y Ciencias, Universidad Diego Portales, Av. Ej\'ercito Libertador 441, Santiago, Chile. 
        \and
            Instituto de Astrof{\'{\i}}sica, Facultad de F{\'{i}}sica, Pontificia Universidad Cat{\'{o}}lica de Chile, Campus San Joaquín, Av. Vicuña Mackenna 4860, Macul Santiago, Chile, 7820436
        \and
            Centro de Astroingenier{\'{\i}}a, Facultad de F{\'{i}}sica, Pontificia Universidad Cat{\'{o}}lica de Chile, Campus San Joaquín, Av. Vicuña Mackenna 4860, Macul Santiago, Chile, 7820436
        \and
            Millennium Institute of Astrophysics, Nuncio Monse{\~{n}}or S{\'{o}}tero Sanz 100, Of 104, Providencia, Santiago, Chile
        \and
            Space Science Institute, 4750 Walnut Street, Suite 205, Boulder, Colorado 80301
        \and
            Department of Astronomy \& Astrophysics, 525 Davey Lab, The Pennsylvania State University, University Park, PA 16802, USA
        \and
            Institute for Gravitation and the Cosmos, The Pennsylvania State University, University Park, PA 16802, USA 
        \and
            Department of Physics, 104 Davey Laboratory, The Pennsylvania State University, University Park, PA 16802, USA
        \and
            Max-Planck-Institut für Astronomie, Königstuhl 17, D-69117 Heidelberg, Germany
             }

   \date{Received XXX; accepted XXX}

 
  \abstract
   {During its calibration and performance verification phase, the eROSITA instrument aboard the SRG satellite performed a uniform wide--area X-ray survey of approximately 140 deg$^{2}$ in a region of the sky known as the eROSITA Final Equatorial Depth Survey (eFEDS).}
   {The primary aim of eFEDS is to demonstrate the scientific performance to be expected at the end of the 8-pass eROSITA all sky survey. This will provide the first focussed image of the whole sky in the hard X-ray ($>2$~keV) bandpass. The expected source population in this energy range is thus of great interest, particularly for AGN studies.}  
   {We use the 2.3--5 keV selection presented by Brunner et al. (2022) to construct a sample of 246 point-like hard X-ray sources for further study and characterization. These are classified as either extragalactic ($\sim 90$~\%) or Galactic ($\sim 10$~\%), with the former consisting overwhelmingly of AGN and the latter active stars. We concentrate our further analysis on the extragalactic/AGN sample, describing their X-ray and multiwavelength properties and comparing them to the eFEDS main AGN sample selected in the softer 0.2-2.3 keV band.}
   {The eROSITA hard band selects a subsample of sources that is a factor $>10$ brighter than the eFEDS main sample. The AGN within the hard population reach up to $z=3.2$ but on the whole are relatively nearby, with median $z$=0.34 compared to $z$=0.94 for the main sample. The hard survey probes typical luminosities in the range $\log L_{\rm X} = 43-46$. X-ray spectral analysis shows significant intrinsic absorption (with $\log N_{\rm H}>21$) in $\sim 20$~\% of the sources, with a hard X-ray power law continuum with mean $<\Gamma>=1.83\pm0.04$, typical of AGN, but slightly harder than the soft-selected eROSITA sample. Around $\sim10$~\% of the hard sample show a significant "soft excess" component. The sampled black hole mass distribution in the eFEDS broad line AGN population is consistent with that of the deeper COSMOS survey that probes a higher redshift populations. On the other hand, the Eddington ratios appear systematically lower, consistent with the idea that the decline in SMBH activity since $z\sim 1$ is due to a reduction in the typical accretion rate, rather than a shift towards activity in lower mass black holes. 
   }
   {The eFEDS hard sample provides a preview of what can be expected from the eRASS final survey in terms of data quality. This pilot survey indicates the power of eROSITA to shed new light on the demographics and evolution of AGN, and the potential for discovery of new and rare populations.}

\authorrunning{K. Nandra et al.}
\titlerunning{eFEDS hard X-ray sample}
   \maketitle
%

\section{Introduction}

The eROSITA instrument \citep{Predehl2021} aboard the Spectrum-RG satellite \citep{Sunyaev2021} consists of seven Wolter-1 type telescopes focussing X-rays onto seven focal plane X-ray cameras. It is designed to survey the sky in the X-ray band, with the driving science being to constrain the cosmological parameters via the evolution of clusters of galaxies \citep{Merloni2012}. A key aim is to detect all clusters in the Universe more massive than about $3 \times 10^{14} M_{\odot}$. This requires all-sky coverage with high surface-brightness sensitivity, combined with angular resolution sufficient to resolve clusters from the much larger population of X-ray point sources in the sky, dominated by active galactic nuclei (AGN) and coronally active stars. To do this efficiently requires an imaging telescope with large field of view and effective area in the soft X-ray band. 

   \begin{figure*}
   \centering
   \includegraphics[width=\textwidth]{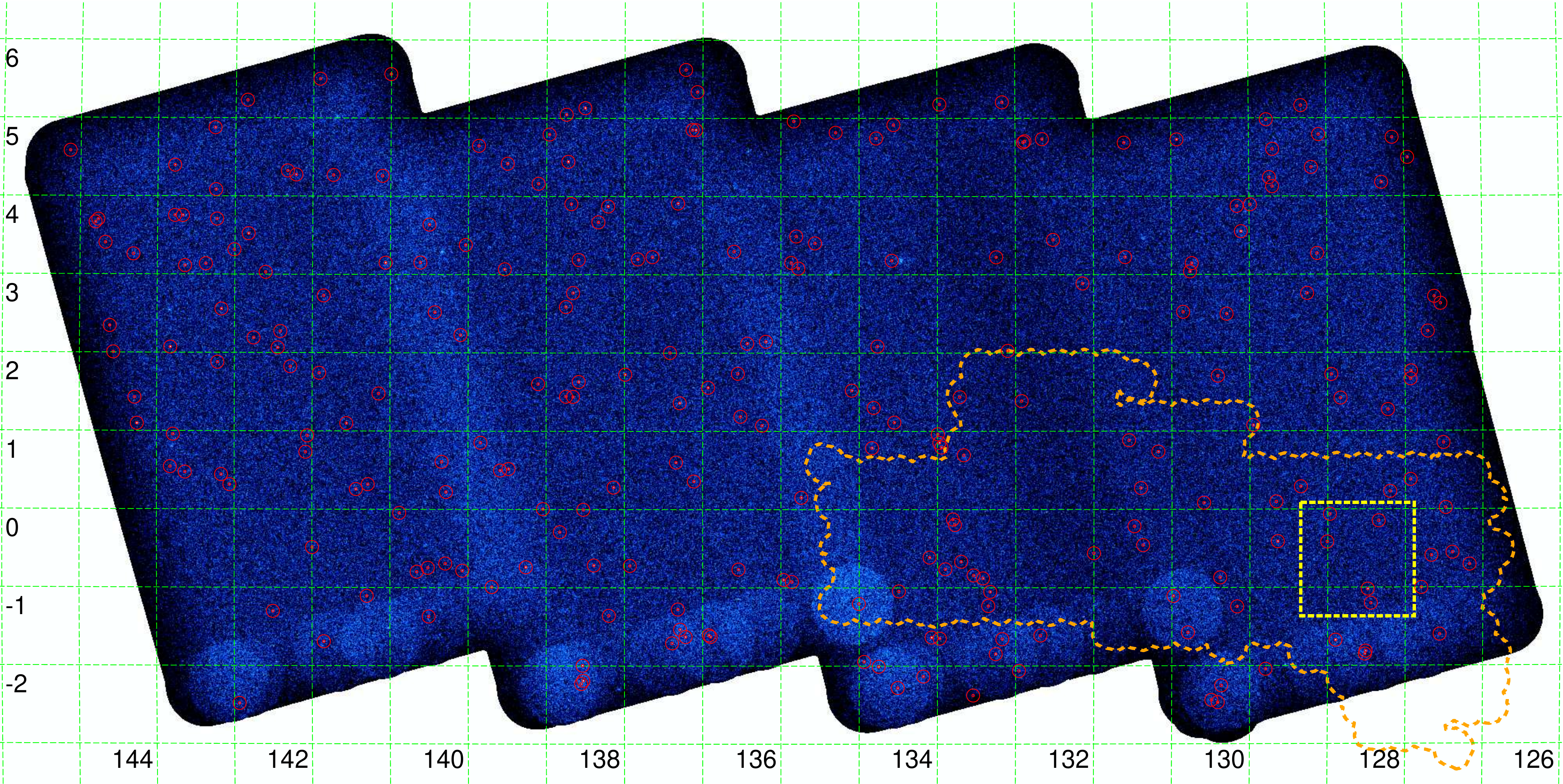}
   \caption{The 2.3--5 keV eFEDS image, smoothed by a Gaussian kernel of $\sigma=16\arcsec$, with each point source overlaid as a red annulus with radii of $60\arcsec$ and $300\arcsec$. Representations of the regions of the XMM-XXL-north (orange) and the COSMOS (yellow) fields, which do not overlap with eFEDS, are placed at the lower right corner for size comparison.}

              \label{fig:efeds}%
    \end{figure*}
    
While the primary driving science of eROSITA is cluster cosmology, the high sensitivity of the instrument and mission profile of SRG enable a vast array of additional science \citep{Merloni2012}. Unlike its predecessor ROSAT \citep{Truemper1982}, the eROSITA telescope system features significant effective area in the harder X-ray band above 2 keV. Together with its sister instrument Mikhail Pavlinsky ART-XC \citep{Pavlinsky2021}, which covers an even harder X-ray band (4--30~keV), it is the first focussing telescope to have performed an all-sky survey at energies $>2$~keV. eROSITA therefore opens up a new parameter space for large area hard X-ray surveys. 

These build on a considerable body of survey work in this energy range, which have covered various regions of area-depth parameter space. One early example was the HEAO-1/A2 survey, which covered the whole sky.  Despite its high flux limit this has nonetheless been very influential as it established a reference set of X-ray bright extragalactic objects \citep{Piccinotti82} which have been the subject of increasingly intense study by subsequent X-ray missions. Numerically, the extragalactic hard X-ray population found by HEAO-1 was dominated by Active Galactic Nuclei (AGN) hosting accreting supermassive black holes, with the next most numerous class being massive clusters of galaxies. 

While soft X-ray surveys, such as those performed by ROSAT and now eROSITA, are arguably more efficient for the selection of galaxy clusters, harder X-ray surveys are of particular interest for AGN studies. One reason for this is that hard X-ray emission can be used to uncover heavily obscured AGN and measure their level of obscuration \citep[e.g.][]{Awaki1991,Turner1997}, as well as their contribution to the X-ray background \citep[e.g.][]{Comastri1995,Gilli2007,Ueda2014}. These factors have motivated numerous surveys with X-ray imaging telescopes with sensitivity above 2 keV, for example with ASCA \citep{Ueda1999}, BeppoSAX \citep{Fiore2001}, Chandra \citep[e.g.,][]{Nandra2015,Civano2016,Luo2017} and XMM-Newton \citep[e.g.][]{Fiore2003,Hasinger2007,Ranalli2013,Pierre2016}. Even harder bandpasses with greater sensitivity to the most heavily obscured AGN have now also been covered by Swift \citep{Oh2018} and NuSTAR \citep{Alexander2013}. These and other similar surveys have formed the backbone for AGN demographic and evolutionary studies over the past two decades \citep[e.g.,][]{Ueda2003,LaFranca2005,Aird2015,Buchner2015}. Despite the great depth of surveys in particular with Chandra and XMM-Newton, they can necessarily only cover limited areas. For example, the largest area surveys with each facility XBootes \citep{Murray2005,Kenter2005} and XMM-XXL \citep{Pierre2016} reach flux limits of around $2 \times 10^{-14}$~erg cm$^{-2}$ s$^{-1}$ in the 2-10 keV band but 
cover only 9 and (two) 25 deg$^2$ contiguous regions respectively. A significant region of parameter space exists in between these Chandra/XMM surveys and the with HEAO-1/A2 \citep{Piccinotti82} and the similar-depth RXTE All-Sky Survey \citep{Revnivtsev2004}, which cover the whole sky but at a flux limit around 3 orders of magnitude brighter. This gap in depth-area coverage can neatly be filled by a wide-field imaging X-ray telescope like eROSITA. 

As a precursor to the eROSITA all-sky survey (eRASS), eROSITA performed a field scan of an equatorial region of approximately 140 deg$^2$ during its calibration and performance verification phase. Aside from all-sky surveys such as ROSAT and the eRASS itself, this survey, known as the eROSITA Final Equatorial Depth Survey (eFEDS) currently represents the largest contiguous dedicated X-ray imaging survey performed thus far, despite taking only approximately 5 days. Occupying a new part of area-depth parameter space for contiguous surveys \citep{Brunner2022}, and with excellent supporting data in numerous other wavebands, eFEDS is capable of yielding new insights, particularly into the properties and evolution of clusters and AGN. It also provides a verification and foretaste of what can be expected ultimately from the full 4-years exposure of the entire sky (eRASS:8). 

In this paper, we present an analysis of the X-ray and optical properties of the hard X-ray (2.3--5 keV) point sources detected in eFEDS. In Section~\ref{sec:methods} we briefly describe the sources of the data and the methodology from which the catalogue products are derived. Section~\ref{sec:sample} presents the basic properties of the sample. In Section \ref{sec:agn} we focus on the subsample of extragalactic sources, predominantly AGN. 
A discussion and outlook for the eventual 8-pass, 4 year eRASS:8 all-sky survey is presented in Section~\ref{sec:disc}. 

Throughout this paper, we adopt a flat $\Lambda$CDM cosmology with $\Omega_{\Lambda} =0.7$, $\Omega_m =0.3$, and $H_0 = 70$ km s$^{-1}$ Mpc$^{-1}$.


\section{Data and Methods}
\label{sec:methods}

\subsection{The eROSITA Final Equatorial Depth Survey}

A full description of the eFEDS X-ray survey is presented in \cite{Brunner2022}, which also contains the X-ray catalogue used to construct the sample presented in this paper. The eFEDS field covers approximately 140 deg$^2$ in an equatorial field chosen to have excellent multi-wavelength supporting data. Of particular note for the current study are: (i) high quality optical imaging and photometry provided by the HSC Wide Area Survey \citep{Aihara2018}; (ii) multi-band photometry from the DESI Legacy Imaging survey \citep{Dey2019} data release 8 (LS8), which also incorporates mid infrared data from the Wide-Field Infrared Spectroscopic Explorer (WISE; \cite{Wright2010}) and (iii) extensive optical spectroscopic coverage mainly from the fourth and fifth incarnations of the Sloan Digital Sky Survey \cite[SDSS;][Kollmeier et al., in preparation]{Blanton17,Kollmeier17}. A fuller description of the multi-wavelength data in this field can be found in \cite{Salvato22}. 

The X-ray data were acquired in a field-scanning mode providing an unusually uniform exposure over the field compared to the "point and stare" observing strategy commonly used for X-ray surveys. The nominal exposure at a given position is typically 2.2ks, making the eFEDS survey representative of, but slightly deeper than, the expectation for the typical equatorial sky position in the full 8-pass eRASS:8.

The X-ray data processing and reduction was performed using the eROSITA Science Analysis Software system (eSASS) using the eROSITA early data release (EDR) version (eSASSusers\_201009). The eSASS procesing chain is described in full in \cite{Brunner2022}. Source
 detection involves the generation of an initial seed source catalogue using a sliding box detection algorithm, and subsequent PSF-fitting to determine source counts and significances using a likelihood ratio approach. An advantage to this method is that the detection can be performed using event files/images in multiple energy bands simultaneously. In practice, the eFEDS main sample is assembled using a soft X-ray selection in the 0.2-2.3 keV band, where the instrument is most sensitive. This main eFEDS source catalogue contains a total of 27,910 sources above a detection likelihood threshold {\tt DET\_LIKE}$>6$ in the detection band, where {\tt DET\_LIKE}$=-\ln P$ where $P$ is the chance probability of the source being a background fluctuation based on a PSF-fitting detection algorithm \citep[see][]{Brunner2022}. The detection algorithm allows for the sources to be extended, also providing a likelihood estimation for this, the quantity {\tt EXT\_LIKE}. Sources with {\tt EXT\_LIKE}$>=6$ are considered candidate galaxy clusters whose properties are presented and analysed in detail by \citet{LiuA2022}. Sources with {\tt EXT\_LIKE}$<6$ are considered point-like and have their \texttt{EXT\_LIKE} fixed to zero in the catalogue of \cite{Brunner2022}. The point sources can have a variety of classifications, and are characterised by a dedicated algorithm \cite[see below and][]{Salvato2022}. 

\subsection{The eFEDS hard X-ray catalogue}

   \begin{figure}
   \centering
    \includegraphics[width=40mm]{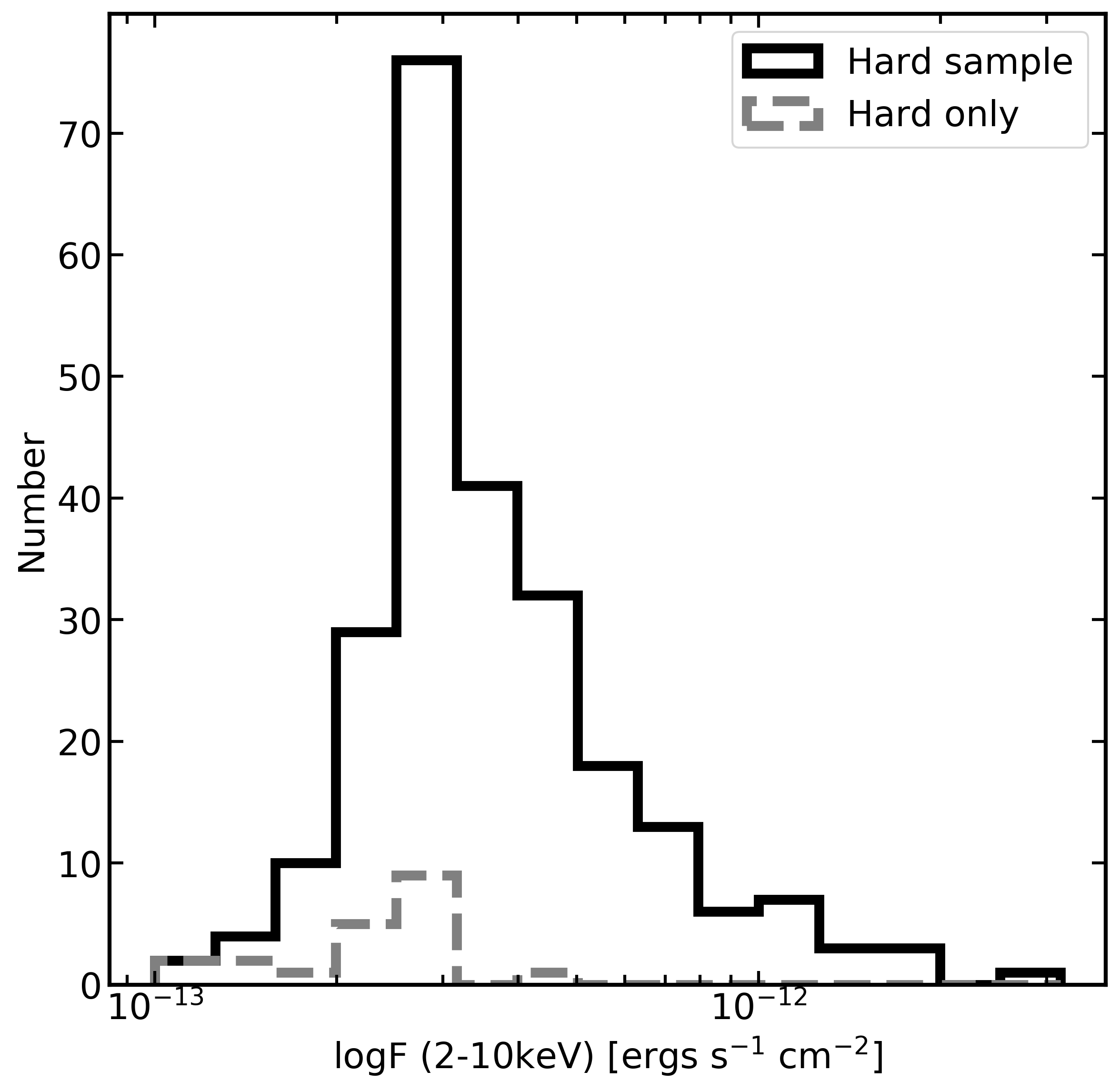}
    \includegraphics[width=40mm]{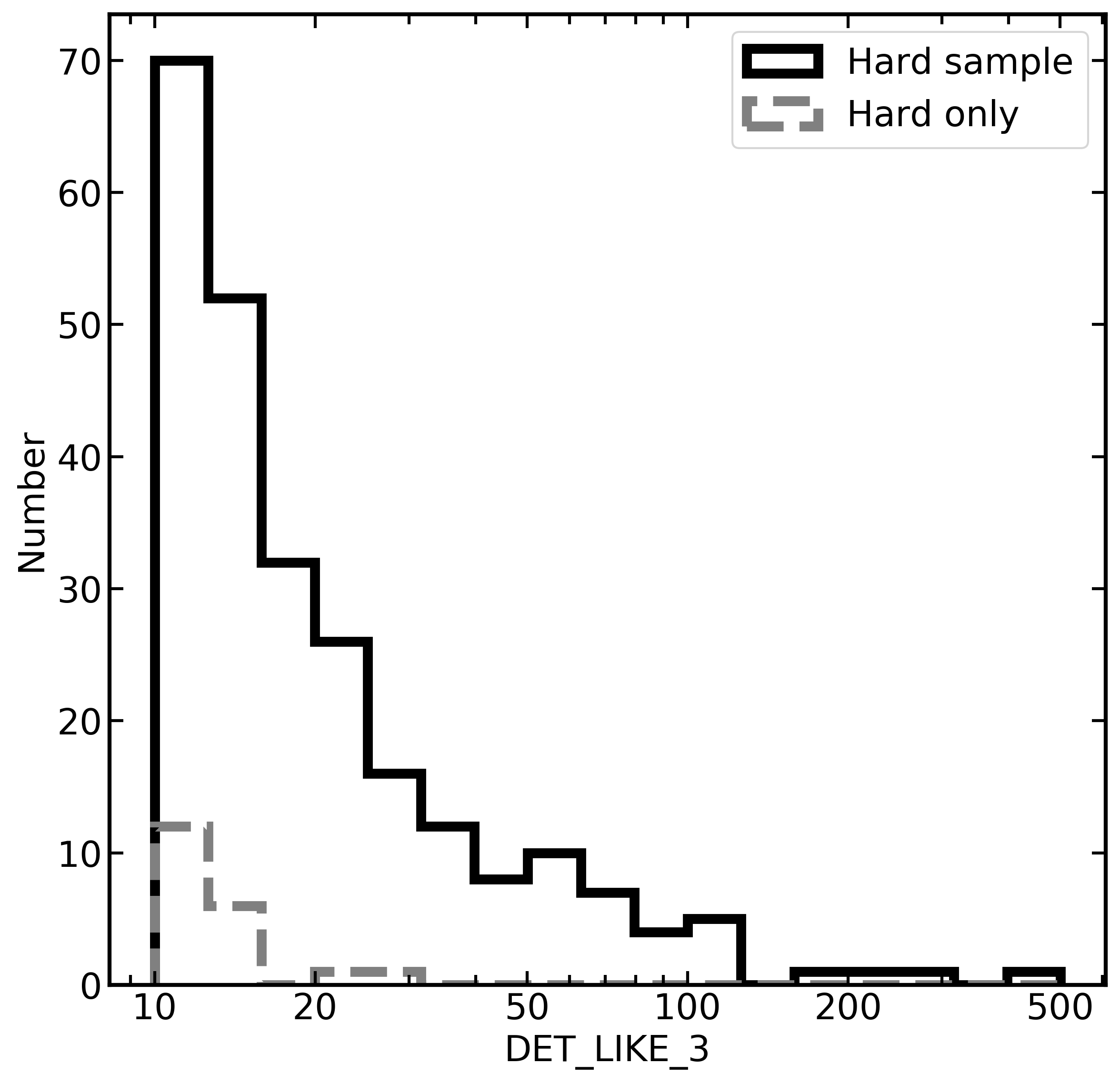}
      \caption{Histograms of the 2-10 keV X-ray flux (left) and detection likelihood {\tt DET\_LIKE} (right) for our sample. Sources detected only in the hard band are shown separately.
              }
         \label{fig:hx_hflux}

   \end{figure}

To supplement the main, soft X-ray source catalogue, \cite{Brunner2022} also present a hard X-ray selected catalogue. This is constructed using a source--detection run performed simultaneously in three energy bands: 0.2--0.6, 0.6--2.3 and 2.3--5 keV \cite[see][for details]{Brunner2022}, which results in a likelihood in each of the these individual bands, as well as a combined likelihood for all three bands. To construct the catalogue presented here, a threshold of the three-band summary likelihood \texttt{DET\_LIKE\_0}$>5$ was adopted. We then selected sources with a detection likelihood in the hardest of those three bands (2.3--5 keV) of \texttt{DET\_LIKE\_3}$>10$. The former threshold is sufficiently loose that it has no impact on the latter, in other words, the \texttt{DET\_LIKE\_0}$>5$ threshold does not remove any potential source with \texttt{DET\_LIKE\_3}$>10$. There are eight extended sources with \texttt{DET\_LIKE\_3}$>10$ and {\tt EXT\_LIKE}$>=6$ which are candidate galaxy clusters \citep{LiuA2022} and thus excluded from this work. The multi-band detection is performed because, as we show below, the vast majority of the hard-selected sources are also detected in the soft band. Inclusion of the soft photons in these cases can substantially improve the determination of the X-ray source position, which will lead to a more accurate determination of the true hard source counts and hence the likelihood of the hard detection. In addition, the better positional accuracy should aid considerably in the counterpart identification. 
According to detailed simulations by \citet{LiuT2022_sim}, the selection threshold imposed for this catalog ensures a low fraction (2.5\%) of spurious sources.

The resulting hard X-ray sample consists of 246 point-like sources, more than two orders of magnitude fewer than the main sample. The sources' positions within the eFEDS field are shown in Fig.~\ref{fig:efeds}, superimposed on the 2.3--5 keV smoothed image. Just 20 of the 246 sources are not also listed in the main catalogue i.e. they are detected in the hard band but not the softer 0.2-2.3 keV band which is used to construct the main catalogue. The lower sensitivity in the hard band can be attributed to two main factors. First, as can be seen in \citep[][,see their Fig. 9]{Predehl2021} there is a pronounced drop in the effective area of eROSITA around $\sim 2.3$ keV due to M-absorption edges in the gold mirror coating. Above this energy the effective area continues to decline due to the relatively short eROSITA focal length of 1.6m (compared, e.g., to \textit{Chandra} or \textit{XMM-Newton}). Second, at harder X-ray energies, the instrumental background starts to become dominant. This is larger than the pre-flight predictions, partly related to the solar cycle during the first few months after launch \citep{Freyberg20,Predehl2021}, and reduces the source detection sensitivity accordingly.

   \begin{figure}
   \centering
    \includegraphics[width=50mm]{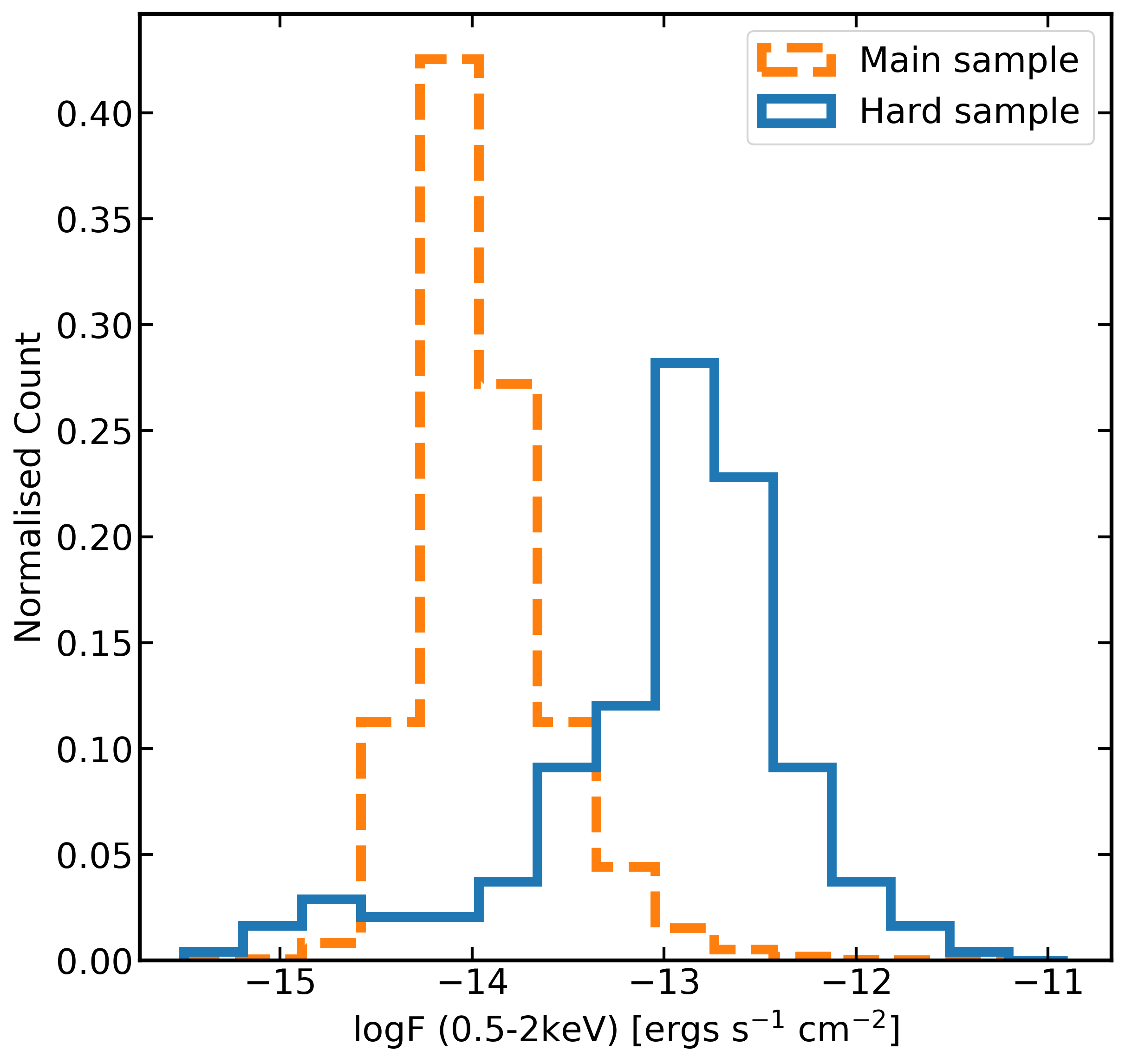}
      \caption{Normalised histogram of the soft X-ray (0.5-2 keV) flux comparing the main sample (selected in the 0.2--2.3 keV band) with the hard sample (selected in the 2.3--5 keV band). Note that the hard-band only sources are included in the blue histogram, and will have large uncertainties in their soft X-ray fluxes. 
              }
         \label{fig:hx_main_sxhist}

   \end{figure}
   
\subsection{Optical identifications, classifications and redshifts}
\label{sec:optical}

Optical counterparts for the hard X-ray sample have been identified by \citet{Salvato2022}. A total of three different matching algorithms were applied: the Bayesian NWAY code \citep{Salvato2018}, a maximum likelihood approach \citep[e.g.,][]{Sutherland1992,Brusa2007} and HamStar, a dedicated matching algorithm designed to identify X-ray emitting stars \citep{Schneider22}. The key datasets to which the X-ray source positions are matched are LS8, which includes forced photometry from unWISE \citep{Lang2014}, and Gaia EDR3. The latter also provides parallax and proper motion information that is crucial in distinguishing between Galactic and extragalactic counterparts to the X-ray sources.

   \begin{figure*}
   \centering

    \includegraphics[width=6.5cm]{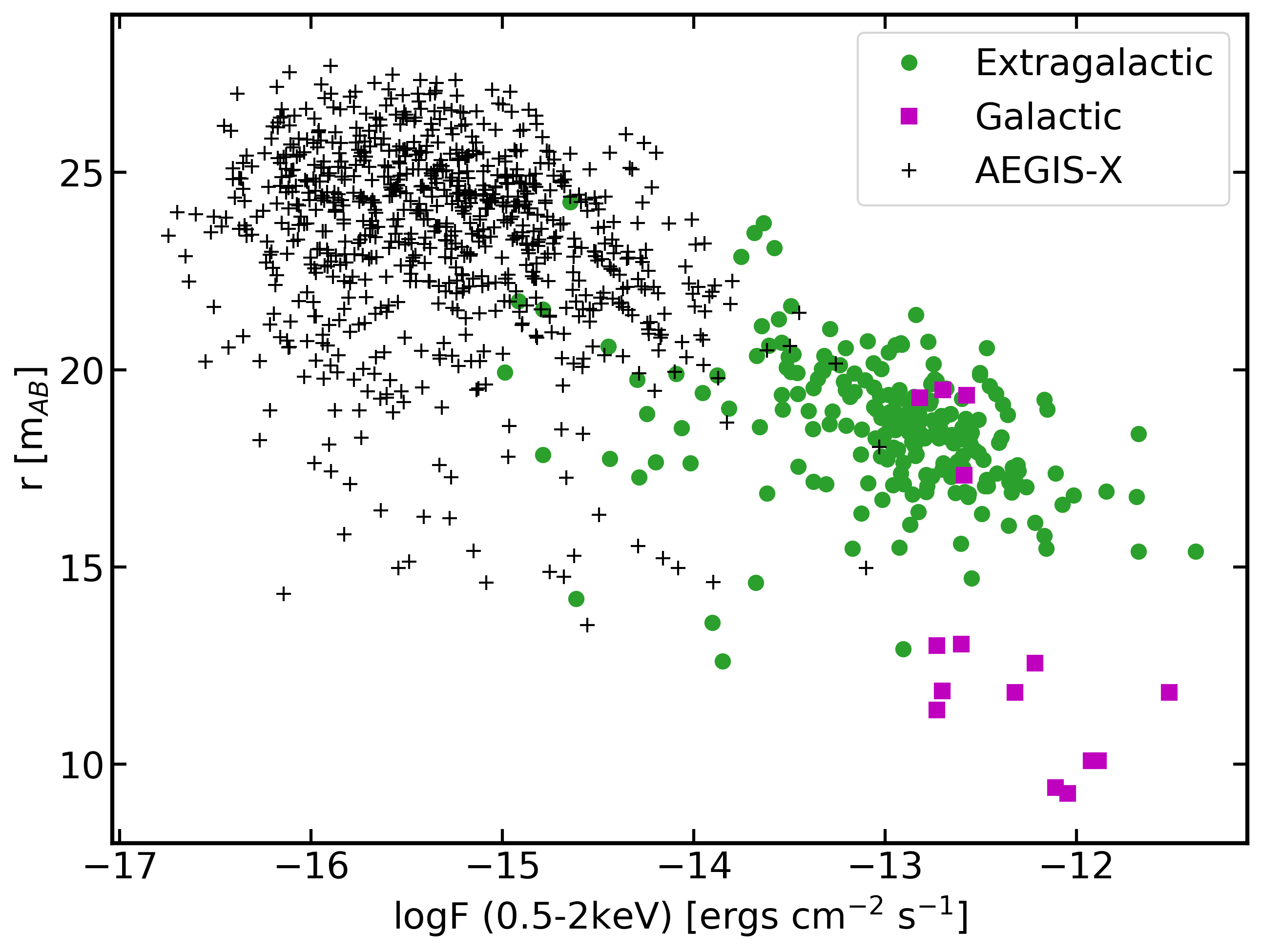}
    \includegraphics[width=6.5cm]{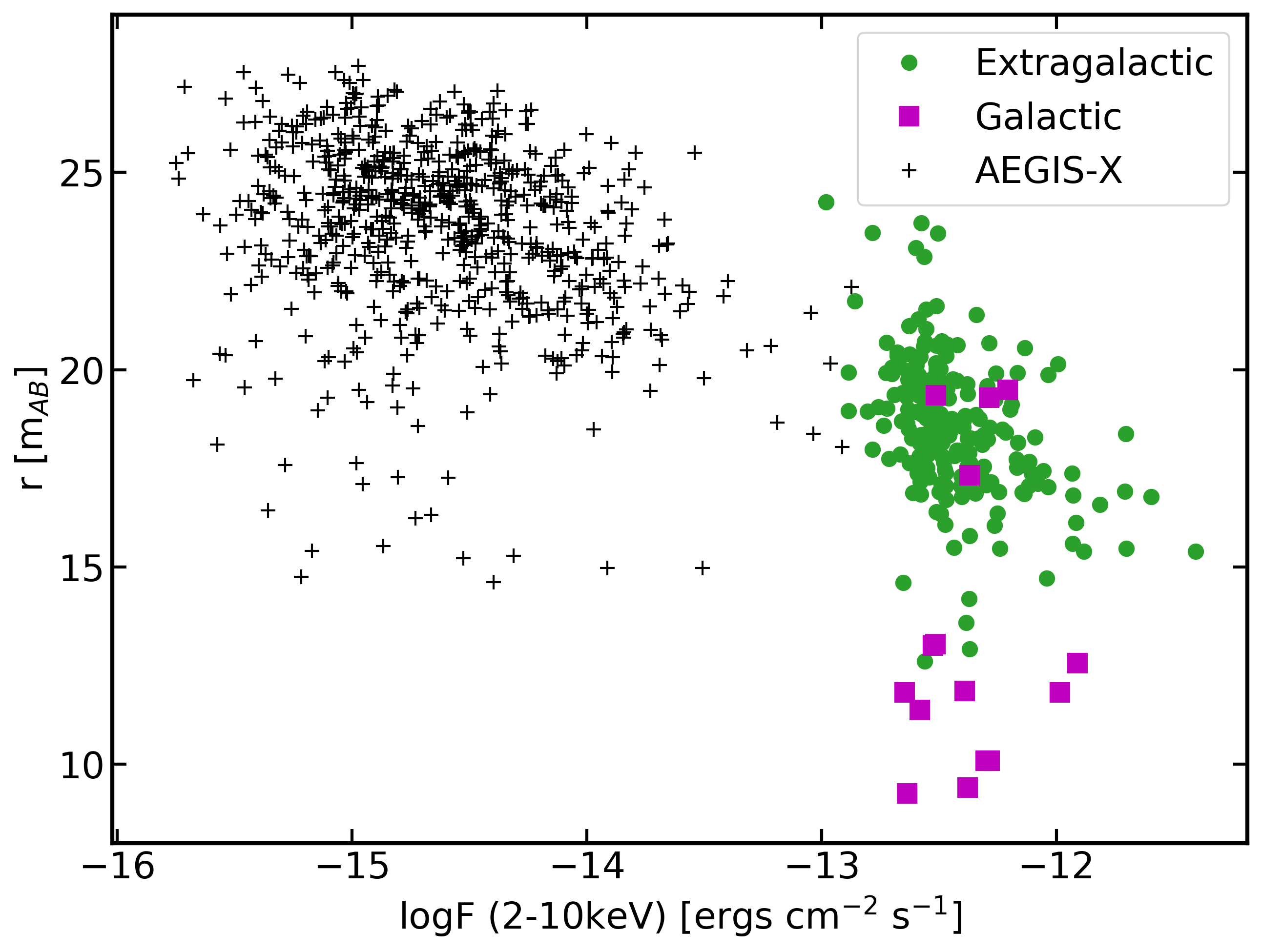}
    \includegraphics[width=6.5cm]{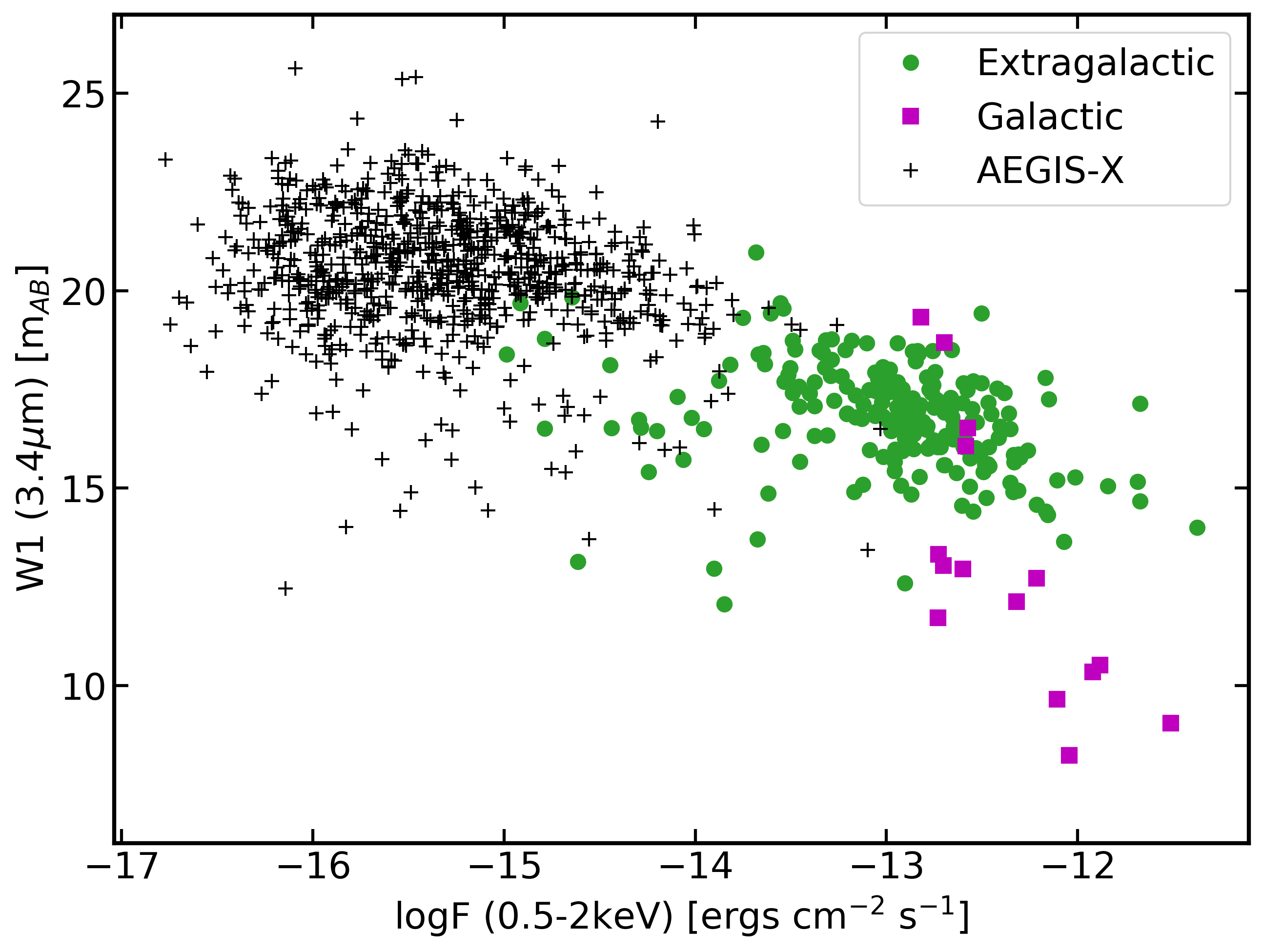}
    \includegraphics[width=6.5cm]{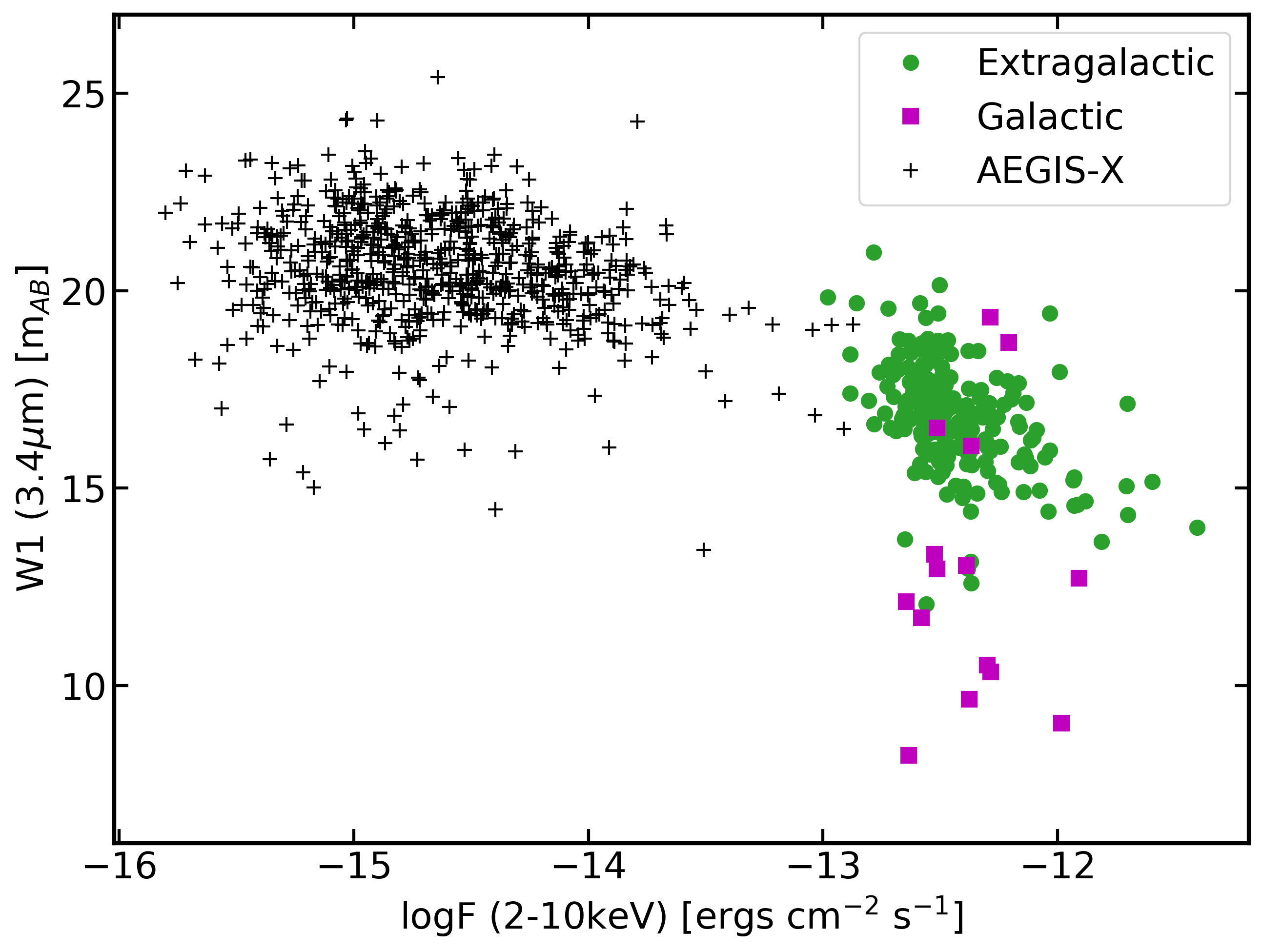}
      \caption{Legacy survey $r$-band (uppers panels) and Wise W1 3.4$\mu m$ (lower panels) magnitude versus the 0.5--2 keV (left panels) and 2-10 keV (right panels) X-ray flux for the eFEDS hard sample. Source classified as being extragalactic (green symbols) and Galactic (purple symbols) are shown separately. As a comparison the data for the AEGIS-XD survey \citep{Nandra2015} are shown (small black symbols), which cover slightly different bands in the optical/IR, specifically Subaru R$_{\rm c}$ and Spitzer/IRAC 3.6$\mu$m.
              }
         \label{fig:fx_fopt}
   \end{figure*}


A well-defined decision tree is applied to find the most likely counterpart, or identify cases where the counterpart is unlikely or ambiguous. This results in the allocation of a counterpart quality flag between 0-4, described in detail in \citet{Salvato2022}. In short, however, counterpart qualities 0 and 1 are considered to be insecure, 3 and 4 to be secure. Counterpart quality 2 represents cases where the counterpart is ambiguous, and/or where more than one source could contribute to the observed X-ray flux. The results for the eFEDS hard sample are given in Section~\ref{sec:results_optical}.

In addition to the counterparts, \cite{Salvato22} present a compilation of optical spectroscopy of the most likely counterparts to our hard sources, along with photometric--redshift determinations. The vast majority of the optical spectra come from the Sloan Digital Sky Survey (SDSS) telescope \citep{Gunn06} using the BOSS spectrographs \citep{Smee13}. Of these, most are from a dedicated program for X-ray followup within SDSS-IV \citep{Blanton17}, the SPIDERS program \citep{Dwelly2017,Comparat2020}. X-ray sources in the eFEDS field have also been targeted explicitly as part of the SDSS-V \citep{Kollmeier17} special plates program. The SDSS-V program greatly increases both the spectroscopic completeness and the quality of the data, given that the exposures for the SDSS-V special plates were generally signigicantly longer. Including a smattering of redshifts from additional programs, a total of 197/246 of our hard sample objects have a spectroscopic redshift as detailed in Table~\ref{tab:specz}. The redshifts used here were derived from the compilation of eFEDS spectroscopic redshifts released as part of SDSS DR18 \citep{Almeida2023}
\footnote{see also \url{https://www.sdss.org/dr18/data_access/value-added-catalogs/?vac_id=10001}}, which have been verified by visual inspection, and superseded those presented by \citet{Salvato2022}. The vast majority of spectroscopic objects are extragalactic, with just 12 being spectroscopic stars. 

\begin{table}
\caption{Optical Spectroscopy. Where multiple spectra exist of the same
object, the table lists the surveys in preference order: (1) \cite{Blanton17}; (2) \cite{Almeida2023}; (3) \cite{Baldry18}; (4) \cite{Jones09}; (5) \cite{Wenger2000}; (6) \cite{Huchra12}; 
\label{tab:specz}}             
\centering          
\begin{tabular}{l | c c c c c}     
\hline\hline       
Origin & Redshifts & Stellar & Ref\\
       & $z$ & ($z<0.001$) & \\
\hline                    
SDSS        & 180 & 2 & (1,2) \\
Gaia-RVS    & 8 & 8 & -- \\
GAMA        & 2 & -- & (3) \\
6dFGS       & 2 & --  & (4) \\
LAMOST      & 2 & 1 & -- \\
Simbad      & 2 & 1 & (5) \\
2mrs        &  1 & -- &  (6) \\

\hline
Total & 197 & 12 & --\\
\hline
\hline                  
\end{tabular}
\end{table}

Most of the spectroscopy in the field yields high--confidence redshifts. Three objects have spectra from which no redshift could be determined, but are detected in the radio in the FIRST survey.
They are considered candidate BL Lac type objects (see Section~\ref{sec:blazars} for more discussion), with the lack of a redshift determination being attributed to a lack of spectral features. 
Reflecting the the fact that the hard sources are significantly brighter, the hard sample has very high spectroscopic completeness (80\%) compared to the main sample(24\%). Several previous hard X-ray surveys have also benefitted from high spectral completeness for the same reason \citep[e.g.][]{Akiyama2003,DellaCeca2004,Eckart2006,Cocchia2007}.

For the relatively small but significant fraction of our sample for which optical spectroscopy is not available, other information must be used to classify the objects. \cite{Salvato2022} present a classification scheme primarily designed to differentiate between Galactic and extragalactic objects, which is particularly important for the current study which focusses on the properties of the latter. Aside from the spectroscopic stars (see above), X-ray source counterparts which have a significant Gaia parallax measurement are considered to be securely Galactic, as are all other objects using the "Hamstar" scheme designed to identify X-ray emitting stars \citep{Schneider22}. Conversely, objects which are extended in their optical images are assumed to be securely extragalactic. For the remainder, when possible a separation scheme based on optical/mid-IR colors is applied \citep{Salvato22} or otherwise an X-ray/mid-IR criterion \citep{Salvato2018}. 

Once secure or likely Galactic sources have been identified, we assign photometric redshift estimates to the extragalactic sources without spectroscopic redshifts. These are again discussed and presented in \cite{Salvato2022}, and are based on template fitting of the IR-optical-UV Spectral Energy Distribution (SED) using the LePhare code \citep{Ilbert2006}. The photometric redshifts are assigned a ``grade'', with a higher grade indicating greater reliability of the photo-z estimate and in particular a lower outlier fraction. Here we adopt {\tt REDSHIFT\_GRADE} $\ge 4$ in the Salvato et al. catalogue, selecting only objects with either spectroscopic or the most reliable photometric redshift measurements. The latter are selected by comparing the SED-fitting photo-z with an independent measurement using machine learning techniques, specifically a multi-layer perceptron \citep[DNNZ;][]{Nishizawa2020}, and adopting only those measurements where both methods agree. For such cases the photometric redshift show a low scatter of $\sigma_{\rm NMAD}=0.043$ and outlier fraction $5.3$~per cent \citep[see][for details]{Salvato2022}.
The redshift information for the eFEDS hard sample is provided in Section~\ref{sec:zandl}.

\subsection{X-ray spectral fitting}

We characterize the spectral properties of our sources both by hardness ratios and full X-ray spectral fitting. We adopt a standard hardness ratio (HR) definition: 
$$HR = (H-S)/(H+S) $$
where $H$ is the flux in the 2.3--5 keV band and $S$ that in the 0.6--2.3 keV band. 

A systematic analysis of the X-ray spectra of the eFEDS main sample has been presented by \citet{LiuT2022_agn}, who fit a variety of spectral models to determine fluxes and luminosities, as well as physical parameters such as the photon index and $N_{\rm H}$.
Although most of the hard sources are included in the main sample, 
the photon-level data can be different. For example the source positions in the multi-band detected catalogue will in general be subtly different to those in the soft band. 
We therefore perform an independent spectral analysis for the hard sample separately from the main sample, but using the same method.

Using the eSASS task \texttt{srctool}, for each source we calculate a signal-to-noise-ratio optimized source extraction circular region and a source-free background extraction annular region, adopting the same settings as used for the main sample \citep{LiuT2022_agn}.
The source and background spectra and the response files in the source regions are then extracted using \texttt{srctool}.
Before fitting the source spectra, each background spectrum is first fit using a model composed of components extracted from a Principal Component Analysis (PCA) of all the background spectra \citep[described in the appendix of][]{Simmonds2018}. This defines a background model specific to the given source. Then we fit the source and background spectra simultaneously in the 0.2-8 keV energy range (unless otherwise noted) using a source spectral model and the dedicated background spectral model. Few source photons are detected above 8~keV due to the decreasing effective area and consequent dominance of the instrumental background at higher energies. 
A Bayesian spectral fit is performed with \texttt{BXA}\footnote{\url{https://github.com/JohannesBuchner/BXA}} \citep{Buchner2014}, which connects \texttt{XSPEC} with the \texttt{UltraNest}\footnote{\url{https://github.com/JohannesBuchner/UltraNest/}} nested sampling package \citep{Buchner2021}. The robust MLFriends nested sampling algorithm \citep{Buchner2016,Buchner2019} implemented in \texttt{UltraNest} explores the model parameter space in a global fashion.

We adopt an absorbed power-law as the baseline model, and modify it according to the requirements of sources with various signal-to-noise ratio, either enriching it with additional components or decreasing the complexity by fixing certain parameters. The models are listed in Table~\ref{tab:models} and described in more detail by \citet{LiuT2022_agn}. Two double-component models (m2: power-law plus a softer power-law and m3: power-law plus a blackbody component) are applied to describe the potential soft excess. Both of these models are phenomenological in describing the soft excess, which may have different origins and thus spectral shapes in different sources. The soft--excess properties of this sample are discussed in further detail by Waddell et al. (2023). Without the soft--excess component (m1), the power-law slope $\Gamma$ should be considered as a general description of the broad-band spectral shape. With the soft excess component (m2 and m3), the primary power-law slope can be taken as an intrinsic property of the X-ray emitting corona.  The $\Gamma$-fixed-powerlaw model (m4) is the same as the single-powerlaw model (m1) but with $\Gamma$ fixed at 1.8. Model 5 (m5) is also the same as model 1 but fitting the 2.3--6~keV spectrum instead of the default 0.2--8~keV band.
In addition to the power-law-based models, a hot plasma model \citep[APEC;][]{Smith2001} (m0), which is appropriate for stars, is used to calculate fluxes of Galactic sources in this sample. 
In all the models, the Galactic absorption is considered, adopting the total Galactic column density measured from the HI4PI survey \citep{HI4PI2016} and the \citet{SFD} dust map using the empirical correlation presented by \citet{Willingale2013} \citep[see][for details]{LiuT2022_agn}.

All the models in Table~\ref{tab:models} are fitted to all the spectra. 
With each model, the observed and absorption-corrected fluxes in the 0.5--2 and 2.3--5~keV band and the intrinsic (absorption-corrected, rest-frame) fluxes and luminosities are calculated in the 0.5--2, 2--10, and a narrow band around 2 keV (1.999--2.001~keV) to provide a monochromatic luminosity at that energy.
In the case of the hard-band fitting (m5), we only calculate the fluxes and luminosities in the hard band ($>2$keV). In the case of the APEC stellar model (m0), we only calculate the observed fluxes. Then we choose the most appropriate luminosity measurements as described in \citet{LiuT2022_agn}. Unlike the main eFEDS sample that contains many extremely faint sources, almost all the sources in this hard sample have at least moderate S/N. The selection of luminosity measurements can therefore be simplified by omitting special treatment of the faintest sources with \texttt{NHclass} of "uninformative", since there are only two such cases. For faint sources with unconstrained $\Gamma$ \citep[selected based on Kullback-Leibler divergence as described in][]{LiuT2022_agn}, we adopt the "$\Gamma$-fixed-powerlaw" model (m4). For the other, brighter sources, we adopt the "powerlaw+blackbody" model (m3) for the unobscured ones and the "single-powerlaw" model (m1) for the obscured ones. The selection is discussed in detail in \citet{LiuT2022_agn}. It provides reasonable but not-uniformly-defined uncertainty measurements. One could also adopt the luminosity measurement using one particular model (e.g., m3) if desired. These are provided in the published catalogue (see Appendix A).

\begin{table*}[hbtp]
  \centering
  \begin{tabular}{lll}
    \hline
Name & Energy Range & Description \\
     & (keV) & \\
\hline
m0  & 0.2-8 & single-temperature APEC model with redshift$=0$, free kT \\
m1  & 0.2-8 & single-powerlaw, free $\Gamma$, $N_{\rm H}$ \\
m2  & 0.2-8 & double-powerlaw, free $\Gamma$s, $N_{\rm H}$  \\
m3  & 0.2-8 & powerlaw plus blackbody, free $\Gamma$, kT, $N_{\rm H}$\\
m4  & 0.2-8 & same as m1 but fixed $\Gamma=1.8$\\
m5  & 2.3-6 & same as m1 but in a narrower band \\
  \hline
    \end{tabular}
    \begin{center}
    \caption{X-ray spectral models applied to the hard sample. In m1, m2, and m3, a log-uniform prior is adopted for $N_\textrm{H}$ in the range $4\times 10^{19}\sim4\times 10^{24}$ cm$^{-2}$, and a Gaussian prior centered at 1.8 with $\sigma=0.5$ is adopted for $\Gamma$ in the range $-2\sim6$. The appropriate Galactic absorption is applied to all models. More details including other parameter prior ranges are described in \citet{LiuT2022_agn}.}
    \end{center}
    \label{tab:models}
\end{table*}

\subsection{Line Measurements and Black Hole Mass estimates}
\label{sec:pyqsofit}

For objects with broad optical/UV emission lines, the central black hole mass can be estimated \citep[e.g.,][]{Kaspi2000}. The high spectroscopic completeness means such estimates are possible for a large fraction of our sample. 

We achieved this using the PyQSOFit code \citep{Guo2018,WuShen2022}, a python package  for $\chi^2$ fiting of optical spectra of quasars. The model consists of several components: a continuum which comprises a power-law, a polynomial component and Balmer continuum; FeII emission in optical and ultraviolet regime; and emission lines that are fitted as single or multiple gaussians.

After host-galaxy decomposition via the PCA method of \citet{Yip_2004} and transformation into the rest-frame, the continuum model is fitted to predefined regions that do not contain strong emission lines. Accurate modeling of the FeII emission is of particular importance, and this was fitted using the following templates in the relevant wavelength range:
\begin{itemize}
    \item $1000-2200$ \AA: \citet{Vestergaard_2001}
    \item $2200-3090$ \AA: \citet{Salviander_2007}
    \item $3090-3500$ \AA: \citet{Tsuzuki_2006}
    \item optical: \citet{1992ApJS...80..109B}
\end{itemize}
The continuum and FeII model is subtracted from the data, leaving only the contribution from emission lines. The emission lines are then fitted using Gaussian functions; the maximum number of Gaussians was predefined for each line component (e.g. 3 Gaussians for broad H$\beta$, 1 gaussian for narrow H$\beta$). The output of the code consists of all parameters used to fit the spectrum and an error estimate for each of them based on Monte Carlo simulation.

Using the results, the black hole mass was calculated via:
\begin{equation}
    \log\left(\frac{M_{BH}}{M_\odot}\right)=A+B \log\left(\frac{\lambda L_\lambda}{10^{44}\text{erg s\textsuperscript{-1}}}\right) + C \log \left(\frac{\text{FWHM}}{\text{km s\textsuperscript{-1}}}\right)
\end{equation}
with the monochromatic luminosity $L_\lambda$ at wavelength $\lambda$, the full width at half maximum FWHM of the broad component of the emission line and the constants A, B and C. The black hole mass was calculated using the H$\beta$ and the MgII emission lines. For the H$\beta$ line, the constants of \citet{Vestergaard_2006} are used with the monochromatic luminosity at 5100\AA; for MgII, the calibration by \citet{Shen_2012} was selected together with the luminosity at 3000\AA. The error of the black--hole mass was calculated via error propagation of the Monte Carlo errors from the fit. These single-epoch black hole mass estimates also carry a systematic uncertainty of around 0.4 dex \citep{Shen2013}. Bolometric luminosities were calculated from the 5100 or 3000\AA\ monochromatic luminosities adopting the bolometric corrections from the SDSS quasar sample \citep{Richards06}. 

\section{Basic properties of the hard sample}
\label{sec:sample}

While eROSITA is significantly more sensitive in the soft band, the eFEDS data show that the instrument is also able to open up new parameter space for large-area, hard X-ray science compared to previous surveys. Histograms of the 2--10 keV flux and {\tt DET\_LIKE} of our sample are shown in Fig.~\ref{fig:hx_hflux}. The typical flux of the hard band sources peaks at around $(1-2) \times 10^{-13}$ erg cm$^{-2}$ s$^{-1}$. This is more than 2 orders of magnitude fainter than the HEAO-1 A2 all-sky hard X-ray sample of \cite{Piccinotti82} and the RXTE all-sky survey of \cite{Revnivtsev2004}, which have similar sensitivites below 10 keV.

A comparison of the soft X-ray (0.5--2) keV fluxes between the hard and main sample is shown in Fig.~\ref{fig:hx_main_sxhist}. The former samples typically much brighter X-ray fluxes, by more than an order of magnitude, given the higher flux limit overall. There is nonetheless a tail to faint soft X-ray fluxes, mainly due to the sources which are detected primarily in the hard band, some of which may be heavily obscured and therefore have weak soft X-ray emission.

\begin{table*}
\caption{Optical/IR counterpart quality (see text) comparing the hard and main samples, as well as the sources detected only in the hard band \label{tab:ctp}}                 
\centering          
\begin{tabular}{l | c | c c c c c | c}     
\hline\hline       
Sample & Total & \multicolumn{5}{c}{{\tt CTP$\_$quality}} & Secure \\
       &        & 0 & 1 & 2 & 3 & 4 & \% \\
\hline                    
Hard        & 246 & 4 & 11 & 3 & 0 & 228 & 92 \\  
Hard only        & 20  &  4   & 7 & 1 & 0 & 8 & 40 \\
Main        & 27,369 &  1225 & 1370 & 2552 & 1379 & 20837 & 81 \\
\hline                  
\end{tabular}
\end{table*}
\subsection{Optical properties}
\label{sec:results_optical}

Turning to the optical identifications, the numbers of counterparts of various types are shown in Table~\ref{tab:ctp}. For our hard sample, we find a very high fraction of good counterparts, with 232 sources ($\sim 94$\%) having a counterpart quality $\ge2$, significantly higher than for the main (soft) sample (81\%). This is most likely due to the fact that the typical main sample source is fainter and hence has a fainter counterpart and additional ambiguity in the association. 

\begin{table}
\caption{Source classifications for the 232 hard-band sources with ${\tt CTP\_quality}$ 2 or higher \label{tab:class} with comparison to classifications of the main sample.}             

\centering          
\begin{tabular}{l | c c c | c c c}     
\hline\hline       
Sample & \multicolumn{3}{c}{Extragalactic} & \multicolumn{3}{c}{Galactic} \\
       & Secure & Likely & All & Secure & Likely & All \\
\hline                    
Hard        & 181 & 36 & 217 & 14 & 1 & 15 \\  
Hard (\%)   & 78.0 & 15.5 & 93.5 & 6.0 & 0.4 & 6.4   \\
Main (\%)   & 48.4 & 46.6 & 95.0 &  3.3 & 1.7 & 5.0 \\
\hline                  
\end{tabular}
\end{table}

%
   \begin{figure}
   \centering
    \includegraphics[width=50mm]{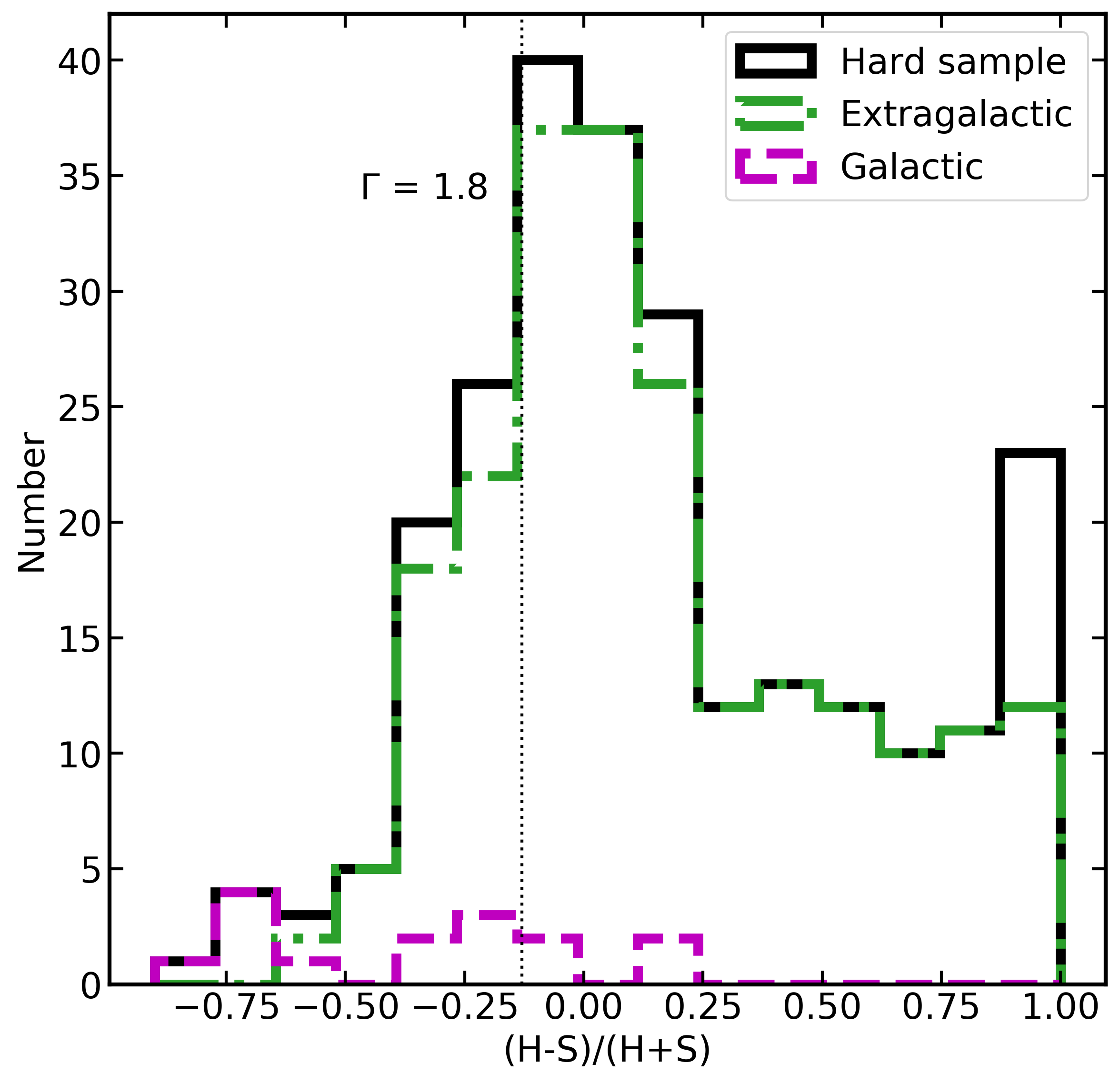}
      \caption{Histogram of the hardness ratios split between sources classified as Galactic and extragalactic, respectively. Note that sources without a high confidence counterpart are not assigned a class in this plot. The extragalactic sources show a tail to large hardness ratio which is suggestive of a population of obscured AGN.  Several of these are hard-band only detections with consequently high values of hardness ratio. The vertical dotted line indicates the hardness ratio expected for an absorbed power law with photon index $\Gamma=1.8$ and Galactic column density typical of eFEDS of $N_{\rm H} = 3 \times 10^{20}$ cm$^{-2}$.
              }
         \label{fig:hx_hr_class}

   \end{figure}

The assigned source classifications as described in Section~\ref{sec:optical}
are shown in Table~\ref{tab:class}. The hard sample has a higher fraction of secure classifications compared to the main sample, reflecting the higher spectroscopic completeness. The fraction of Galactic counterparts is also higher, most likely because on average the hard sample sources are brighter, and the stellar content of hard X-ray samples typically increases with flux. 

Fig.~\ref{fig:fx_fopt} shows a comparison between the X-ray flux and optical $R$ and mid-IR W1 magnitudes for the sample, compared to the equivalent data for a representative {\it Chandra} deep survey, AEGIS-XD \citep{Nandra2015}. The full optical magnitude range for our sample is extremely broad, spanning the range $R_{AB}=9-24$. The bulk of the AGN population in eFEDS covers a narrower range from $R_{AB}=15-22$. As is typical of X-ray surveys, a locus is traced by the AGN with a rough anti-correlation between X-ray flux and optical magnitude, with large scatter.  X-ray emitting stars are relatively optically-bright and can be distinguished by their low X-ray/optical ratios which stand out from the AGN locus. Interestingly, a few of the objects identified as Galactic show ratios more typical of AGN. These could be accretion-powered binaries with enhanced X-ray emission compared to normal stars. Conversely there are some extragalactic objects which are optically as bright as stars. 

\subsection{Hard-band only sources and catalogue integrity}
\label{sec:hbo}

As noted earlier, of the 246 sources in the hard sample, 226 are also detected and listed in the main, soft-selected X-ray catalogue of \citet{Brunner2022}. Thus 20 of our sources have no significant detection in the softer bands, hereafter referred to as hard-band only (HBO) sources. If real, these sources must be quite unusual and interesting spectrally, most likely being very heavily obscured, with their soft X-ray emission suppressed by photoelectric absorption hence evading detection in the highly sensitive eROSITA soft bandpass. 

On the other hand, the purity of our sample is not expected to be 100\%. Specifically, at our chosen likelihood threshold of {\tt DET\_LIKE\_3}$>10$, the simulations of \citet{LiuT2022_sim} suggest a spurious source fraction of approximately 2.5\%. Based on this we would expect around 6 of the 246 sources in the hard band not to reflect astrophysical X-ray emission and instead could be random background flucuations. These spurious sources are more likely to be found in the HBO subsample, as the detection in the softer band(s) for the others increases the probability that the X-ray emission has an astrophysical origin. It can be seen from Fig.~\ref{fig:hx_hflux} that the HBO sources tend to be clustered at low detection likelihood (albeit not at particularly faint hard X-ray flux) compared to the remainder of the sample. 

The suspicion that some of the HBO sources may be spurious is reinforced by the identification statistics. As can be seen in Table~\ref{tab:ctp} the fraction of the HBO sources without a valid counterpart is much higher than for the whole sample. Indeed, considering the 226 sources which are detected in both the hard and softer bands, $>98\%$ of them have a valid counterpart ({\tt CTP\_QUALITY}$\geq2$) while only $\sim 40\%$ of the HBO sources do. Indeed these statistics are entirely consistent with \textit{all} of the predicted $\sim 6$ spurious sources in our sample being HBO sources. On the other hand, the 9 HBO sources \textit{with} valid optical counterparts seem likely to be real, with the association of the optical source with the X-ray emission lending additional confidence to the X-ray detection itself. We note further that these sources all have an extragalactic classification, which is reassuring given that coronally-emitting stars are typically very strong soft X-ray emitters. 

   \begin{figure*}
   \centering
   \includegraphics[width=8cm]{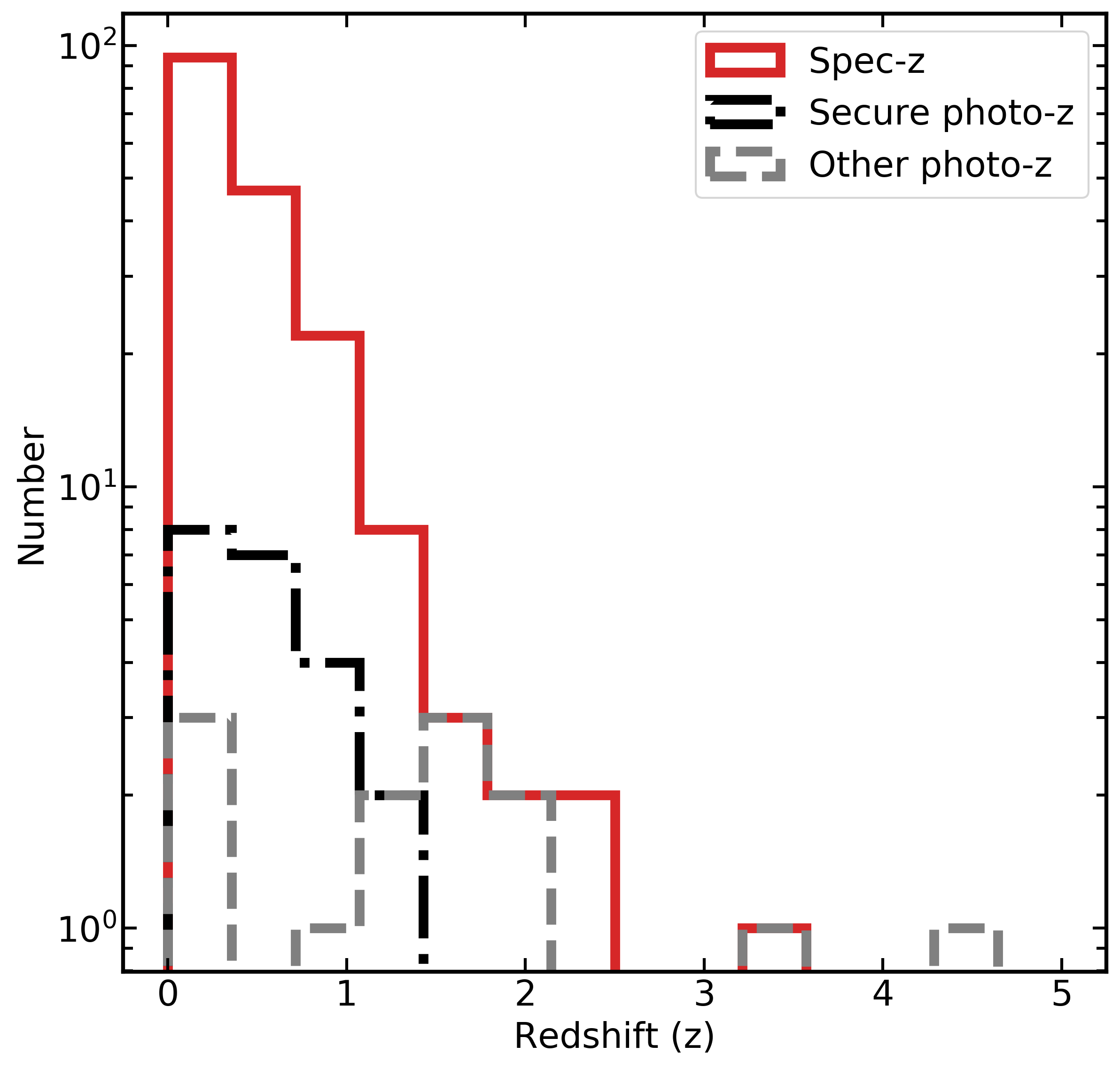}
   \includegraphics[width=8cm]{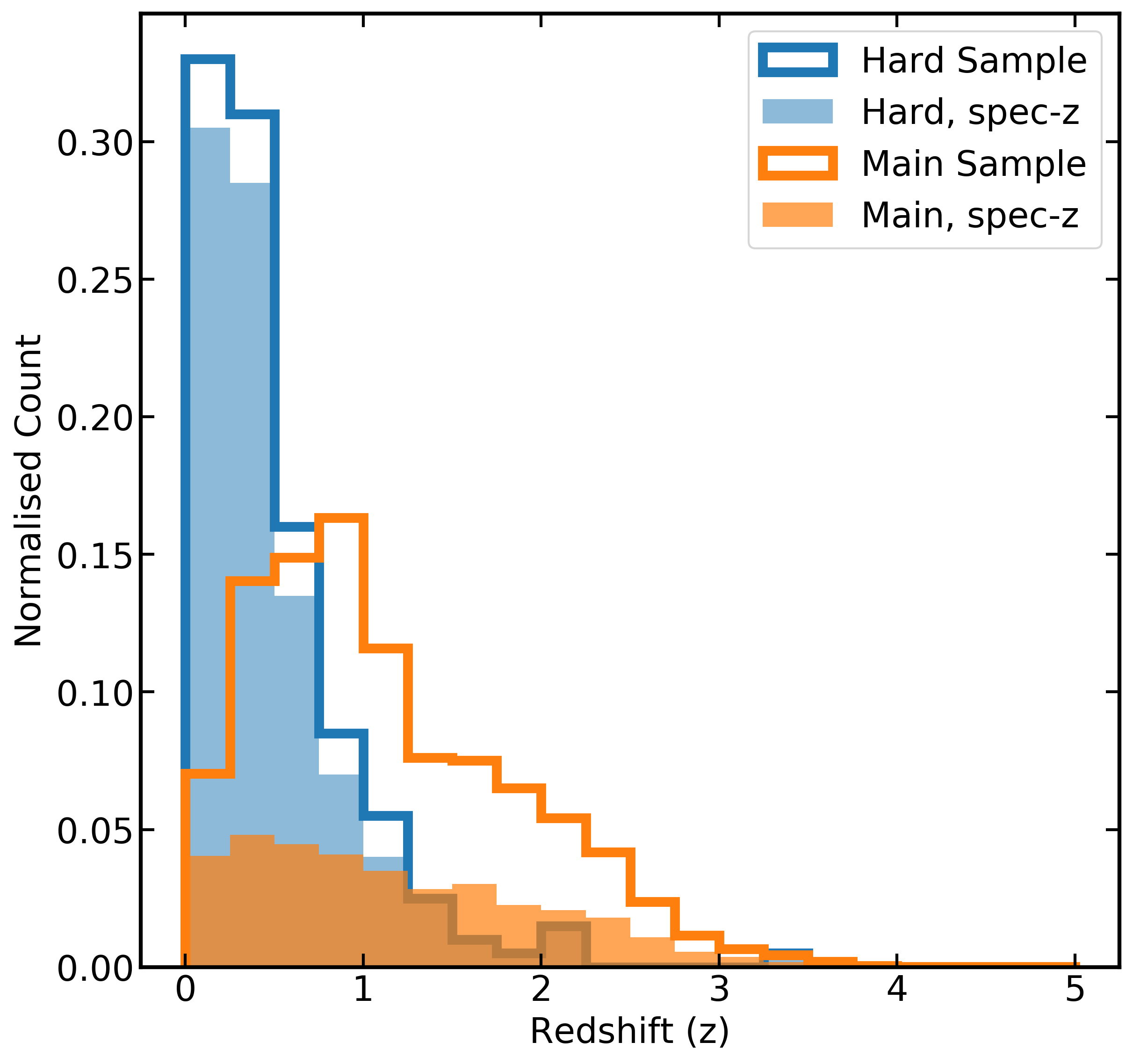}     
   \caption{(left panel) Redshift distribution for the extragalactic sample. Sources with spectroscopic redshift and high reliability photometric redshift are shown separately. (right panel) Normalised redshift histograms for spectroscopic and highly reliable photo-z comparing the hard and soft samples. The hard X-ray selection yields a relatively local AGN sample, with a median redshift $<z>=0.34$, compared to the main sample which has $<z>=0.94$.
              }
         \label{fig:xgal_zhist}
   \end{figure*}

\subsection{Basic spectral properties}

Fig.~\ref{fig:hx_hr_class} shows the hardness ratio distribution of our sample, split into Galactic and extragalactic sources for sources with reliable classifications. Both span a reasonably wide range of hardness ratio, which in part will reflect the uncertainties in the fluxes from which the HR is calculated. The extragalactic sources show a peak hardness ratio corresponding to an unobscured power law with $\Gamma \sim 1.7-1.8$, typical of AGN, but it can be seen clearly that the extragalactic sources show a tail of hard X-ray sources which is not present in the stellar sample. This is entirely expected because the extragalactic sample will contain some fraction of obscured AGN (see Section~\ref{sec:obscuration}), while the stars will generally show soft emission. A very hard X-ray source with a stellar identification would tend to be indicative of an absorbed binary system, where the ID is of a star, and the X-ray emission from accretion onto a compact companion. 

\section{The AGN sub-sample}
\label{sec:agn}

In the remainder of this paper we restrict our discussion to a subsample of the sources which meet all the following criteria: 

\begin{itemize}
\item Are contained within the 90\%-area region\footnote{This exposure depth thresholding results in 90\% of the total area of eFEDS excluding the edges of the field, which have a strong exposure gradient \citep{Brunner2022}, and provides a relatively clean selection function for subsequent statistical analysis such as the X-ray luminosity function (Buchner et al., in preparation)}, where the exposure is at least 500s (241/246 sources).
\item Have counterpart quality 2 or higher (228/241 sources).
\item Are considered secure or likely extragalactic sources (213/228 sources)
\end{itemize}

This sample is likely to consist overwhelmingly of AGN and we examine their properties in further detail below.

\subsection{Redshifts and Luminosities}
\label{sec:zandl}

The compilation of spectroscopic redshifts in this field is summarised for the hard sample in Table~\ref{tab:specz}. A total of 179 out of 213 candidate extragalactic sources have a secure spec-z. On top of this spectroscopic sample, an additional 21 sources have a highly reliable photometric redshift determination. We focus our further attention on this subsample of 200 sources, excluding the 13 where the redshift determination is less certain. Fig.~\ref{fig:xgal_zhist} (left panel) shows a histogram of the redshift distribution separating the sources into these three categories.  The source with the highest reliable redshift measurement in the hard X-ray sample is at $z=3.2$, spectroscopically confirmed and likely a blazar (See Section~\ref{sec:blazars}). Two objects show high photometric redshift estimates ($z>3$), but come from the subsample with less reliable redshifts, and we do not consider them to be secure. Henceforth, unless otherwise noted, we plot and consider only the 200/213 (94\%) sources which have a spectroscopic or highly reliable photometric redshift. 

The right panel of Fig.~\ref{fig:xgal_zhist} shows the hard sample redshift distribution compared to that of the main sample. The hard selection clearly samples a significantly lower redshift range than the main sample, with a median redshift $<z>=0.34$ for the former compared to $<z>=0.94$ for the latter. The eROSITA hard selection thus provides an interesting low-$z$ AGN sample for comparison to the higher redshift populations seen in deeper surveys (see Section~\ref{sec:bhmass}). Together with the Swift-BAT survey \citep{Koss2022a} eFEDS thus provides information about the low redshift X-ray selected AGN population, which is poorly sampled by deep X-ray surveys due to their limited area and hence sampled cosmological volume at low-z.

\begin{figure}
   \centering
   \includegraphics[width=\hsize]{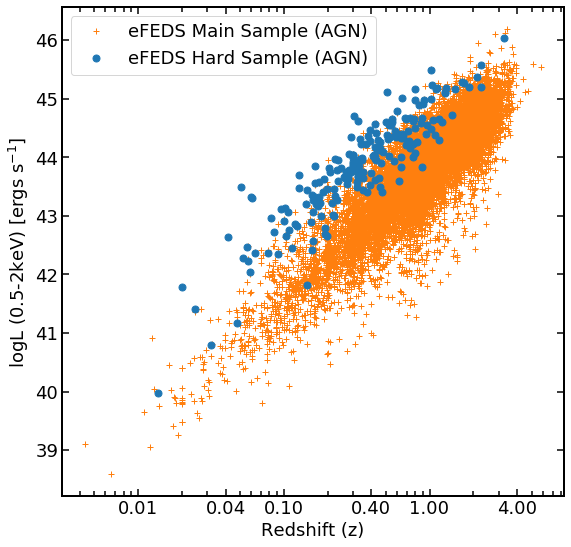}
      \caption{Soft X-ray (0.5--2 keV) luminosity-redshift relation for the eFEDS hard sample compared to the main sample. The luminosities are corrected for absorption and show the subsamples with either spectroscopic redshift or high quality photometric redshift (see text). 
              }
         \label{fig:hx_main_lz}
   \end{figure}

Luminosities for our sample have been derived from the X-ray spectral fits, as described above and in \citet{LiuT2022_agn} and are thus absorption-corrected. The luminosity-redshift relation in Fig.~\ref{fig:hx_main_lz} is shown in the soft X-ray (0.5--2 keV) band to facilitate comparison to the main sample. The vast majority of the sample covers the range $\log L_{\rm X}=42-45$ with a few lower luminosity sources at very low redshift and the ``standout'' $z=3.2$ object at $\log L_{\rm X}=46$ being amongst the most luminous even in comparison to the main sample. As expected, the hard selection samples higher luminosities at a given redshift, given the higher flux limit compared to the main sample.

\subsection{Optical spectroscopic classifications}
\label{sec:spectroscopy}

For the sub-sample of our objects which have optical spectroscopy, we are able to provide a basic optical spectral classification. This is most easily achieved by fitting objects with SDSS spectra for which we have access the spectra themselves. We have visually inspected the 172 spectra with the primary intention of identifying objects with broad lines, predominantly $H\alpha$, $H\beta$, MgII or CIV. A total of 159 objects ($\sim$92\% of the sample) show evidence for broad emission in at least one of these lines, with the majority of these having prominent broad $H\beta$, MgII and/or CIV. A subset of these objects (25 in total), however, show only broad wings to H$\alpha$, sometimes marginally visible, and with no evidence for broad H$\beta$ which we henceforth consider them as optical type 1.9, following the definition thereof by \cite{Koss2022a}. Of the 13 objects that show no visual evidence for broad emission in any line, 12 show narrow emission lines and one is an absorption-line galaxy. These objects will be either type 2 AGN or host-galaxy dominated.  Thus, despite the hard X-ray selection being sensitive to obscured objects, more than 90\% of our sample appear to be optical broad-lined AGN of some kind.   

\subsection{X-ray spectral properties}
\label{sec:xrayspec} 

   \begin{figure}
   \centering
   \includegraphics[width=70mm]{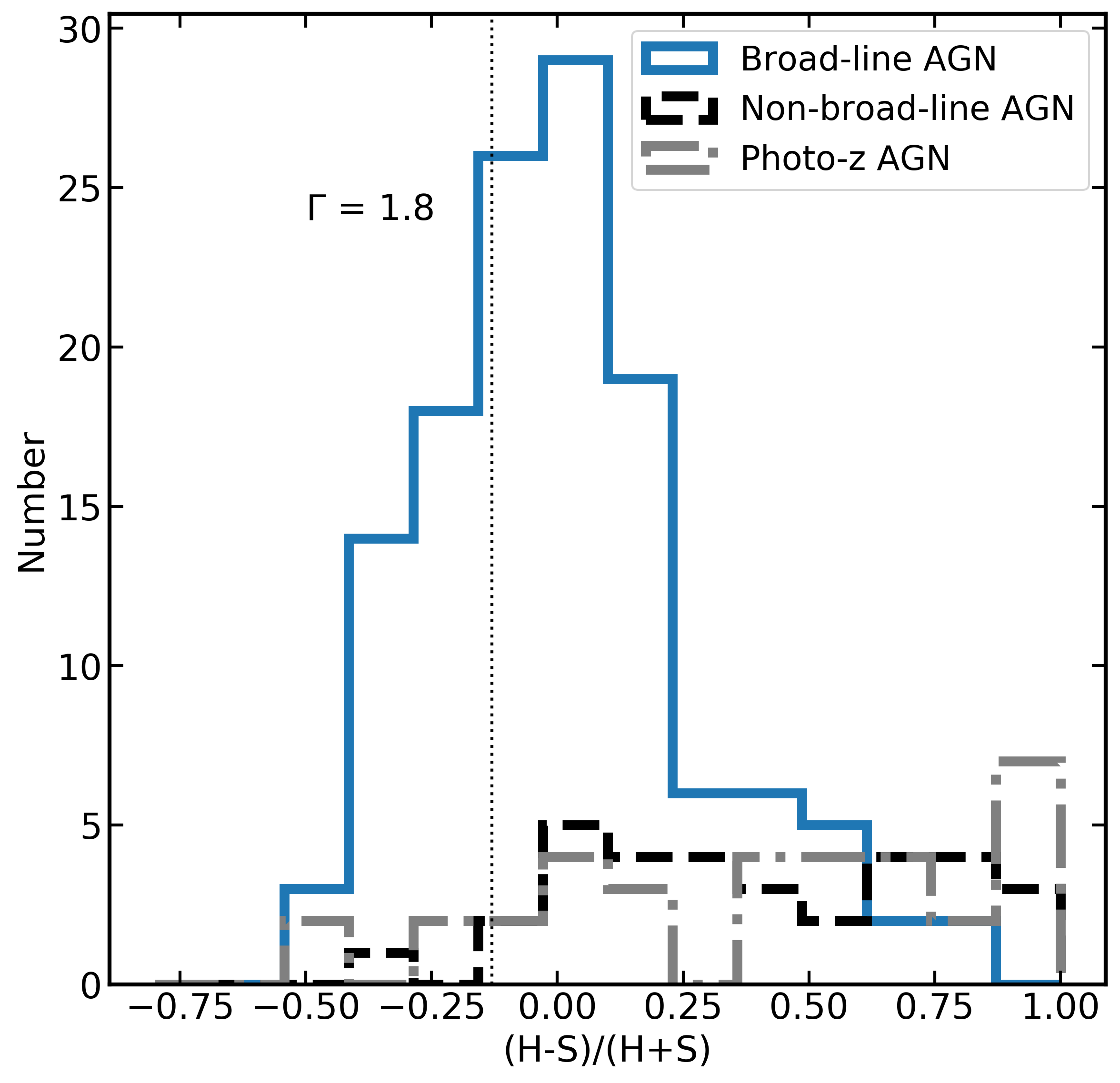}
     \caption{Hardness ratio distribution for the extragalactic/AGN subsample as a function of spectroscopic classification from SDSS. The vertical dotted line indicates the hardness ratio expected for an absorbed power law with photon index $\Gamma=1.8$ and Galactic column density typical of eFEDS of $N_{\rm H} = 3 \times 10^{20}$ cm$^{-2}$. Objects with broad optical lines (Type 1 classification) dominate the sample, but some of these show large hardness ratios indicative of X-ray absorption. AGN without broad lines or without spectroscopic classifications show systematically larger values of the hardness ratio, indicating that they are obscured. 
              }
         \label{fig:agn_hr_class}
   \end{figure}

The hard X-ray selection and relative brightness of the current sample enable analysis of the spectra with relatively complex models. In particular, the existence of significant hard X-ray counts makes it much easier to infer whether there is line-of-sight absorption of the soft X-rays, or if there is excess emission at low energies above the extrapolation of the power law continuum in the hard X-ray band (a.k.a the ``soft excess''). As a first test, we plot in Fig.~\ref{fig:agn_hr_class} the distribution of hardness ratios for three different classes based on the SDSS spectrum. We separate the objects with a measured broad optical line and those without, and plot in addition those sources without optical spectroscopy. 

The sources without an optical broad line detection are more likely to exhibit large hardness ratios indicative of absorption, as are those objects without a spectroscopic redshift. Broad-line objects can also nonetheless exhibit large hardness ratios, and are discussed in Section~\ref{sec:obstype1}.
We note that the HR is a relatively crude measure of the spectral properties and turn our further attention to the physical properties based on spectral fitting. 

\subsubsection{Continuum properties}
\begin{figure}
\centering
\includegraphics[width=0.98\linewidth]{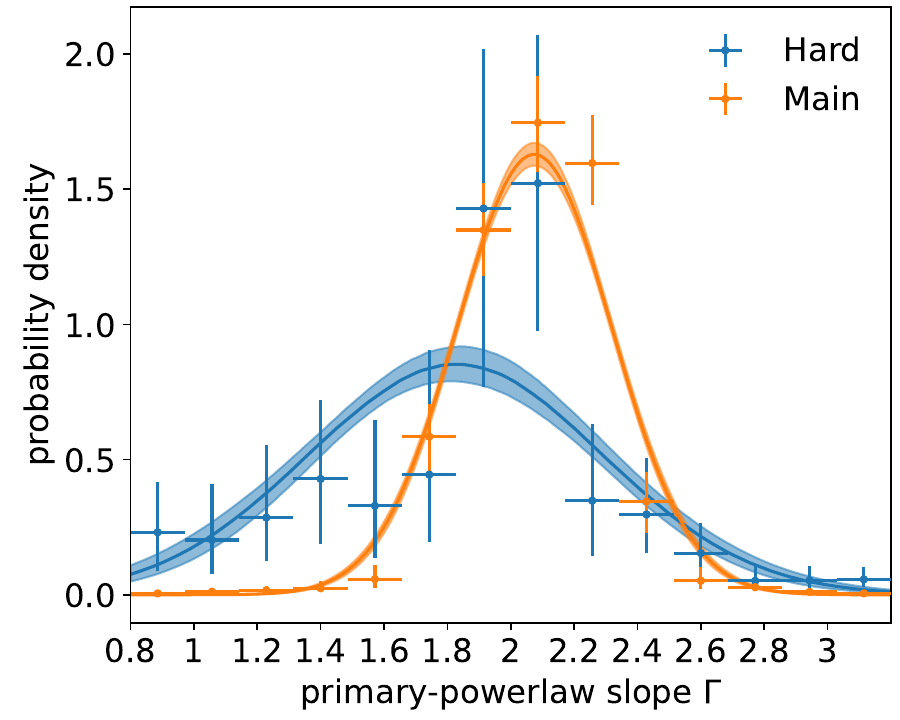} %
\caption{Distributions of the primary-powerlaw slope in the "power-law plus blackbody" model (m3) for the hard (blue) and main (orange) AGN sample, obtained using the HBM method adopting histogram (points with 1-$\sigma$ error bar) and Gaussian model (lines with 1-$\sigma$ range).
\label{fig:gammadist}}%
\includegraphics[width=0.98\linewidth]{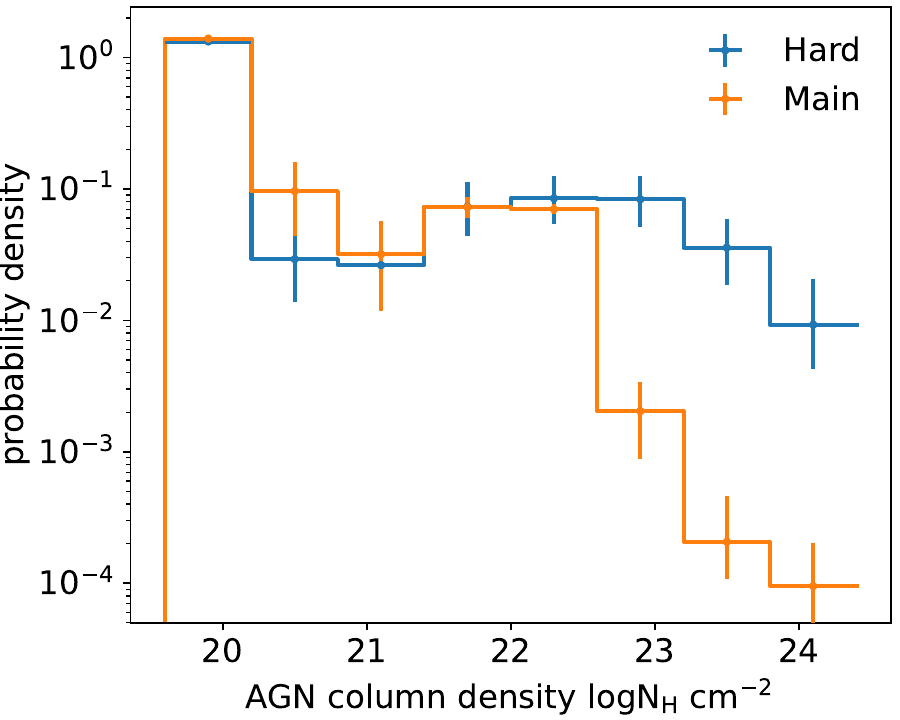}
\caption{Distributions of the obscuring column density in the "single-powerlaw" model (m1) for the hard AGN sample (blue) and the main AGN sample (orange), obtained using the HBM method. The 1-$\sigma$ error bar is plotted on the histogram.
\label{fig:nhdist}}
\end{figure}

The photon index distribution of the hard sample was determined based on a fit to the spectrum in the full energy range with the "powerlaw+blackbody" model (m3 in Table~\ref{tab:models}). The posterior distribution of $\Gamma$ is obtained for each source. We combine the posteriors of all the ($209$) AGN with good redshifts (redshift measurement quality \texttt{CTP\_REDSHIFT\_GRADE}$\geqslant3$) with a Hierarchical Bayesian Model (HBM) to obtain a sample distribution. Following the approach described in \citet{Baronchelli2020} and \citet{LiuT2022_agn}, we adopt both a Gaussian model with a mean $\mu$ and a standard deviation $\sigma$ and a non-parametric, histogram model for the sample distribution. The posterior distribution of the parameters of the sample distribution can be inferred using the PosteriorStacker tool\footnote{\url{https://github.com/JohannesBuchner/PosteriorStacker}}, which uses the nested sampling code \texttt{UltraNest}.
The hierarchical model implicitly allows poorly constrained posteriors to be informed by better-constrained posteriors, as the same model is assumed to describe the distribution for all objects.
Fig.~\ref{fig:gammadist} shows the photon index distribution comparing the main and hard samples. 

The main sample, selected in the soft band, has inferred Gaussian distribution centered at $\Gamma=2.07\pm0.006$ with a intrinsic dispersion (accounting for measurement errors) of $0.24\pm0.006$ \citep{LiuT2022_agn}. For the hard sample, the Gaussian model is found to be centered at $1.83\pm0.04$, significantly harder than the main sample, with a larger dispersion of $0.47\pm0.04$. This is easily understood as sources with very steep intrinsic continua are likely to be missed by the hard selection and vice versa. 
The histogram model shows that the $\Gamma$ distribution peaks at $1.9-2$ and has a tail at lower $\Gamma$ values. This low-$\Gamma$ tail may not reflect the true intrinsic distribution and instead be due to additional spectral complexity (e.g., complex or ionized absorption) that is not considered in our modelling \citep[but see][]{Waddell2023}. Hence the mean intrinsic index in the hard sample may be somewhat softer than that indicated by the mean of the Gaussian HBM. 

\subsubsection{Obscuration}
\label{sec:obscuration}

%
%

We use the same HBM method as described above to infer the sample distribution of AGN column density $N_\textrm{H}$ from the posteriors measured with the "single-powerlaw" model (m1) of all the AGN with good redshifts. We adopt only the histogram model and not the Gaussian model because in practice the shape of the distribution is very different from a Gaussian. We use the "single-powerlaw" model rather than the double-component models because in the obscured cases a soft--excess component can be strongly degenerate with the AGN obscuration, and our data quality does not allow constraints on both simultaneously.
Figure~\ref{fig:nhdist} compares the $N_\textrm{H}$ distribution for the main and hard samples.
 In both cases, the distributions show a dominant bin at the lower boundary $\sim10^{20}\mathrm{cm}^{-2}$, i.e. the majority of sources in both hard and main samples are unobscured.
 
Both the main and the hard sample $N_\textrm{H}$ distribution show a bi-modality. This can be quantified by drawing a line between the fractions in the $\sim10^{20}\mathrm{cm}^{-2}$ bin and the $\sim10^{22}\mathrm{cm}^{-2}$ bin, and noticing that both the $\sim10^{20.5}\mathrm{cm}^{-2}$ and the $\sim10^{21}\mathrm{cm}^{-2}$ bins are $10\sigma$ lower, indicating a dip and thus multi-modality  \citep{Hartigan1985DipTest}. Therefore, we have evidence that there is a low-column density mode $<10^{20.5}\mathrm{cm}^{-2}$ and a high-column density mode $\sim10^{22}\mathrm{cm}^{-2}$. The latter can be associated with nuclear or host galaxy obscuration \citep{Buchner2017a,Buchner2017b}.

The distributions diverge at high $N_\textrm{H}$.
The main sample, selected in the soft band (0.2--2.3~keV), has a very low fraction of obscured sources with $N_\textrm{H}>10^{22}$ cm$^{-2}$ \citep[4\%;][]{LiuT2022_agn} in the HBM-inferred distribution, reflecting the strong selection bias against obscured sources.
For the hard sample, the fraction above $10^{22}$ cm$^{-2}$ is much higher (13\%).
Even more highly obscured sources with $N_\textrm{H}>10^{23}$ cm$^{-2}$ are almost absent from the main sample, while in the hard sample the numbers are significant, with a probability of around 4\% of having such a high $N_\textrm{H}$. This is roughly in line with expectations. At the median redshift of $z=0.34$ a column density of $10^{22}$~cm$^{-2}$ only suppresses the 2.3--5~keV flux by around 5~per cent, whereas the flux is reduced by 40~per cent at a column of $N_{\rm H}=10^{23}$~cm$^{-2}$, which will start impacting significantly on source detectability.  

To define an ``obscured'' subsample, we adopt the same definition as \citet{LiuT2022_agn}.
Based on the posterior PDF for N$_\textrm{H}$ obtained with the single-powerlaw model (m1), \citet{LiuT2022_agn}, we divide the N$_\textrm{H}$ measurements into four classes (\texttt{NHclass}): 1) uninformative, 2) unobscured, 3) mildly-measured, and 4) well-measured. The posterior median N$_\textrm{H}$ value (\texttt{lognH\_Med\_m1}) is invalid in the first two cases. Following \citet{LiuT2022_agn}, we adopt the criterion \texttt{NHclass}$\geqslant3$ and \texttt{lognH\_Med\_m1}$>21.5$ and select $36$ obscured AGN, which comprises 18\% of the hard AGN sample ($200$). The largest measured value us \texttt{lognH\_Med\_m1}$=24.0$

\subsubsection{Soft excesses}

The hard-selected sample has by selection a defined hard X-ray continuum and, combined with the high soft X-ray sensitivity of eROSITA in the soft X-ray band, is ideal to search for so-called "soft excess" emission. The origin of this component is still debated, as it can arise e.g., from a second, cooler coronal component, ionized disk reflection, or even be an artefact of complex soft X-ray absorption. In the present work we restrict ourselves to relatively simple modelling of the soft excess emission.

The standard fits shown in Table~\ref{tab:models} include both a double power law (model 2) and a power law plus black body. We can therefore identify sources with a soft excess by comparing the Bayesian evidence of these models to a single power law. Specifically we use the criterion 
 $logZ_{m2}-logZ_{m1}>1$, which selects 17 sources better fit with a double power law, or $logZ_{m3}-logZ_{m1}>1$ which yields 7 sources better fit with a powerlaw plus blackbody. The 7 sources with apparent "blackbody" soft excesses are a subset of the 17 power law soft excesses, indicating that the latter is a better model and more sensitive for soft excess detection. Sources with a soft excess comprise about 10\% of the AGN/extragalactic sample.

   \begin{figure*}
   \centering

    \includegraphics[width=7.5cm]{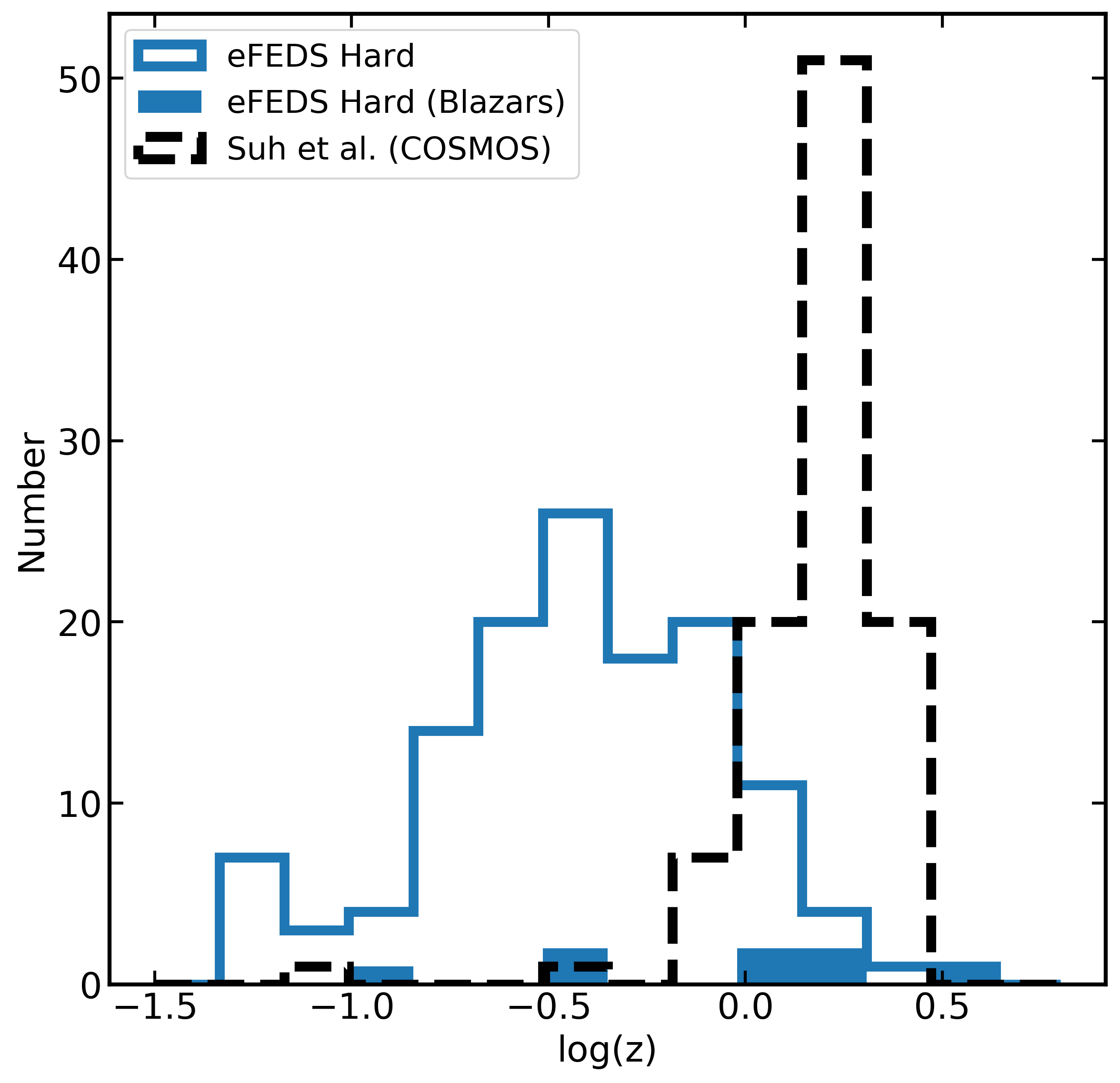}
    \includegraphics[width=7.5cm]{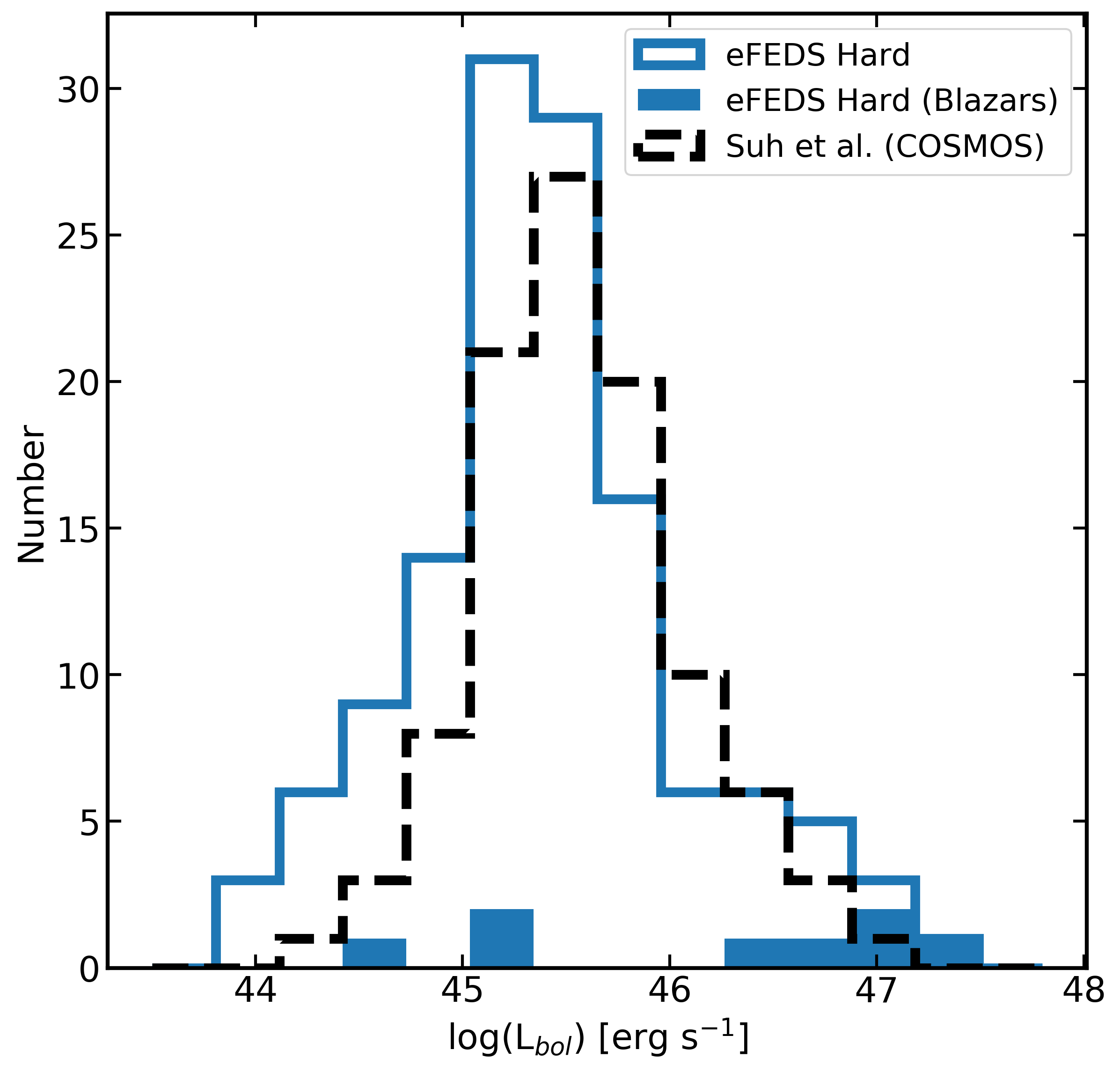}
    \includegraphics[width=7.5cm]{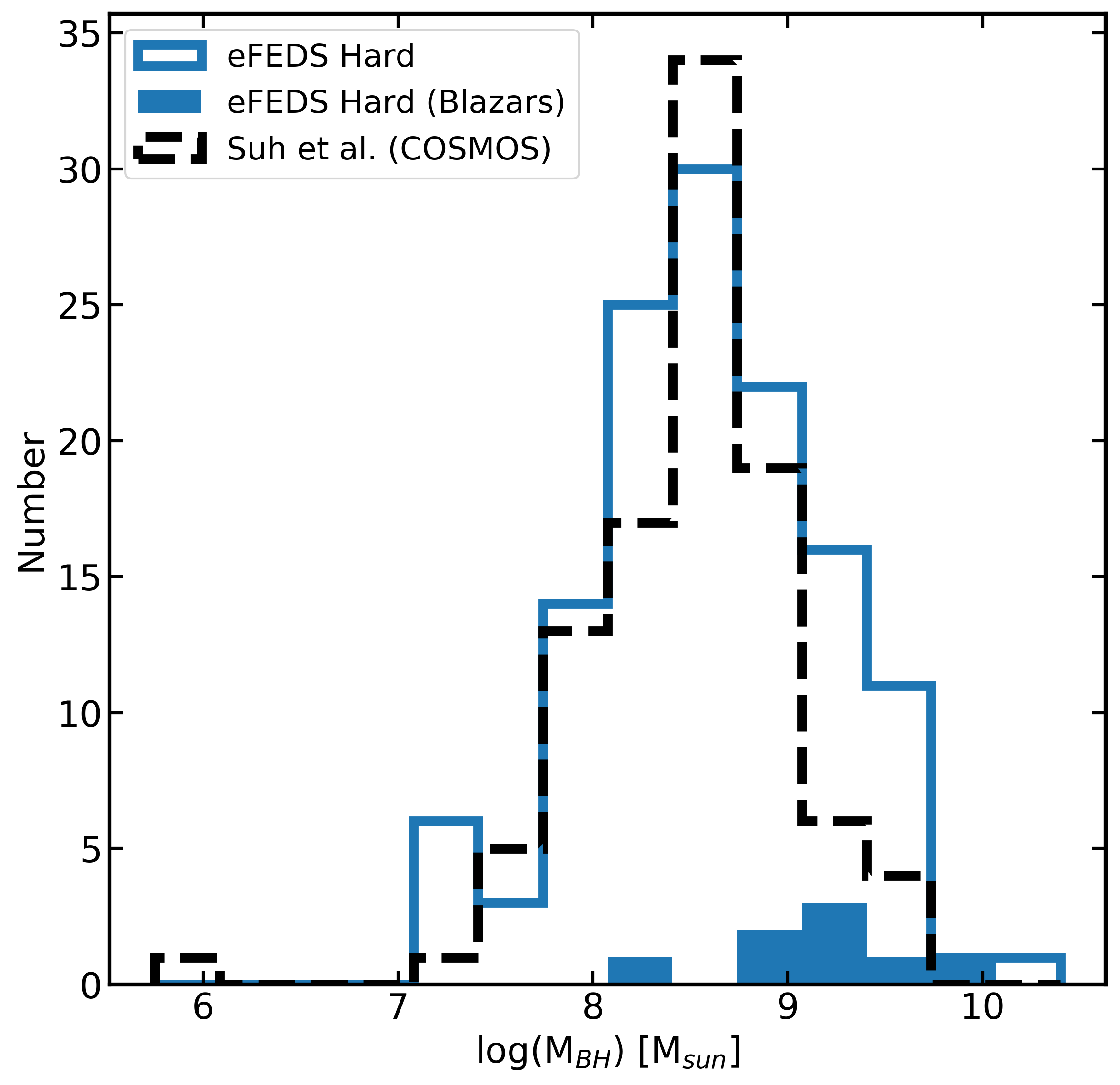}
    \includegraphics[width=7.5cm]{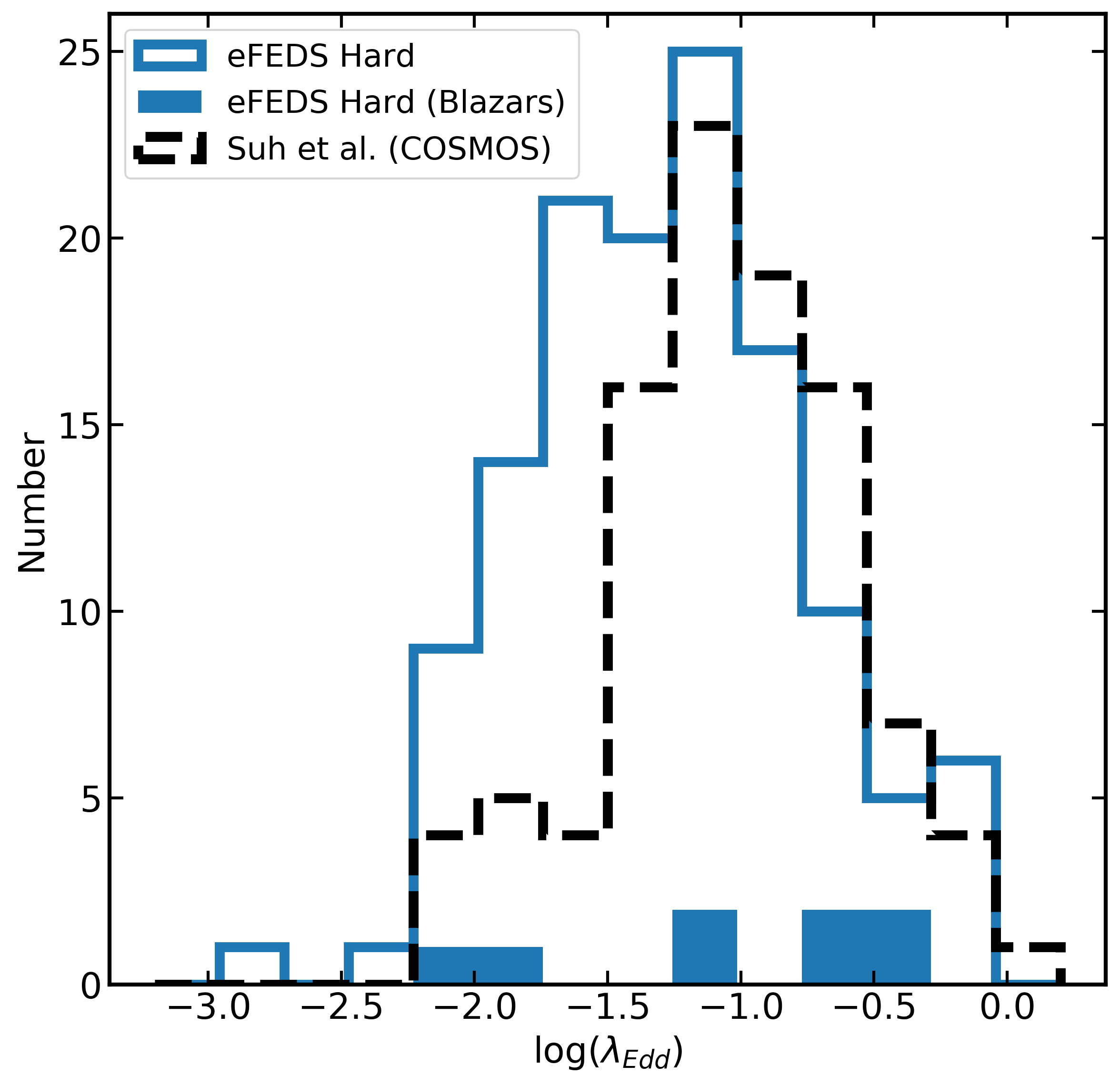}
      \caption{Distribution of the redshift (top left), bolometric luminosity (top right), black hole mass 
      (bottom left) and Eddington ratio (bottom right) for spectroscopically confirmed type 1 AGN in the eFEDS hard sample, shown in blue. The dotted black histogram shows a comparison sample from the {\it Chandra} deep survey in the COSMOS field \citep{Suh2020}.   
              }
         \label{fig:bhmass_cosmos}
   \end{figure*}

\subsection{Black hole masses and Eddington ratios}
\label{sec:bhmass}

As discussed above, black hole masses have been estimated using the single-epoch virial method for the objects in our sample exhibiting a broad H$\beta$, MgII or C{\sc iv} line, in other words the spectroscopic type-1 AGN. 
For the 134 AGN which show broad emission at H$\beta$ or blueward thereof, we calculate the black hole masses and other derived parameters with optical spectral fitting using pyQSOFit (see Section~\ref{sec:bhmass}. We do not consider H$\alpha$ based masses \citep[e.g.,][]{GreeneHo2005} because there are indications that these may be biased and/or unreliable in Seyfert 1 type AGN.

Of the 200 AGN in the sample with reliable redshift, 179 are spectroscopic. We restrict our black holes analysis to the 172 objects with spectra originating from SDSS, to which we have full access and for which we have optical spectral fitting results based on PyQSOFit as described in \cite{WuShen2022} and Section~\ref{sec:pyqsofit}.

As described there, for type 1 objects the black hole masses and bolometric luminosities can be estimated using the line widths and optical/UV continuum, and the accretion rate relative to Eddington can also be estimated from these quantities. As we are interested in the latter derived quantity, we follow Waddell et al. (2023) in sub-selecting only objects with a constrained Eddington ratio. This results in a subsample of 154 sources. The distributions of redshift, black hole mass, bolometric luminosity and Eddington ratio for this BH mass subsample are shown in Fig.~\ref{fig:bhmass_cosmos}. This subsample has a median redshift of $z_{\rm med}=0.39$ and median luminosity $\log L_{\rm bol,med}=45.3$. 

The black hole masses cover $\log M_{\rm BH}=7-10$, peaking in the middle of this range with a median $\log M_{\rm BH,med}=8.6$. The objects with Mg II-derived masses are at the top end of the mass range. This is unsurprising, as they are the highest redshift and hence amongst the highest luminosity sources in this flux-limited sample. The Eddington ratios cover a very broad range, from $\sim 10^{-3}$ up to the Eddington limit. Certainly, eFEDS samples objects at the high ends of both the mass and accretion rate distributions. We note, however, that there are significant uncertainties on our estimated quantities, so the most extreme cases should be treated cautiously. In addition, the objects with the highest bolometric luminosities in the eFEDS sample are the handful of blazars (see Section~\ref{sec:blazars} and, e.g., Fig.~\ref{fig:bhmass_cosmos}). We caution that in these objects, the optical/UV continuum may be contaminated by jet emission which could lead to an overestimate of the bolometric luminosity and quantities derived therefrom (mass and accretion rate). As there are relatively few of these in our sample, however, we do not consider this a major effect. 

To place the eFEDS results in context, we show in Fig~\ref{fig:bhmass_cosmos} the corresponding distributions from the work of \cite{Suh2020}, from the {\it Chandra} COSMOS legacy survey. The latter samples a significantly higher redshift range, with $z_{\rm med}=1.58$, but due to the higher sensitivity of the deep {\it Chandra} observations in this region the two surveys sample a similar limiting luminosity at their median redshift, around $\log L_{\rm 2-10}=44$ in the rest frame. It is therefore of interest to compare the two samples in relation to possible evolution of the AGN population. 

From Fig~\ref{fig:bhmass_cosmos} we can see that the black hole mass range in the two samples is rather similar, covering the same range and with a remarkably similar median, with $\log M_{\rm BH,med}=8.55$ in COSMOS compared to 8.63 for non-blazars in eFEDS. A K-S test reveals no significant difference between the distributions, with false probability $p=0.08$. In contrast, the Eddington ratio distributions are dramatically different with the K-S probability of them being the same $<4 \times 10^{-5}$ (once again excluding blazars). The median Eddington ratio for eFEDS ($\lambda_{\rm Edd,med}=0.03$) is about three times lower than that in COSMOS ($\lambda_{\rm Edd,med}=0.09$).

\subsection{X-ray absorbed type 1 AGN}
\label{sec:obstype1}

   \begin{figure}
   \centering
   \includegraphics[width=\hsize]{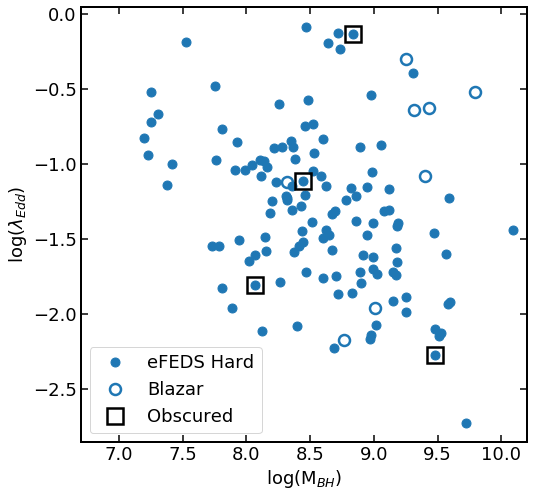}
      \caption{Black hole mass versus Eddington ratio for the type 1 AGN sample. Type 1 AGN that show evidence for significant absorption, having an NHclass of 3 or 4 and a column density greater than 10$^{21.5}$~cm$^{-2}$, are indicated with black squares.
              }
         \label{fig:agn_mledd}
   \end{figure}

With a combination of relatively deep and sensitive soft X-ray data, as well as comprehensive follow-up optical spectroscopy, we are able simultaneously to classify the objects in our sample optically and measure their absorption in the X-ray. A particularly interesting population in this regard are X-ray obscured type 1 AGN \citep[e.g.,][]{Page2001,Brusa2003,Page2011}, i.e., objects which show broad optical lines but exhibit soft X-ray absorption. The demographics of this population have been explored in, e.g., \citet{Merloni2014}. 

In the very simplest unification scenarios, this is not expected, because the same gas which absorbs the X-rays should contain dust which extinguishes the optical broad lines. On the other hand, there is evidence that the obscuring material is clumpy \citep[e.g.,][]{Nenkova2008,Markowitz2014}, that the broad line clouds can have a complex geometry \citep{Maiolino2010} and that there may be substantial departures from normal gas-to-dust ratios \citep{Maiolino2001}. The broad lines and AGN continuum may also be seen only in scattered light \citep{AntonucciMiller1985,Alexandroff2018,Assef2020}. An interesting possibility is that the X-ray absorbing gas is dust-free, which would be the case if it is close to the central engine within the dust sublimation radius. If so, the gas may be significantly ionized and in the form of ``warm absorbers'' \citep{Halpern84,Yaqoob89} which are, in turn, associated with ionized winds \citep[e.g.,][]{Kaastra2000}. Evidence for significant X-ray absorption in type 1 AGN has also been found in broad absorption line quasars \citep[e.g.][]{Gallagher2006,Gibson2009}, which have direct evidence for strong outflows.

Our hard X-ray sample is dominated by broad-line AGN, but nonetheless contains a number of objects with significant soft X-ray absorption. Considering the 154 objects in the "black hole mass" sample as defined in Section~\ref{sec:bhmass} - which by definition must have a visible and constrained broad optical line - we find that 4 also satisfy the criterion of being obscured by a column of $\log N_{\rm H}>21.5$ as defined in Section~\ref{sec:obscuration}. These obscured type 1 AGN are identified in black hole mass-accretion rate parameter space in Fig.~\ref{fig:agn_mledd}. Naively, one might expect these absorbers to be seen in AGN with high accretion rates where, close to the Eddington ratio, radiation pressure drives an outflow \citep[e.g.,][]{Fabian2008,Ricci2017}. In practice they show a wide range of Eddington ratios. None of the objects shows BAL signatures in the Sloan spectra, but but only one (at $z$=2.25) is within the redsfhit range where classical UV BAL signatures would be evident. 

In addition to the black hole mass subsample, as noted in Section~\ref{sec:spectroscopy}, Seyfert 1.9 galaxies comprise a substantial subset of our sample. Several of these (6/25) also show significant soft X-ray absorption but this is unsurprising given the evidence for line-of-sight obscuration based on the suppression of optical broad lines blueward of H$\alpha$. 

\subsection{Blazars}
\label{sec:blazars}

Due to the large area of the eFEDS survey it is sensitive to rare AGN such as those at the highest luminosities. In some fraction of these, the luminosity is enhanced due to the presence of a relativistic jet whose emission is Doppler boosted for sightlines close to the jet axis. Such blazars can be identified via strong radio and/or Gamma-ray emission. A more detailed study of the blazars in eFEDS is given in Collmar et al. 2024 (in preparation), but here we highlight the sources in the hard X-ray sample which are likely to host strong emission from a relativistic jet. Confirmed blazars and blazar candidates have been identified by cross-correlating the hard sample with catalogues of blazars identified via their radio and/or gamma-ray emission. 

As input catalogs we used the following: the Fermi Large Area Telescope Fourth Source Catalog \citep[4GL;][]{Abdollahi2020}; the ROMA-BZCAT Multifrequency Catalog of Blazars \citep{Massaro2015}; the 3HSP catalog of extreme and high-synchrotron peaked blazars \citep{Chang2019};the Swift-XRT catalog of blazars \citep{Giommi2019}; the Blazar Radio and Optical Survey \citep[BROS;][]{Itoh2020} and the WISE Blazar-like Radio-Loud Sources catalog \citep{DAbrusco2019}. 

To select blazars or blazar candidates we searched for a positional coincidence of $<15$ arcseconds and cross-checked the identifications with the eFEDS counterpart catalog of \citet{Salvato2022}. 

There are 15 such objects in our sample, as detailed in Table~\ref{tab:blazar}, of which two are outside the 90\% area region, both BL Lac type objects. Of the remaining 13, 7 are flat spectrum radio quasars, all spectroscopically confirmed, two are BL Lac type objects and four are candidate blazars or of unknown type. Our own visual examination of the SDSS spectra largely confirms the literature classifications. The four BL Lac type objects show largely featureless blue continua, with only one (4FGL J0831.8+0429) showing weak lines sufficient for a redshift determination. The Flat Spectrum radio quasars generally exhibit QSO broad emission lines, but two show optical properties indicating obscuration, with 5BZQ J0924+0309 having the optical spectrum of a Seyfert 1.9 and WIBRaLS2 J092203.20-004443.3, listed as a FSRQ candidate, having narrow lines indicative of a type-II QSO classification. A final noteworthy object spectroscopically is 3HSPJ093303.0+045235, which presents as an absorption-line galaxy with no AGN emission lines, but a possibly enhanced blue continuum. This is plausibly a BL-Lac type object in which the non-thermal jet emission is subdominant in the optical compared to the host galaxy.   

%
\begin{table*}
\caption{Blazars and blazar candidates identified in the hard sample.
$^{a}$ Outside Area90.
$^{b}$ bll= BL Lac; fsrq = Flat Spectrum Radio Quasar; fsrqc = fsrq candidate; bcu = Unclassified blazar; bcuc = Unclassified blazar candidate. 
\label{tab:blazar}}             

\centering          
\centering          
\begin{tabular}{l | c | c | l c c c}     
\hline\hline       
Name & SRCID & DET\_LIKE\_3 & Name & z & z\_orig & Class$^{b}$ \\
\hline                    
eFEDS J083148.9+042939$^{a}$ & 65 & 29.8  & 4FGL J0831.8+0429 & 0.174 & SDSS & bll \\
eFEDS J083949.5+010427 & 112 & 57.1  & 4FGL J0839.8+0105 & 1.124 & SDSS & fsrq \\
eFEDS J085301.2-015049 & 148 & 41.7 & 5BZQ J0853-0150 & 1.498 & Simbad & fsrq \\
eFEDS J085920.5+004711 & 19 & 45.4 & 4FGL J0859.2+0047 & 1.47 & photz & bll \\
eFEDS J090111.8+044900 & 1116 & 10.8 & 5BZQ J0901+0448 & 1.862 & SDSS & fsrq \\
eFEDS J090910.2+012135 & 25 & 68.4 & 4FGL J0909.1+0121 & 1.024 & SDSS & fsrq \\
eFEDS J090915.8+035442 & 78 & 75.1 & BROS J0909.2+0354 & 3.262 & SDSS & bcuc \\
eFEDS J090939.9+020005 & 354 & 15.4 & 4FGL J0909.6+0159 & 0.15 & photz & bll \\
eFEDS J091408.2-015944 & 10 & 61.9 & 4FGL J0914.1-0202 & 1.15 & photz & bcu \\
eFEDS J091437.8+024558 & 55 & 41.9 & 4FGL J0914.4+0249 & 0.427 & SDSS & fsrq \\
eFEDS J092203.5-004442 & 16185 & 11.3 & WIBRaLS2 J092203.20-004443.3 & 0.576 & SDSS & fsrqc \\
eFEDS J092400.9+053345$^{a}$ & 20 & 50.4 & 4FGL J0924.0+0534 & 1.99 & photz & bll \\
eFEDS J092414.7+030859 & 23 & 114.1 & 5BZQ J0924+0309 & 0.128 & SDSS & fsrq \\
eFEDS J092507.8+001913 & 457 & 19.9 & 5BZQ J0925+0019 & 1.721 & SDSS & fsrq \\
eFEDS J093303.3+045235 & 41 & 27.9 & 3HSPJ093303.0+045235 & 0.378 & SDSS & bcu \\
\hline
\hline                  
\end{tabular}
\end{table*}

A particularly extreme object amongst these and indeed the eFEDS hard sample as a whole is eFEDS J090915.8+035442, identified with the blazar candidate BROS J0909.2+0354 \citep{Itoh2020}. This is both the highest redshift and the highest luminosity source in the sample. It has optical broad lines, but a measured absorption column of $\log N_{\rm H}=10^{21.4}$~cm$^{-2}$, almost satisfying our criterion for being obscured. This absorption may not be intrinsic to the quasar, but instead may originate in the intergalactic medium \citep{Arcodia18}.

\section{Discussion}
\label{sec:disc}

In this paper we have presented the sample of hard X-ray (2.3--5 keV) selected objects detected in the eFEDS survey. At the time the data were acquired, the 140 deg$^{2}$ of coverage provided by eFEDS constituted the largest contiguous X-ray survey above 2 keV, now surpassed only by the all-sky eRASSs (Merloni et al. 2023, in preparation). The large area and cosmological volume probed, together with comprehensive supporting data particularly in terms of spectroscopy, have yielded new insights into the hard X-ray source populations and this hard X-ray parent sample can also be used to select objects or subsamples or particular interest for further detalied study \citep[e.g.][]{Brusa2022,Waddell2023}. 

\subsection{The hard X-ray sky seen by eROSITA}

Due to the short focal length and hence soft response of eROSITA, the hard X-ray sensitivity is considerably less than in the main (0.2-2.3 keV) eROSITA detection band, with the sample size being around 100 times smaller. The 246 hard X-ray sources in our sample have been identified optically and classified according to whether they are Galactic or extragalactic. The vast majority (90~\%) are in the latter category and are overwhelmingly AGN. Extensive spectroscopic coverage in the field thanks mainly to the SDSS-IV and SDSS-V surveys yields a high spectroscopic redshift completeness for our sample of around 80\%, with the remaining sources having photometric redshift estimates. This high redshift completeness at bright fluxes is in part what distinguishes eFEDS from larger, non-contiguous wide field surveys such as Champ, Chandra Source Catalogue and 4XMM \citep{Green2004,Evans2010,Webb2020}. The redshift distribution shows that the eFEDS hard survey yields a relatively nearby population of extragalactic sources, with median redshift $<z>=0.34$, compared to the eFEDS soft X-ray sample or deeper pencil-beam surveys with {\it Chandra} and {\it XMM-Newton}. The hard survey therefore provides an interesting lower-redshift baseline for evolutionary studies of AGN. The extensive optical spectroscopy in the field shows that the overwhelming majority of the objects in the eFEDS hard X-ray sample are type 1 AGN, despite the hard X-ray selection being sensitive to at least moderately obscured objects. Similar effects have been found also in previous hard X-ray surveys at relatively bright fluxes \citep[e.g.]{LaFranca2002,DellaCeca2008}

An interesting sub-class of objects are the optically bright extragalactic sources which stand out somewhat from the general population in Fig.~\ref{fig:fx_fopt}. For example our extragalactic/AGN subsample contains six objects with $r<15$. Optical spectroscopy show that these are all very low redshift galaxies with $z<0.05$. In principle these may be normal galaxies whose X-ray emission is dominated by binary populations and hot gas emission \citep[see][for a discussion of the normal galaxies in eFEDS]{Vulic2022} but in reality most of these hard X-ray selected sources show AGN signatures including broad H$\alpha$ and their X-ray luminosities are also indicative of AGN activity, with $\log L_{\rm X}>41$. One such object, eFEDS J085547.9+004747, is a hard-band only detected object and one of the most obscured objects in the entire sample, with $\log N_{\rm H}=23.3$ for an absorbed power law fit. Despite our concern that many hard-band only sources could be spurious, this object is very well detected with {\tt DET$\_$LIKE$\_$3=30.6} and a confirmatory detection in the Swift 105-month catalogue \citep{Oh2018}. There is a hint in Fig~\ref{fig:fx_fopt} that the optical magnitude distribution of the hard-band selected objects is bimodal, and it will be interesting to see if this is borne out in the eRASS. 

\subsection{X-ray spectral properties of the hard sample}

X-ray spectral analysis of the hard sources shows that a significant fraction exhibit soft X-ray absorption, with the survey having some sensitivity to column densities as high as $N_{\rm H}=10^{24}\mathrm{cm}^{-2}$. eROSITA hard samples can therefore be used to explore the demographics of the (Compton-thin) obscured populations. We observe a dip in the column density distribution near $N_{\rm H}\sim10^{20.5}\mathrm{cm}^{-2}$. This may indicate that the medium around AGN is not a single, continous density distribution from $10^{20}$ to $10^{23}\mathrm{cm}^{-2}$ and above, but composed of discrete clumps of a significant, typical column density $N_{\rm H}\sim10^{22}\mathrm{cm}^{-2}$ \citep[see, for example][]{Buchner2017b,Buchner2019}, with unobscured sight-lines in between. A bi-modality has also been seen in the optical obscuration \citep{Assef2013}.

At column densities of $N_{\rm H}>10^{22}$~$cm^{-2}$, the selection function needs to be taken into account carefully, given that the survey will become increasingly incomplete for higher absorbing columns. After accounting for absorption, the hard sample exhibits similar intrinsic photon indices as the eROSITA soft sample, but with a tail to flat photon indices that might be indicative of unmodelled soft X-ray spectral complexity, including e.g. ionized absorbers which may not be correctly matched by the cold absorption model employed here. 

This conclusion is supported by the fact that a small but significant fraction of our sources exhibit soft X-ray absorption, while being optical type-1s \citep[see, e.g.][]{Brusa2003,Page2011,Merloni2014}. One explanation for this is that the absorber is ionized and dust-free, presumably being located close to the origin of the nuclear ionizing radiation and well within the dust sublimation radius. Ionized or "warm" absorbers have been shown via high resolution X-ray spectroscopy to take the form of outflows \citep[e.g.][]{Kaastra2000,Kaspi2002} and such outflows could be an important component of the accreting system as well as having wider influence. Naively, one might expect these winds to be launched at the highest accretion rates, but as shown in Fig.~\ref{fig:agn_mledd} the absorbed type 1 AGN are preferentially seen at more modest accretion rates. This is in concurrence with the conclusions of \cite{TLiu2018} based on the XMM-XXL field and the more detailed analysis of the eFEDS hard sample data presented in \cite{Waddell2023}. It should also be noted that the exact relationship between accretion rate and the launching of winds could be complicated by factors such as geometry and the presence or absence of dust \citep[e.g.,][]{Fabian2008,Ricci2017}

Soft X-ray spectral complexity of the opposite kind is also evident in the form of a soft excess, detected in around 10\% of the extragalactic sources. Our modeling indicates that this has a power law, rather than blackbody form. The origin of the soft excess emission in AGN is still debated, with possibilities including a secondary corona, complex absorption or relativistic X-ray reflection from an ionized accretion disk. The spectral modeling favours the warm corona interpretation. Again, the above results confirm the more sophisticated analysis of the soft X-ray complexity of the eFEDS hard X-ray sample excesses presented in \cite{Waddell2023}. The most likely overall explanation for this is that at the highest accretion rates, the warm absorber becomes over-ionized decreasing the efficiency of radiative line-driving \citep{Proga2000}.
 
\subsection{AGN evolution and downsizing}

The relatively local nature of the eFEDS hard sample offers an interesting baseline to be compared to deeper pencil beam surveys, with the large area of eFEDS offering sufficient cosmological volume to provide a fair sample of the low redshift Universe. In this work, in particular, we have compared the black hole mass and accretion rate distributions between eFEDS and the deeper COSMOS survey. Fortuitously, the luminosity limit of eFEDS and COSMOS observations at their respective median redshift is similar, as is their rest frame selection bandpass, and they also yield similar sample sizes.  

We find a remarkable similarity between the black hole mass distributions, but with a clear difference in the accretion rate distributions, in the sense that the typical black hole growth rate relative to Eddington is about a factor 3 lower in eFEDS than in COSMOS. This result is of particular interest given that the global accretion rate density has declined substantially between the two cosmic epochs sampled by the COSMOS and eFEDS samples. Under the assumption that the populations sampled by these surveys are representative of the overall AGN populations and hence the global accretion rate it suggests that the decline in accretion rate is due primarily to a reduction in the typical fuelling rate of black holes, rather than a "downsizing" effect where the sites of active accretion occur in objects with lower typical black holes mass. 

This general effect is similar to that seen in work which analyses the evolution of the specific accretion rate distribution of AGN over cosmic time, as traced by the ratio of the X-ray luminosity to host stellar mass
\citep[e.g.][]{Aird2012,Bongiorno2016,Georgakakis2017}. Based on the extensive optical spectroscopy in our field, it has been possible here to calculate the accretion rate more directly via the line width and black hole mass estimates. Our finding of a lack of "downsizing" effect in the overall black hole populations contrasts somewhat with some previous studies of optical quasars, which do see such an effect \citep[e.g.][]{KellyShen13,Schulze15}. The difference may be due to the selection band, but full understanding requires careful modeling of the selection effects \citep{Trump15,Jones2016}.

   \begin{figure}
   \centering
   \includegraphics[width=\hsize]{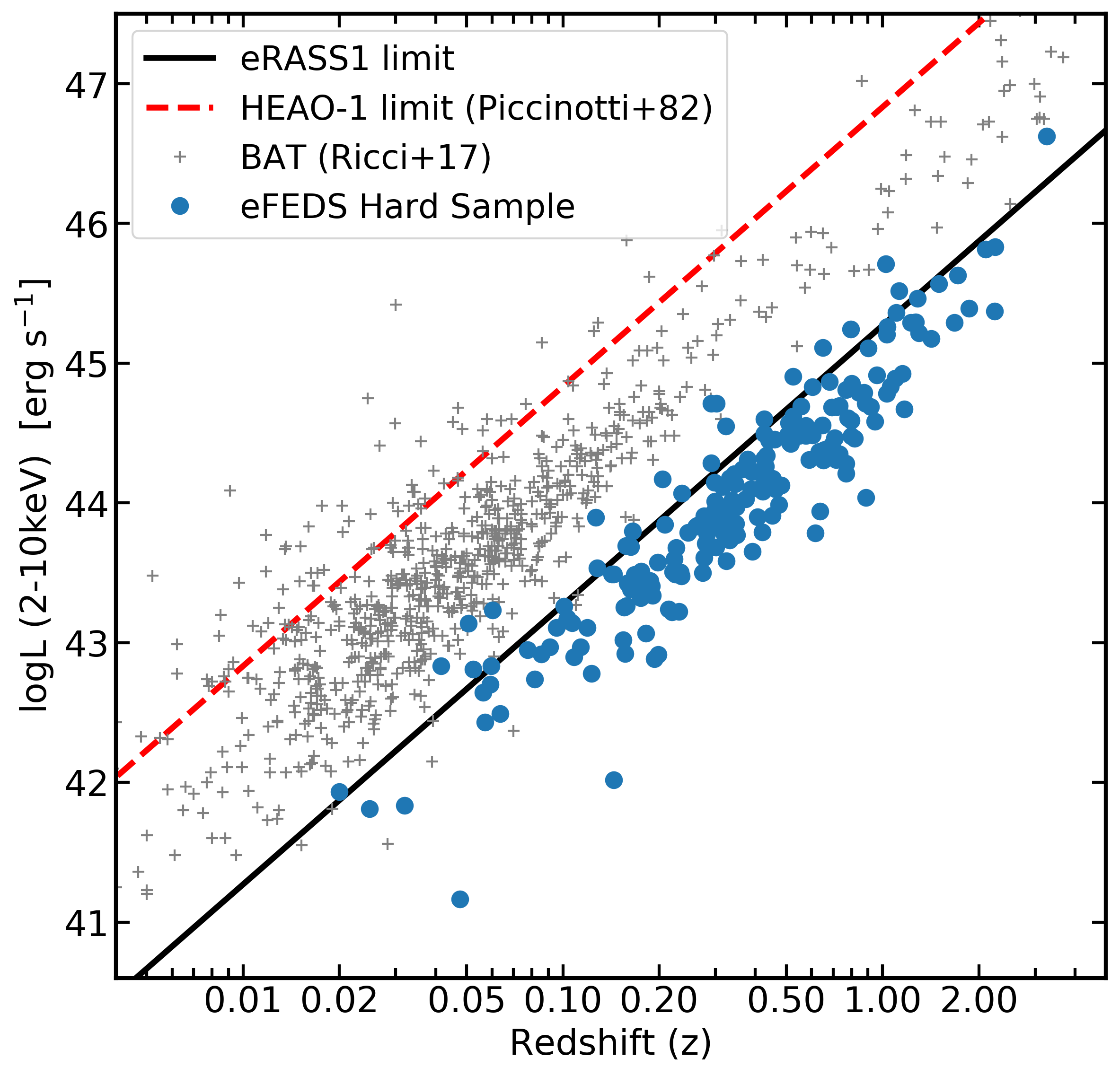}
      \caption{Luminosity-redshift relation for the eFEDS hard X-ray sample compared to the Swift-BAT sample \citep{Ricci2017} and approximate flux limit for the HEAO-1 A2 all-sky survey \citep{Piccinotti82}. The eFEDS luminosities are absorption-corrected and dervied from the spectra (see Secttion~\ref{sec:zandl}). eFEDS probes a similar redshift range to Swift-BAT but samples luminosities around an order of magnitude fainter. 
              }
         \label{fig:hx_lz}
   \end{figure}

\section{Conclusions and outlook}

We have presented the properties of a sample of X-ray sources selected in the relatively hard 2.3--5 keV band from the eROSITA eFEDS survey. The sample of 246 sources is around two orders of magnitude smaller than the main, soft X-ray selected eROSITA catalogue in this region, reflecting the lower effective area and higher background of the instrument in this band. The hard sample consists overwhelmingly of extragalactic sources and more specifically AGN, with a small fraction of stars. Being relatively bright and benefiting from extensive dedicated spectroscopy from the SDSS-IV and SDSS-V surveys, the sample has very high spectroscopic completeness (80\%) which increases to $\sim$94\% when only the extragalctic/AGN subsample is considered, and enables a number of properties to be explored in detail:

\begin{enumerate}
  \item The hard-selected AGN cover a wide luminosity and redshift range up to z$\sim3$, but the selection predominantly results in a relatively nearby sample with $<z>\sim 0.3$, which is interesting to compare to higher redshift objects in the main sample and in deeper surveys. 
  \item The intrinsic power law index appears similar, if perhaps slightly harder, than the main eFEDS sample.
  \item The hard selection yields a small but significant subsample of highly obscured AGN, with $\log N_{\rm H}>23$. We characterize the column density distribution from $\log N_{\rm H}=20-22$ in fine detail.
  \item The hardest sources in the sample tend to be optically unclassified spectroscopically and are probably obscured, type 2 AGN. On the other hand a significant subsample ($\sim$10\%) of the optical broad-line AGN are also X-ray obscured. These may be "warm absorbers" with an outflowing nuclear wind, and are observed preferentially at low Eddington ratios \citep[see also][]{Waddell2023}. 
  \item Several of the brightest and most luminous sources in our sample are blazars, predominantly flat-spectrum radio quasars many of which are also $\gamma$-ray emitters.
  \item For the objects with optical broad lines, we are able to estimate black hole masses and accretion rates. Comparing to a similar sample of high redshift objects in the COSMOS field, we find a very similar distribution of black hole masses, but significantly lower Eddington ratios (by a factor $\sim 3$) in the lower redshift eFEDS sample. This provides important clues as to the origin of the decrease in the overall accretion luminosity of the Universe since $z\sim1-2$. 
   \end{enumerate}

As mentioned earlier, eFEDS was designed as a precursor to the 8-pass eRASS:8 survey, intended to demonstrate the power and potential of the full survey. With the exception of the XMM-Slew survey \citep{Saxton2008}, previous all-sky or near all-sky surveys in the hard band have had to rely on non-imaging instruments. Notable examples include the HEAO-1 A2 all sky survey catalogue \citep{Rothschild79}, the RXTE all-sky survey \citep{Revnivtsev2004} and the Swift-BAT survey \citep[BASS;][]{Oh2018,Koss2022a}. With resulting AGN catalogues presented by e.g. \cite{Piccinotti82} for HEAO-1 A2, \cite{Sazonov2004} for RXTE and, for BASS,\cite{Ricci2017} and \cite{Koss2022b}. The luminosity-redshift relation of eFEDS is compared to some of these surveys in Fig.~\ref{fig:hx_lz}. This clearly shows the superior sensitivity even of the single-pass eRASS1 survey Merloni et al. (2024) compared to previous hard X-ray all-sky surveys, being effectively around 2 orders of magnitude deeper than HEAO-1 (or the similar RXTE all-sky survey) in the standard 2-10 keV bandpass.

Arguably the most interesting comparison survey from the point of view of AGN science is the Swift-BAT survey \citep{Oh2018} which is highly spectroscopically complete \citep{Koss2017,Koss2022a} and has given important new insights into the local AGN population. The harder bandpass of Swift BAT means it is much more complete in terms of obscured AGN, but Fig.~\ref{fig:hx_lz} shows that the eFEDS hard sample, and by extension the eRASS, clearly probes to significantly fainter fluxes, being a true imaging survey, and hence luminosities an order of magnitude lower at any given redshift.  A combination of these surveys will undoubtedly yield new insights into the nature of the AGN population at low redshift. 

Overall the eFEDS survey illustrates the high potential for the full eROSITA all sky survey for studies of AGN physics and evolution, with the all-sky survey yielding unprecedented samples of hard X-ray AGN. A first taste of this is presented in Waddell et al. (2024), who present the hard-selected sample from the first eROSITA sky survey eRASS1 (Merloni et al. 2024). 
Based on eFEDS and applying approximate corrections for the sky area and the slightly deeper exposure for compared to the expectation for the eRASS:8 survey, we would expected approximately 50,000 hard X-ray selected AGN all-sky. While these will be mostly unobscured, the eFEDS analysis shows that objects with absorbing columns up to N$_{\rm H} \sim 10^{24}$~cm$^{-2}$ are sampled in the survey, so provided obscuration bias can be accurately characterised via the selection function, the eRASS has the potential to add significantly to AGN demographic studies. The information content of the resulting dataset will be very substantial, given the excellent eROSITA spectral response, especially when augmented by massive optical ground-based spectroscopy from SDSS and, eventually, 4MOST \citep{deJong2019,Merloni2019}, both of which have dedicated eROSITA follow-up programs. 
   
\begin{acknowledgements}

FEB acknowledges support from ANID-Chile BASAL CATA FB210003, FONDECYT Regular 1200495,
and Millennium Science Initiative Program  – ICN12\_009.

RJA was supported by FONDECYT grant number 123171 and by the ANID BASAL project FB210003.

MB, AG and BM are supported by the European Union's Innovative Training  Network (ITN) "BiD4BEST", funded by the  Marie  Sklodowska-Curie Actions in Horizon 2020 (GA N. 860744).

MK acknowledges support from DLR grant FKZ 50 OR 2307.

This work is based on data from eROSITA, the soft X-ray instrument aboard SRG, a joint Russian-German science mission supported by the Russian Space Agency (Roskosmos), in the interests of the Russian Academy of Sciences represented by its Space Research Institute (IKI), and the Deutsches Zentrum für Luft- und Raumfahrt (DLR). The SRG spacecraft was built by Lavochkin Association (NPOL) and its subcontractors, and is operated by NPOL with support from the Max Planck Institute for Extraterrestrial Physics (MPE).

The development and construction of the eROSITA X-ray instrument was led by MPE, with contributions from the Dr. Karl Remeis Observatory Bamberg \& ECAP (FAU Erlangen-Nuernberg), the University of Hamburg Observatory, the Leibniz Institute for Astrophysics Potsdam (AIP), and the Institute for Astronomy and Astrophysics of the University of Tübingen, with the support of DLR and the Max Planck Society. The Argelander Institute for Astronomy of the University of Bonn and the Ludwig Maximilians Universität Munich also participated in the science preparation for eROSITA.

The eROSITA data shown here were processed using the eSASS/NRTA software system developed by the German eROSITA consortium.
The Hyper Suprime-Cam (HSC) collaboration includes the astronomical communities of Japan and Taiwan, and Princeton University.  The HSC instrumentation and software were developed by the National Astronomical Observatory of Japan(NAOJ), the Kavli Institute for the Physics and Mathematics of the Universe (Kavli IPMU), the University of Tokyo, the High Energy Accelerator Research Organization (KEK), the Academia Sinica Institute for Astronomy and Astrophysics in Taiwan (ASIAA), and Princeton University.  Funding was contributed by the FIRST program from Japanese Cabinet Office, the Ministry of Education, Culture, Sports, Science and Technology (MEXT), the Japan Society for the Promotion of Science (JSPS), Japan Science and Technology Agency (JST),the Toray Science Foundation, NAOJ, Kavli IPMU, KEK,ASIAA, and Princeton University.
Funding for the Sloan Digital Sky Survey IV has been provided by the Alfred P. Sloan Foundation, the U.S. Department of Energy Office of Science, and the Participating Institutions. SDSS acknowledges support and resources from the Center for High-Performance Computing at the University of Utah. The SDSS web site is \url{www.sdss.org}.

Funding for the Sloan Digital Sky Survey V has been provided by the Alfred P. Sloan Foundation, the Heising-Simons Foundation, the National Science Foundation, and the Participating Institutions. SDSS acknowledges support and resources from the Center for High-Performance Computing at the University of Utah. The SDSS web site is \url{www.sdss.org}.

SDSS is managed by the Astrophysical Research Consortium for the Participating Institutions of the SDSS Collaboration, including the Carnegie Institution for Science, Chilean National Time Allocation Committee (CNTAC) ratified researchers, the Gotham Participation Group, Harvard University, Heidelberg University, The Johns Hopkins University, L’Ecole polytechnique f\'{e}d\'{e}rale de Lausanne (EPFL), Leibniz-Institut f\"{u}r Astrophysik Potsdam (AIP), Max-Planck-Institut f\"{u}r Astronomie (MPIA Heidelberg), Max-Planck-Institut f\"{u}r Extraterrestrische Physik (MPE), Nanjing University, National Astronomical Observatories of China (NAOC), New Mexico State University, The Ohio State University, Pennsylvania State University, Smithsonian Astrophysical Observatory, Space Telescope Science Institute (STScI), the Stellar Astrophysics Participation Group, Universidad Nacional Aut\'{o}noma de M\'{e}xico, University of Arizona, University of Colorado Boulder, University of Illinois at Urbana-Champaign, University of Toronto, University of Utah, University of Virginia, Yale University, and Yunnan University.

\end{acknowledgements}

%
%

\bibliographystyle{aa} 
\bibliography{efeds_hx_sample}

\begin{thebibliography}{140}
\expandafter\ifx\csname natexlab\endcsname\relax\def\natexlab#1{#1}\fi

\bibitem[{{Abdollahi} {et~al.}(2020){Abdollahi}, {Acero}, {Ackermann},
  {Ajello}, {Atwood}, {Axelsson}, {Baldini}, {Ballet}, {Barbiellini},
  {Bastieri}, {Becerra Gonzalez}, {Bellazzini}, {Berretta}, {Bissaldi},
  {Blandford}, {Bloom}, {Bonino}, {Bottacini}, {Brandt}, {Bregeon}, {Bruel},
  {Buehler}, {Burnett}, {Buson}, {Cameron}, {Caputo}, {Caraveo}, {Casandjian},
  {Castro}, {Cavazzuti}, {Charles}, {Chaty}, {Chen}, {Cheung}, {Chiaro},
  {Ciprini}, {Cohen-Tanugi}, {Cominsky}, {Coronado-Bl{\'a}zquez}, {Costantin},
  {Cuoco}, {Cutini}, {D'Ammando}, {DeKlotz}, {de la Torre Luque}, {de Palma},
  {Desai}, {Digel}, {Di Lalla}, {Di Mauro}, {Di Venere}, {Dom{\'\i}nguez},
  {Dumora}, {Fana Dirirsa}, {Fegan}, {Ferrara}, {Franckowiak}, {Fukazawa},
  {Funk}, {Fusco}, {Gargano}, {Gasparrini}, {Giglietto}, {Giommi}, {Giordano},
  {Giroletti}, {Glanzman}, {Green}, {Grenier}, {Griffin}, {Grondin}, {Grove},
  {Guiriec}, {Harding}, {Hayashi}, {Hays}, {Hewitt}, {Horan},
  {J{\'o}hannesson}, {Johnson}, {Kamae}, {Kerr}, {Kocevski}, {Kovac'evic'},
  {Kuss}, {Landriu}, {Larsson}, {Latronico}, {Lemoine-Goumard}, {Li},
  {Liodakis}, {Longo}, {Loparco}, {Lott}, {Lovellette}, {Lubrano}, {Madejski},
  {Maldera}, {Malyshev}, {Manfreda}, {Marchesini}, {Marcotulli},
  {Mart{\'\i}-Devesa}, {Martin}, {Massaro}, {Mazziotta}, {McEnery}, {Mereu},
  {Meyer}, {Michelson}, {Mirabal}, {Mizuno}, {Monzani}, {Morselli},
  {Moskalenko}, {Negro}, {Nuss}, {Ojha}, {Omodei}, {Orienti}, {Orlando},
  {Ormes}, {Palatiello}, {Paliya}, {Paneque}, {Pei}, {Pe{\~n}a-Herazo},
  {Perkins}, {Persic}, {Pesce-Rollins}, {Petrosian}, {Petrov}, {Piron}, {Poon},
  {Porter}, {Principe}, {Rain{\`o}}, {Rando}, {Razzano}, {Razzaque}, {Reimer},
  {Reimer}, {Remy}, {Reposeur}, {Romani}, {Saz Parkinson}, {Schinzel},
  {Serini}, {Sgr{\`o}}, {Siskind}, {Smith}, {Spandre}, {Spinelli}, {Strong},
  {Suson}, {Tajima}, {Takahashi}, {Tak}, {Thayer}, {Thompson}, {Tibaldo},
  {Torres}, {Torresi}, {Valverde}, {Van Klaveren}, {van Zyl}, {Wood},
  {Yassine}, \& {Zaharijas}}]{Abdollahi2020}
{Abdollahi}, S., {Acero}, F., {Ackermann}, M., {et~al.} 2020, \apjs, 247, 33

\bibitem[{{Aihara} {et~al.}(2018){Aihara}, {Arimoto}, {Armstrong}, {Arnouts},
  {Bahcall}, {Bickerton}, {Bosch}, {Bundy}, {Capak}, {Chan}, {Chiba}, {Coupon},
  {Egami}, {Enoki}, {Finet}, {Fujimori}, {Fujimoto}, {Furusawa}, {Furusawa},
  {Goto}, {Goulding}, {Greco}, {Greene}, {Gunn}, {Hamana}, {Harikane},
  {Hashimoto}, {Hattori}, {Hayashi}, {Hayashi}, {He{\l}miniak}, {Higuchi},
  {Hikage}, {Ho}, {Hsieh}, {Huang}, {Huang}, {Ikeda}, {Imanishi}, {Inoue},
  {Iwasawa}, {Iwata}, {Jaelani}, {Jian}, {Kamata}, {Karoji}, {Kashikawa},
  {Katayama}, {Kawanomoto}, {Kayo}, {Koda}, {Koike}, {Kojima}, {Komiyama},
  {Konno}, {Koshida}, {Koyama}, {Kusakabe}, {Leauthaud}, {Lee}, {Lin}, {Lin},
  {Lupton}, {Mandelbaum}, {Matsuoka}, {Medezinski}, {Mineo}, {Miyama},
  {Miyatake}, {Miyazaki}, {Momose}, {More}, {More}, {Moritani}, {Moriya},
  {Morokuma}, {Mukae}, {Murata}, {Murayama}, {Nagao}, {Nakata}, {Niida},
  {Niikura}, {Nishizawa}, {Obuchi}, {Oguri}, {Oishi}, {Okabe}, {Okamoto},
  {Okura}, {Ono}, {Onodera}, {Onoue}, {Osato}, {Ouchi}, {Price}, {Pyo}, {Sako},
  {Sawicki}, {Shibuya}, {Shimasaku}, {Shimono}, {Shirasaki}, {Silverman},
  {Simet}, {Speagle}, {Spergel}, {Strauss}, {Sugahara}, {Sugiyama}, {Suto},
  {Suyu}, {Suzuki}, {Tait}, {Takada}, {Takata}, {Tamura}, {Tanaka}, {Tanaka},
  {Tanaka}, {Tanaka}, {Terai}, {Terashima}, {Toba}, {Tominaga}, {Toshikawa},
  {Turner}, {Uchida}, {Uchiyama}, {Umetsu}, {Uraguchi}, {Urata}, {Usuda},
  {Utsumi}, {Wang}, {Wang}, {Wong}, {Yabe}, {Yamada}, {Yamanoi}, {Yasuda},
  {Yeh}, {Yonehara}, \& {Yuma}}]{Aihara2018}
{Aihara}, H., {Arimoto}, N., {Armstrong}, R., {et~al.} 2018, \pasj, 70, S4

\bibitem[{{Aird} {et~al.}(2015){Aird}, {Coil}, {Georgakakis}, {Nandra},
  {Barro}, \& {P{\'e}rez-Gonz{\'a}lez}}]{Aird2015}
{Aird}, J., {Coil}, A.~L., {Georgakakis}, A., {et~al.} 2015, \mnras, 451, 1892

\bibitem[{{Aird} {et~al.}(2012){Aird}, {Coil}, {Moustakas}, {Blanton},
  {Burles}, {Cool}, {Eisenstein}, {Smith}, {Wong}, \& {Zhu}}]{Aird2012}
{Aird}, J., {Coil}, A.~L., {Moustakas}, J., {et~al.} 2012, \apj, 746, 90

\bibitem[{{Akiyama} {et~al.}(2003){Akiyama}, {Ueda}, {Ohta}, {Takahashi}, \&
  {Yamada}}]{Akiyama2003}
{Akiyama}, M., {Ueda}, Y., {Ohta}, K., {Takahashi}, T., \& {Yamada}, T. 2003,
  \apjs, 148, 275

\bibitem[{{Alexander} {et~al.}(2013){Alexander}, {Stern}, {Del Moro},
  {Lansbury}, {Assef}, {Aird}, {Ajello}, {Ballantyne}, {Bauer}, {Boggs},
  {Brandt}, {Christensen}, {Civano}, {Comastri}, {Craig}, {Elvis},
  {Grefenstette}, {Hailey}, {Harrison}, {Hickox}, {Luo}, {Madsen}, {Mullaney},
  {Perri}, {Puccetti}, {Saez}, {Treister}, {Urry}, {Zhang}, {Bridge},
  {Eisenhardt}, {Gonzalez}, {Miller}, \& {Tsai}}]{Alexander2013}
{Alexander}, D.~M., {Stern}, D., {Del Moro}, A., {et~al.} 2013, \apj, 773, 125

\bibitem[{{Alexandroff} {et~al.}(2018){Alexandroff}, {Zakamska}, {Barth},
  {Hamann}, {Strauss}, {Krolik}, {Greene}, {P{\^a}ris}, \&
  {Ross}}]{Alexandroff2018}
{Alexandroff}, R.~M., {Zakamska}, N.~L., {Barth}, A.~J., {et~al.} 2018, \mnras,
  479, 4936

\bibitem[{{Almeida} {et~al.}(2023){Almeida}, {Anderson},
  {Argudo-Fern{\'a}ndez}, {Badenes}, {Barger}, {Barrera-Ballesteros}, {Bender},
  {Benitez}, {Besser}, {Bizyaev}, {Blanton}, {Bochanski}, {Bovy}, {Brandt},
  {Brownstein}, {Buchner}, {Bulbul}, {Burchett}, {Cano D{\'\i}az}, {Carlberg},
  {Casey}, {Chandra}, {Cherinka}, {Chiappini}, {Coker}, {Comparat}, {Conroy},
  {Contardo}, {Cortes}, {Covey}, {Crane}, {Cunha}, {Dabbieri}, {Davidson},
  {Davis}, {De Lee}, {M{\'e}ndez Delgado}, {Demasi}, {Di Mille}, {Donor},
  {Dow}, {Dwelly}, {Eracleous}, {Eriksen}, {Fan}, {Farr}, {Frederick}, {Fries},
  {Frinchaboy}, {Gaensicke}, {Ge}, {Gonz{\'a}lez {\'A}vila}, {Grabowski},
  {Grier}, {Guiglion}, {Gupta}, {Hall}, {Hawkins}, {Hayes}, {Hermes},
  {Hern{\'a}ndez-Garc{\'\i}a}, {Hogg}, {Holtzman}, {Ibarra-Medel}, {Ji},
  {Jofre}, {Johnson}, {Jones}, {Kinemuchi}, {Kluge}, {Koekemoer}, {Kollmeier},
  {Kounkel}, {Krishnarao}, {Krumpe}, {Lacerna}, {Jakson Assuncao Lago},
  {Laporte}, {Liu}, {Liu}, {Liu}, {Lopes}, {Macktoobian}, {Malanushenko},
  {Maoz}, {Masseron}, {Masters}, {Matijevic}, {McBride}, {Medan}, {Merloni},
  {Morrison}, {Myers}, {M{\'e}sz{\'a}ros}, {Negrete}, {Nidever}, {Nitschelm},
  {Oravetz}, {Oravetz}, {Pan}, {Peng}, {Pinsonneault}, {Pogge}, {Qiu},
  {Queiroz}, {Ramirez}, {Rix}, {Fern{\'a}ndez Rosso}, {Runnoe}, {Salvato},
  {Sanchez}, {Santana}, {Saydjari}, {Sayres}, {Schlaufman}, {Schneider},
  {Schwope}, {Serna}, {Shen}, {Sobeck}, {Song}, {Souto}, {Spoo}, {Stassun},
  {Steinmetz}, {Straumit}, {Stringfellow}, {S{\'a}nchez-Gallego},
  {Taghizadeh-Popp}, {Tayar}, {Thakar}, {Tissera}, {Tkachenko}, {Hernandez
  Toledo}, {Trakhtenbrot}, {Fernandez Trincado}, {Troup}, {Trump}, {Tuttle},
  {Ulloa}, {Vazquez-Mata}, {Alfaro}, {Villanova}, {Wachter}, {Weijmans},
  {Wheeler}, {Wilson}, {Wojno}, {Wolf}, {Xue}, {Ybarra}, {Zari}, \&
  {Zasowski}}]{Almeida2023}
{Almeida}, A., {Anderson}, S.~F., {Argudo-Fern{\'a}ndez}, M., {et~al.} 2023,
  arXiv e-prints, arXiv:2301.07688

\bibitem[{{Antonucci} \& {Miller}(1985)}]{AntonucciMiller1985}
{Antonucci}, R.~R.~J. \& {Miller}, J.~S. 1985, \apj, 297, 621

\bibitem[{{Arcodia} {et~al.}(2018){Arcodia}, {Campana}, {Salvaterra}, \&
  {Ghisellini}}]{Arcodia18}
{Arcodia}, R., {Campana}, S., {Salvaterra}, R., \& {Ghisellini}, G. 2018, \aap,
  616, A170

\bibitem[{{Assef} {et~al.}(2020){Assef}, {Brightman}, {Walton}, {Stern},
  {Bauer}, {Blain}, {D{\'\i}az-Santos}, {Eisenhardt}, {Hickox}, {Jun},
  {Psychogyios}, {Tsai}, \& {Wu}}]{Assef2020}
{Assef}, R.~J., {Brightman}, M., {Walton}, D.~J., {et~al.} 2020, \apj, 897, 112

\bibitem[{{Assef} {et~al.}(2013){Assef}, {Stern}, {Kochanek}, {Blain},
  {Brodwin}, {Brown}, {Donoso}, {Eisenhardt}, {Jannuzi}, {Jarrett}, {Stanford},
  {Tsai}, {Wu}, \& {Yan}}]{Assef2013}
{Assef}, R.~J., {Stern}, D., {Kochanek}, C.~S., {et~al.} 2013, \apj, 772, 26

\bibitem[{{Awaki} {et~al.}(1991){Awaki}, {Koyama}, {Inoue}, \&
  {Halpern}}]{Awaki1991}
{Awaki}, H., {Koyama}, K., {Inoue}, H., \& {Halpern}, J.~P. 1991, \pasj, 43,
  195

\bibitem[{{Baldry} {et~al.}(2018){Baldry}, {Liske}, {Brown}, {Robotham},
  {Driver}, {Dunne}, {Alpaslan}, {Brough}, {Cluver}, {Eardley}, {Farrow},
  {Heymans}, {Hildebrandt}, {Hopkins}, {Kelvin}, {Loveday}, {Moffett},
  {Norberg}, {Owers}, {Taylor}, {Wright}, {Bamford}, {Bland-Hawthorn},
  {Bourne}, {Bremer}, {Colless}, {Conselice}, {Croom}, {Davies}, {Foster},
  {Grootes}, {Holwerda}, {Jones}, {Kafle}, {Kuijken}, {Lara-Lopez},
  {L{\'o}pez-S{\'a}nchez}, {Meyer}, {Phillipps}, {Sutherland}, {van Kampen}, \&
  {Wilkins}}]{Baldry18}
{Baldry}, I.~K., {Liske}, J., {Brown}, M.~J.~I., {et~al.} 2018, \mnras, 474,
  3875

\bibitem[{{Baronchelli} {et~al.}(2020){Baronchelli}, {Nandra}, \&
  {Buchner}}]{Baronchelli2020}
{Baronchelli}, L., {Nandra}, K., \& {Buchner}, J. 2020, \mnras, 498, 5284

\bibitem[{{Blanton} {et~al.}(2017){Blanton}, {Bershady}, {Abolfathi},
  {Albareti}, {Allende Prieto}, {Almeida}, {Alonso-Garc{\'\i}a}, {Anders},
  {Anderson}, {Andrews}, {Aquino-Ort{\'\i}z}, {Arag{\'o}n-Salamanca},
  {Argudo-Fern{\'a}ndez}, {Armengaud}, {Aubourg}, {Avila-Reese}, {Badenes},
  {Bailey}, {Barger}, {Barrera-Ballesteros}, {Bartosz}, {Bates}, {Baumgarten},
  {Bautista}, {Beaton}, {Beers}, {Belfiore}, {Bender}, {Berlind}, {Bernardi},
  {Beutler}, {Bird}, {Bizyaev}, {Blanc}, {Blomqvist}, {Bolton}, {Boquien},
  {Borissova}, {van den Bosch}, {Bovy}, {Brandt}, {Brinkmann}, {Brownstein},
  {Bundy}, {Burgasser}, {Burtin}, {Busca}, {Cappellari}, {Delgado Carigi},
  {Carlberg}, {Carnero Rosell}, {Carrera}, {Chanover}, {Cherinka}, {Cheung},
  {G{\'o}mez Maqueo Chew}, {Chiappini}, {Choi}, {Chojnowski}, {Chuang},
  {Chung}, {Cirolini}, {Clerc}, {Cohen}, {Comparat}, {da Costa}, {Cousinou},
  {Covey}, {Crane}, {Croft}, {Cruz-Gonzalez}, {Garrido Cuadra}, {Cunha},
  {Damke}, {Darling}, {Davies}, {Dawson}, {de la Macorra}, {Dell'Agli}, {De
  Lee}, {Delubac}, {Di Mille}, {Diamond-Stanic}, {Cano-D{\'\i}az}, {Donor},
  {Downes}, {Drory}, {du Mas des Bourboux}, {Duckworth}, {Dwelly}, {Dyer},
  {Ebelke}, {Eigenbrot}, {Eisenstein}, {Emsellem}, {Eracleous}, {Escoffier},
  {Evans}, {Fan}, {Fern{\'a}ndez-Alvar}, {Fernandez-Trincado}, {Feuillet},
  {Finoguenov}, {Fleming}, {Font-Ribera}, {Fredrickson}, {Freischlad},
  {Frinchaboy}, {Fuentes}, {Galbany}, {Garcia-Dias},
  {Garc{\'\i}a-Hern{\'a}ndez}, {Gaulme}, {Geisler}, {Gelfand},
  {Gil-Mar{\'\i}n}, {Gillespie}, {Goddard}, {Gonzalez-Perez}, {Grabowski},
  {Green}, {Grier}, {Gunn}, {Guo}, {Guy}, {Hagen}, {Hahn}, {Hall}, {Harding},
  {Hasselquist}, {Hawley}, {Hearty}, {Gonzalez Hern{\'a}ndez}, {Ho}, {Hogg},
  {Holley-Bockelmann}, {Holtzman}, {Holzer}, {Huehnerhoff}, {Hutchinson},
  {Hwang}, {Ibarra-Medel}, {da Silva Ilha}, {Ivans}, {Ivory}, {Jackson},
  {Jensen}, {Johnson}, {Jones}, {J{\"o}nsson}, {Jullo}, {Kamble}, {Kinemuchi},
  {Kirkby}, {Kitaura}, {Klaene}, {Knapp}, {Kneib}, {Kollmeier}, {Lacerna},
  {Lane}, {Lang}, {Law}, {Lazarz}, {Lee}, {Le Goff}, {Liang}, {Li}, {Li},
  {Lian}, {Lima}, {Lin}, {Lin}, {Bertran de Lis}, {Liu}, {de Icaza Lizaola},
  {Long}, {Lucatello}, {Lundgren}, {MacDonald}, {Deconto Machado}, {MacLeod},
  {Mahadevan}, {Geimba Maia}, {Maiolino}, {Majewski}, {Malanushenko},
  {Malanushenko}, {Manchado}, {Mao}, {Maraston}, {Marques-Chaves}, {Masseron},
  {Masters}, {McBride}, {McDermid}, {McGrath}, {McGreer}, {Medina Pe{\~n}a},
  {Melendez}, {Merloni}, {Merrifield}, {Meszaros}, {Meza}, {Minchev},
  {Minniti}, {Miyaji}, {More}, {Mulchaey}, {M{\"u}ller-S{\'a}nchez}, {Muna},
  {Munoz}, {Myers}, {Nair}, {Nandra}, {Correa do Nascimento}, {Negrete},
  {Ness}, {Newman}, {Nichol}, {Nidever}, {Nitschelm}, {Ntelis}, {O'Connell},
  {Oelkers}, {Oravetz}, {Oravetz}, {Pace}, {Padilla}, {Palanque-Delabrouille},
  {Alonso Palicio}, {Pan}, {Parejko}, {Parikh}, {P{\^a}ris}, {Park}, {Patten},
  {Peirani}, {Pellejero-Ibanez}, {Penny}, {Percival}, {Perez-Fournon},
  {Petitjean}, {Pieri}, {Pinsonneault}, {Pisani}, {Poleski}, {Prada},
  {Prakash}, {Queiroz}, {Raddick}, {Raichoor}, {Barboza Rembold}, {Richstein},
  {Riffel}, {Riffel}, {Rix}, {Robin}, {Rockosi}, {Rodr{\'\i}guez-Torres},
  {Roman-Lopes}, {Rom{\'a}n-Z{\'u}{\~n}iga}, {Rosado}, {Ross}, {Rossi}, {Ruan},
  {Ruggeri}, {Rykoff}, {Salazar-Albornoz}, {Salvato}, {S{\'a}nchez}, {Aguado},
  {S{\'a}nchez-Gallego}, {Santana}, {Santiago}, {Sayres}, {Schiavon}, {da Silva
  Schimoia}, {Schlafly}, {Schlegel}, {Schneider}, {Schultheis}, {Schuster},
  {Schwope}, {Seo}, {Shao}, {Shen}, {Shetrone}, {Shull}, {Simon}, {Skinner},
  {Skrutskie}, {Slosar}, {Smith}, {Sobeck}, {Sobreira}, {Somers}, {Souto},
  {Stark}, {Stassun}, {Stauffer}, {Steinmetz}, {Storchi-Bergmann},
  {Streblyanska}, {Stringfellow}, {Su{\'a}rez}, {Sun}, {Suzuki}, {Szigeti},
  {Taghizadeh-Popp}, {Tang}, {Tao}, {Tayar}, {Tembe}, {Teske}, {Thakar},
  {Thomas}, {Thompson}, {Tinker}, {Tissera}, {Tojeiro}, {Hernandez Toledo}, {de
  la Torre}, {Tremonti}, {Troup}, {Valenzuela}, {Martinez Valpuesta},
  {Vargas-Gonz{\'a}lez}, {Vargas-Maga{\~n}a}, {Vazquez}, {Villanova}, {Vivek},
  {Vogt}, {Wake}, {Walterbos}, {Wang}, {Weaver}, {Weijmans}, {Weinberg},
  {Westfall}, {Whelan}, {Wild}, {Wilson}, {Wood-Vasey}, {Wylezalek}, {Xiao},
  {Yan}, {Yang}, {Ybarra}, {Y{\`e}che}, {Zakamska}, {Zamora}, {Zarrouk},
  {Zasowski}, {Zhang}, {Zhao}, {Zheng}, {Zheng}, {Zhou}, {Zhou}, {Zhu},
  {Zoccali}, \& {Zou}}]{Blanton17}
{Blanton}, M.~R., {Bershady}, M.~A., {Abolfathi}, B., {et~al.} 2017, \aj, 154,
  28

\bibitem[{{Bongiorno} {et~al.}(2016){Bongiorno}, {Schulze}, {Merloni},
  {Zamorani}, {Ilbert}, {La Franca}, {Peng}, {Piconcelli}, {Mainieri},
  {Silverman}, {Brusa}, {Fiore}, {Salvato}, \& {Scoville}}]{Bongiorno2016}
{Bongiorno}, A., {Schulze}, A., {Merloni}, A., {et~al.} 2016, \aap, 588, A78

\bibitem[{{Boroson} \& {Green}(1992)}]{1992ApJS...80..109B}
{Boroson}, T.~A. \& {Green}, R.~F. 1992, \apjs, 80, 109

\bibitem[{{Brunner} {et~al.}(2022){Brunner}, {Liu}, {Lamer}, {Georgakakis},
  {Merloni}, {Brusa}, {Bulbul}, {Dennerl}, {Friedrich}, {Liu}, {Maitra},
  {Nandra}, {Ramos-Ceja}, {Sanders}, {Stewart}, {Boller}, {Buchner}, {Clerc},
  {Comparat}, {Dwelly}, {Eckert}, {Finoguenov}, {Freyberg}, {Ghirardini},
  {Gueguen}, {Haberl}, {Kreykenbohm}, {Krumpe}, {Osterhage}, {Pacaud},
  {Predehl}, {Reiprich}, {Robrade}, {Salvato}, {Santangelo}, {Schrabback},
  {Schwope}, \& {Wilms}}]{Brunner2022}
{Brunner}, H., {Liu}, T., {Lamer}, G., {et~al.} 2022, \aap, 661, A1

\bibitem[{{Brusa} {et~al.}(2003){Brusa}, {Comastri}, {Mignoli}, {Fiore},
  {Ciliegi}, {Vignali}, {Severgnini}, {Cocchia}, {La Franca}, {Matt}, {Perola},
  {Maiolino}, {Baldi}, \& {Molendi}}]{Brusa2003}
{Brusa}, M., {Comastri}, A., {Mignoli}, M., {et~al.} 2003, \aap, 409, 65

\bibitem[{{Brusa} {et~al.}(2022){Brusa}, {Urrutia}, {Toba}, {Buchner}, {Li},
  {Liu}, {Perna}, {Salvato}, {Merloni}, {Musiimenta}, {Nandra}, {Wolf},
  {Arcodia}, {Dwelly}, {Georgakakis}, {Goulding}, {Matsuoka}, {Nagao},
  {Schramm}, {Silverman}, \& {Terashima}}]{Brusa2022}
{Brusa}, M., {Urrutia}, T., {Toba}, Y., {et~al.} 2022, \aap, 661, A9

\bibitem[{{Brusa} {et~al.}(2007){Brusa}, {Zamorani}, {Comastri}, {Hasinger},
  {Cappelluti}, {Civano}, {Finoguenov}, {Mainieri}, {Salvato}, {Vignali},
  {Elvis}, {Fiore}, {Gilli}, {Impey}, {Lilly}, {Mignoli}, {Silverman}, {Trump},
  {Urry}, {Bender}, {Capak}, {Huchra}, {Kneib}, {Koekemoer}, {Leauthaud},
  {Lehmann}, {Massey}, {Matute}, {McCarthy}, {McCracken}, {Rhodes}, {Scoville},
  {Taniguchi}, \& {Thompson}}]{Brusa2007}
{Brusa}, M., {Zamorani}, G., {Comastri}, A., {et~al.} 2007, \apjs, 172, 353

\bibitem[{{Buchner}(2016)}]{Buchner2016}
{Buchner}, J. 2016, Statistics and Computing, 26, 383

\bibitem[{{Buchner}(2019)}]{Buchner2019}
{Buchner}, J. 2019, \pasp, 131, 108005

\bibitem[{{Buchner}(2021)}]{Buchner2021}
{Buchner}, J. 2021, arXiv e-prints, arXiv:2101.09604

\bibitem[{{Buchner} \& {Bauer}(2017)}]{Buchner2017b}
{Buchner}, J. \& {Bauer}, F.~E. 2017, \mnras, 465, 4348

\bibitem[{{Buchner} {et~al.}(2015){Buchner}, {Georgakakis}, {Nandra},
  {Brightman}, {Menzel}, {Liu}, {Hsu}, {Salvato}, {Rangel}, {Aird}, {Merloni},
  \& {Ross}}]{Buchner2015}
{Buchner}, J., {Georgakakis}, A., {Nandra}, K., {et~al.} 2015, \apj, 802, 89

\bibitem[{{Buchner} {et~al.}(2014){Buchner}, {Georgakakis}, {Nandra}, {Hsu},
  {Rangel}, {Brightman}, {Merloni}, {Salvato}, {Donley}, \&
  {Kocevski}}]{Buchner2014}
{Buchner}, J., {Georgakakis}, A., {Nandra}, K., {et~al.} 2014, \aap, 564, A125

\bibitem[{{Buchner} {et~al.}(2017){Buchner}, {Schulze}, \&
  {Bauer}}]{Buchner2017a}
{Buchner}, J., {Schulze}, S., \& {Bauer}, F.~E. 2017, \mnras, 464, 4545

\bibitem[{{Chang} {et~al.}(2019){Chang}, {Arsioli}, {Giommi}, {Padovani}, \&
  {Brandt}}]{Chang2019}
{Chang}, Y.~L., {Arsioli}, B., {Giommi}, P., {Padovani}, P., \& {Brandt}, C.~H.
  2019, \aap, 632, A77

\bibitem[{{Civano} {et~al.}(2016){Civano}, {Marchesi}, {Comastri}, {Urry},
  {Elvis}, {Cappelluti}, {Puccetti}, {Brusa}, {Zamorani}, {Hasinger},
  {Aldcroft}, {Alexander}, {Allevato}, {Brunner}, {Capak}, {Finoguenov},
  {Fiore}, {Fruscione}, {Gilli}, {Glotfelty}, {Griffiths}, {Hao}, {Harrison},
  {Jahnke}, {Kartaltepe}, {Karim}, {LaMassa}, {Lanzuisi}, {Miyaji}, {Ranalli},
  {Salvato}, {Sargent}, {Scoville}, {Schawinski}, {Schinnerer}, {Silverman},
  {Smolcic}, {Stern}, {Toft}, {Trakhtenbrot}, {Treister}, \&
  {Vignali}}]{Civano2016}
{Civano}, F., {Marchesi}, S., {Comastri}, A., {et~al.} 2016, \apj, 819, 62

\bibitem[{{Cocchia} {et~al.}(2007){Cocchia}, {Fiore}, {Vignali}, {Mignoli},
  {Brusa}, {Comastri}, {Feruglio}, {Baldi}, {Carangelo}, {Ciliegi}, {D'Elia},
  {La Franca}, {Maiolino}, {Matt}, {Molendi}, {Perola}, \&
  {Puccetti}}]{Cocchia2007}
{Cocchia}, F., {Fiore}, F., {Vignali}, C., {et~al.} 2007, \aap, 466, 31

\bibitem[{{Comastri} {et~al.}(1995){Comastri}, {Setti}, {Zamorani}, \&
  {Hasinger}}]{Comastri1995}
{Comastri}, A., {Setti}, G., {Zamorani}, G., \& {Hasinger}, G. 1995, \aap, 296,
  1

\bibitem[{{Comparat} {et~al.}(2020){Comparat}, {Merloni}, {Dwelly}, {Salvato},
  {Schwope}, {Coffey}, {Wolf}, {Arcodia}, {Liu}, {Buchner}, {Nandra},
  {Georgakakis}, {Clerc}, {Brusa}, {Brownstein}, {Schneider}, {Pan}, \&
  {Bizyaev}}]{Comparat2020}
{Comparat}, J., {Merloni}, A., {Dwelly}, T., {et~al.} 2020, \aap, 636, A97

\bibitem[{{D'Abrusco} {et~al.}(2019){D'Abrusco}, {{\'A}lvarez Crespo},
  {Massaro}, {Campana}, {Chavushyan}, {Landoni}, {La Franca}, {Masetti},
  {Milisavljevic}, {Paggi}, {Ricci}, \& {Smith}}]{DAbrusco2019}
{D'Abrusco}, R., {{\'A}lvarez Crespo}, N., {Massaro}, F., {et~al.} 2019, \apjs,
  242, 4

\bibitem[{{de Jong} {et~al.}(2019){de Jong}, {Agertz}, {Berbel}, {Aird},
  {Alexander}, {Amarsi}, {Anders}, {Andrae}, {Ansarinejad}, {Ansorge},
  {Antilogus}, {Anwand-Heerwart}, {Arentsen}, {Arnadottir}, {Asplund}, {Auger},
  {Azais}, {Baade}, {Baker}, {Baker}, {Balbinot}, {Baldry}, {Banerji},
  {Barden}, {Barklem}, {Barth{\'e}l{\'e}my-Mazot}, {Battistini}, {Bauer},
  {Bell}, {Bellido-Tirado}, {Bellstedt}, {Belokurov}, {Bensby}, {Bergemann},
  {Bestenlehner}, {Bielby}, {Bilicki}, {Blake}, {Bland-Hawthorn}, {Boeche},
  {Boland}, {Boller}, {Bongard}, {Bongiorno}, {Bonifacio}, {Boudon}, {Brooks},
  {Brown}, {Brown}, {Br{\"u}ggen}, {Brynnel}, {Brzeski}, {Buchert},
  {Buschkamp}, {Caffau}, {Caillier}, {Carrick}, {Casagrande}, {Case}, {Casey},
  {Cesarini}, {Cescutti}, {Chapuis}, {Chiappini}, {Childress}, {Christlieb},
  {Church}, {Cioni}, {Cluver}, {Colless}, {Collett}, {Comparat}, {Cooper},
  {Couch}, {Courbin}, {Croom}, {Croton}, {Daguis{\'e}}, {Dalton}, {Davies},
  {Davis}, {de Laverny}, {Deason}, {Dionies}, {Disseau}, {Doel}, {D{\"o}scher},
  {Driver}, {Dwelly}, {Eckert}, {Edge}, {Edvardsson}, {Youssoufi}, {Elhaddad},
  {Enke}, {Erfanianfar}, {Farrell}, {Fechner}, {Feiz}, {Feltzing}, {Ferreras},
  {Feuerstein}, {Feuillet}, {Finoguenov}, {Ford}, {Fotopoulou}, {Fouesneau},
  {Frenk}, {Frey}, {Gaessler}, {Geier}, {Gentile Fusillo}, {Gerhard},
  {Giannantonio}, {Giannone}, {Gibson}, {Gillingham},
  {Gonz{\'a}lez-Fern{\'a}ndez}, {Gonzalez-Solares}, {Gottloeber}, {Gould},
  {Grebel}, {Gueguen}, {Guiglion}, {Haehnelt}, {Hahn}, {Hansen}, {Hartman},
  {Hauptner}, {Hawkins}, {Haynes}, {Haynes}, {Heiter}, {Helmi}, {Aguayo},
  {Hewett}, {Hinton}, {Hobbs}, {Hoenig}, {Hofman}, {Hook}, {Hopgood},
  {Hopkins}, {Hourihane}, {Howes}, {Howlett}, {Huet}, {Irwin}, {Iwert},
  {Jablonka}, {Jahn}, {Jahnke}, {Jarno}, {Jin}, {Jofre}, {Johl}, {Jones},
  {J{\"o}nsson}, {Jordan}, {Karovicova}, {Khalatyan}, {Kelz}, {Kennicutt},
  {King}, {Kitaura}, {Klar}, {Klauser}, {Kneib}, {Koch}, {Koposov},
  {Kordopatis}, {Korn}, {Kosmalski}, {Kotak}, {Kovalev}, {Kreckel}, {Kripak},
  {Krumpe}, {Kuijken}, {Kunder}, {Kushniruk}, {Lam}, {Lamer}, {Laurent},
  {Lawrence}, {Lehmitz}, {Lemasle}, {Lewis}, {Li}, {Lidman}, {Lind}, {Liske},
  {Lizon}, {Loveday}, {Ludwig}, {McDermid}, {Maguire}, {Mainieri}, {Mali},
  {Mandel}, {Mandel}, {Mannering}, {Martell}, {Martinez Delgado}, {Matijevic},
  {McGregor}, {McMahon}, {McMillan}, {Mena}, {Merloni}, {Meyer}, {Michel},
  {Micheva}, {Migniau}, {Minchev}, {Monari}, {Muller}, {Murphy},
  {Muthukrishna}, {Nandra}, {Navarro}, {Ness}, {Nichani}, {Nichol}, {Nicklas},
  {Niederhofer}, {Norberg}, {Obreschkow}, {Oliver}, {Owers}, {Pai},
  {Pankratow}, {Parkinson}, {Paschke}, {Paterson}, {Pecontal}, {Parry},
  {Phillips}, {Pillepich}, {Pinard}, {Pirard}, {Piskunov}, {Plank},
  {Pl{\"u}schke}, {Pons}, {Popesso}, {Power}, {Pragt}, {Pramskiy}, {Pryer},
  {Quattri}, {Queiroz}, {Quirrenbach}, {Rahurkar}, {Raichoor}, {Ramstedt},
  {Rau}, {Recio-Blanco}, {Reiss}, {Renaud}, {Revaz}, {Rhode}, {Richard},
  {Richter}, {Rix}, {Robotham}, {Roelfsema}, {Romaniello}, {Rosario},
  {Rothmaier}, {Roukema}, {Ruchti}, {Rupprecht}, {Rybizki}, {Ryde}, {Saar},
  {Sadler}, {Sahl{\'e}n}, {Salvato}, {Sassolas}, {Saunders}, {Saviauk},
  {Sbordone}, {Schmidt}, {Schnurr}, {Scholz}, {Schwope}, {Seifert}, {Shanks},
  {Sheinis}, {Sivov}, {Sk{\'u}lad{\'o}ttir}, {Smartt}, {Smedley}, {Smith},
  {Smith}, {Sorce}, {Spitler}, {Starkenburg}, {Steinmetz}, {Stilz}, {Storm},
  {Sullivan}, {Sutherland}, {Swann}, {Tamone}, {Taylor}, {Teillon}, {Tempel},
  {ter Horst}, {Thi}, {Tolstoy}, {Trager}, {Traven}, {Tremblay}, {Tresse},
  {Valentini}, {van de Weygaert}, {van den Ancker}, {Veljanoski}, {Venkatesan},
  {Wagner}, {Wagner}, {Walcher}, {Waller}, {Walton}, {Wang}, {Winkler},
  {Wisotzki}, {Worley}, {Worseck}, {Xiang}, {Xu}, {Yong}, {Zhao}, {Zheng},
  {Zscheyge}, \& {Zucker}}]{deJong2019}
{de Jong}, R.~S., {Agertz}, O., {Berbel}, A.~A., {et~al.} 2019, The Messenger,
  175, 3

\bibitem[{{Della Ceca} {et~al.}(2008){Della Ceca}, {Caccianiga}, {Severgnini},
  {Maccacaro}, {Brunner}, {Carrera}, {Cocchia}, {Mateos}, {Page}, \&
  {Tedds}}]{DellaCeca2008}
{Della Ceca}, R., {Caccianiga}, A., {Severgnini}, P., {et~al.} 2008, \aap, 487,
  119

\bibitem[{{Della Ceca} {et~al.}(2004){Della Ceca}, {Maccacaro}, {Caccianiga},
  {Severgnini}, {Braito}, {Barcons}, {Carrera}, {Watson}, {Tedds}, {Brunner},
  {Lehmann}, {Page}, {Lamer}, \& {Schwope}}]{DellaCeca2004}
{Della Ceca}, R., {Maccacaro}, T., {Caccianiga}, A., {et~al.} 2004, \aap, 428,
  383

\bibitem[{{Dey} {et~al.}(2019){Dey}, {Schlegel}, {Lang}, {Blum}, {Burleigh},
  {Fan}, {Findlay}, {Finkbeiner}, {Herrera}, {Juneau}, {Landriau}, {Levi},
  {McGreer}, {Meisner}, {Myers}, {Moustakas}, {Nugent}, {Patej}, {Schlafly},
  {Walker}, {Valdes}, {Weaver}, {Y{\`e}che}, {Zou}, {Zhou}, {Abareshi},
  {Abbott}, {Abolfathi}, {Aguilera}, {Alam}, {Allen}, {Alvarez}, {Annis},
  {Ansarinejad}, {Aubert}, {Beechert}, {Bell}, {BenZvi}, {Beutler}, {Bielby},
  {Bolton}, {Brice{\~n}o}, {Buckley-Geer}, {Butler}, {Calamida}, {Carlberg},
  {Carter}, {Casas}, {Castander}, {Choi}, {Comparat}, {Cukanovaite}, {Delubac},
  {DeVries}, {Dey}, {Dhungana}, {Dickinson}, {Ding}, {Donaldson}, {Duan},
  {Duckworth}, {Eftekharzadeh}, {Eisenstein}, {Etourneau}, {Fagrelius},
  {Farihi}, {Fitzpatrick}, {Font-Ribera}, {Fulmer}, {G{\"a}nsicke},
  {Gaztanaga}, {George}, {Gerdes}, {Gontcho}, {Gorgoni}, {Green}, {Guy},
  {Harmer}, {Hernandez}, {Honscheid}, {Huang}, {James}, {Jannuzi}, {Jiang},
  {Joyce}, {Karcher}, {Karkar}, {Kehoe}, {Kneib}, {Kueter-Young}, {Lan},
  {Lauer}, {Le Guillou}, {Le Van Suu}, {Lee}, {Lesser}, {Perreault Levasseur},
  {Li}, {Mann}, {Marshall}, {Mart{\'\i}nez-V{\'a}zquez}, {Martini}, {du Mas des
  Bourboux}, {McManus}, {Meier}, {M{\'e}nard}, {Metcalfe},
  {Mu{\~n}oz-Guti{\'e}rrez}, {Najita}, {Napier}, {Narayan}, {Newman}, {Nie},
  {Nord}, {Norman}, {Olsen}, {Paat}, {Palanque-Delabrouille}, {Peng},
  {Poppett}, {Poremba}, {Prakash}, {Rabinowitz}, {Raichoor}, {Rezaie},
  {Robertson}, {Roe}, {Ross}, {Ross}, {Rudnick}, {Safonova}, {Saha},
  {S{\'a}nchez}, {Savary}, {Schweiker}, {Scott}, {Seo}, {Shan}, {Silva},
  {Slepian}, {Soto}, {Sprayberry}, {Staten}, {Stillman}, {Stupak}, {Summers},
  {Sien Tie}, {Tirado}, {Vargas-Maga{\~n}a}, {Vivas}, {Wechsler}, {Williams},
  {Yang}, {Yang}, {Yapici}, {Zaritsky}, {Zenteno}, {Zhang}, {Zhang}, {Zhou}, \&
  {Zhou}}]{Dey2019}
{Dey}, A., {Schlegel}, D.~J., {Lang}, D., {et~al.} 2019, \aj, 157, 168

\bibitem[{{Dwelly} {et~al.}(2017){Dwelly}, {Salvato}, {Merloni}, {Brusa},
  {Buchner}, {Anderson}, {Boller}, {Brandt}, {Budav{\'a}ri}, {Clerc}, {Coffey},
  {Del Moro}, {Georgakakis}, {Green}, {Jin}, {Menzel}, {Myers}, {Nandra},
  {Nichol}, {Ridl}, {Schwope}, \& {Simm}}]{Dwelly2017}
{Dwelly}, T., {Salvato}, M., {Merloni}, A., {et~al.} 2017, \mnras, 469, 1065

\bibitem[{{Eckart} {et~al.}(2006){Eckart}, {Stern}, {Helfand}, {Harrison},
  {Mao}, \& {Yost}}]{Eckart2006}
{Eckart}, M.~E., {Stern}, D., {Helfand}, D.~J., {et~al.} 2006, \apjs, 165, 19

\bibitem[{{Evans} {et~al.}(2010){Evans}, {Primini}, {Glotfelty}, {Anderson},
  {Bonaventura}, {Chen}, {Davis}, {Doe}, {Evans}, {Fabbiano}, {Galle}, {Gibbs},
  {Grier}, {Hain}, {Hall}, {Harbo}, {He}, {Houck}, {Karovska}, {Kashyap},
  {Lauer}, {McCollough}, {McDowell}, {Miller}, {Mitschang}, {Morgan},
  {Mossman}, {Nichols}, {Nowak}, {Plummer}, {Refsdal}, {Rots}, {Siemiginowska},
  {Sundheim}, {Tibbetts}, {Van Stone}, {Winkelman}, \& {Zografou}}]{Evans2010}
{Evans}, I.~N., {Primini}, F.~A., {Glotfelty}, K.~J., {et~al.} 2010, \apjs,
  189, 37

\bibitem[{{Fabian} {et~al.}(2008){Fabian}, {Vasudevan}, \&
  {Gandhi}}]{Fabian2008}
{Fabian}, A.~C., {Vasudevan}, R.~V., \& {Gandhi}, P. 2008, \mnras, 385, L43

\bibitem[{{Fiore} {et~al.}(2003){Fiore}, {Brusa}, {Cocchia}, {Baldi},
  {Carangelo}, {Ciliegi}, {Comastri}, {La Franca}, {Maiolino}, {Matt},
  {Molendi}, {Mignoli}, {Perola}, {Severgnini}, \& {Vignali}}]{Fiore2003}
{Fiore}, F., {Brusa}, M., {Cocchia}, F., {et~al.} 2003, \aap, 409, 79

\bibitem[{{Fiore} {et~al.}(2001){Fiore}, {Giommi}, {Vignali}, {Comastri},
  {Matt}, {Perola}, {La Franca}, {Molendi}, {Tamburelli}, \&
  {Antonelli}}]{Fiore2001}
{Fiore}, F., {Giommi}, P., {Vignali}, C., {et~al.} 2001, \mnras, 327, 771

\bibitem[{{Freyberg} {et~al.}(2020){Freyberg}, {Perinati}, {Pacaud}, {Eraerds},
  {Churazov}, {Dennerl}, {Predehl}, {Merloni}, {Meidinger}, {Bulbul},
  {Friedrich}, {Gilfanov}, {Tenzer}, {Pommranz}, {Eckert}, {Schmitt}, {Brusa},
  \& {Santangelo}}]{Freyberg20}
{Freyberg}, M., {Perinati}, E., {Pacaud}, F., {et~al.} 2020, in Society of
  Photo-Optical Instrumentation Engineers (SPIE) Conference Series, Vol. 11444,
  Space Telescopes and Instrumentation 2020: Ultraviolet to Gamma Ray, ed.
  J.-W.~A. {den Herder}, S.~{Nikzad}, \& K.~{Nakazawa}, 114441O

\bibitem[{{Gallagher} {et~al.}(2006){Gallagher}, {Brandt}, {Chartas},
  {Priddey}, {Garmire}, \& {Sambruna}}]{Gallagher2006}
{Gallagher}, S.~C., {Brandt}, W.~N., {Chartas}, G., {et~al.} 2006, \apj, 644,
  709

\bibitem[{{Georgakakis} {et~al.}(2017){Georgakakis}, {Aird}, {Schulze},
  {Dwelly}, {Salvato}, {Nandra}, {Merloni}, \& {Schneider}}]{Georgakakis2017}
{Georgakakis}, A., {Aird}, J., {Schulze}, A., {et~al.} 2017, \mnras, 471, 1976

\bibitem[{{Gibson} {et~al.}(2009){Gibson}, {Jiang}, {Brandt}, {Hall}, {Shen},
  {Wu}, {Anderson}, {Schneider}, {Vanden Berk}, {Gallagher}, {Fan}, \&
  {York}}]{Gibson2009}
{Gibson}, R.~R., {Jiang}, L., {Brandt}, W.~N., {et~al.} 2009, \apj, 692, 758

\bibitem[{{Gilli} {et~al.}(2007){Gilli}, {Comastri}, \& {Hasinger}}]{Gilli2007}
{Gilli}, R., {Comastri}, A., \& {Hasinger}, G. 2007, \aap, 463, 79

\bibitem[{{Giommi} {et~al.}(2019){Giommi}, {Brandt}, {Barres de Almeida},
  {Pollock}, {Arneodo}, {Chang}, {Civitarese}, {De Angelis}, {D'Elia}, {Del Rio
  Vera}, {Di Pippo}, {Middei}, {Penacchioni}, {Perri}, {Ruffini}, {Sahakyan},
  \& {Turriziani}}]{Giommi2019}
{Giommi}, P., {Brandt}, C.~H., {Barres de Almeida}, U., {et~al.} 2019, \aap,
  631, A116

\bibitem[{{Green} {et~al.}(2004){Green}, {Silverman}, {Cameron}, {Kim},
  {Wilkes}, {Barkhouse}, {LaCluyz{\'e}}, {Morris}, {Mossman}, {Ghosh},
  {Grimes}, {Jannuzi}, {Tananbaum}, {Aldcroft}, {Baldwin}, {Chaffee}, {Dey},
  {Dosaj}, {Evans}, {Fan}, {Foltz}, {Gaetz}, {Hooper}, {Kashyap}, {Mathur},
  {McGarry}, {Romero-Colmenero}, {Smith}, {Smith}, {Smith}, {Torres},
  {Vikhlinin}, \& {Wik}}]{Green2004}
{Green}, P.~J., {Silverman}, J.~D., {Cameron}, R.~A., {et~al.} 2004, \apjs,
  150, 43

\bibitem[{{Greene} \& {Ho}(2005)}]{GreeneHo2005}
{Greene}, J.~E. \& {Ho}, L.~C. 2005, \apj, 630, 122

\bibitem[{{Gunn} {et~al.}(2006){Gunn}, {Siegmund}, {Mannery}, {Owen}, {Hull},
  {Leger}, {Carey}, {Knapp}, {York}, {Boroski}, {Kent}, {Lupton}, {Rockosi},
  {Evans}, {Waddell}, {Anderson}, {Annis}, {Barentine}, {Bartoszek}, {Bastian},
  {Bracker}, {Brewington}, {Briegel}, {Brinkmann}, {Brown}, {Carr},
  {Czarapata}, {Drennan}, {Dombeck}, {Federwitz}, {Gillespie}, {Gonzales},
  {Hansen}, {Harvanek}, {Hayes}, {Jordan}, {Kinney}, {Klaene}, {Kleinman},
  {Kron}, {Kresinski}, {Lee}, {Limmongkol}, {Lindenmeyer}, {Long}, {Loomis},
  {McGehee}, {Mantsch}, {Neilsen}, {Neswold}, {Newman}, {Nitta}, {Peoples},
  {Pier}, {Prieto}, {Prosapio}, {Rivetta}, {Schneider}, {Snedden}, \&
  {Wang}}]{Gunn06}
{Gunn}, J.~E., {Siegmund}, W.~A., {Mannery}, E.~J., {et~al.} 2006, \aj, 131,
  2332

\bibitem[{{Guo} {et~al.}(2018){Guo}, {Shen}, \& {Wang}}]{Guo2018}
{Guo}, H., {Shen}, Y., \& {Wang}, S. 2018, {PyQSOFit: Python code to fit the
  spectrum of quasars}, Astrophysics Source Code Library

\bibitem[{{Halpern}(1984)}]{Halpern84}
{Halpern}, J.~P. 1984, \apj, 281, 90

\bibitem[{Hartigan \& Hartigan(1985)}]{Hartigan1985DipTest}
Hartigan, J.~A. \& Hartigan, P.~M. 1985, The Annals of Statistics, 13, 70

\bibitem[{{Hasinger} {et~al.}(2007){Hasinger}, {Cappelluti}, {Brunner},
  {Brusa}, {Comastri}, {Elvis}, {Finoguenov}, {Fiore}, {Franceschini}, {Gilli},
  {Griffiths}, {Lehmann}, {Mainieri}, {Matt}, {Matute}, {Miyaji}, {Molendi},
  {Paltani}, {Sanders}, {Scoville}, {Tresse}, {Urry}, {Vettolani}, \&
  {Zamorani}}]{Hasinger2007}
{Hasinger}, G., {Cappelluti}, N., {Brunner}, H., {et~al.} 2007, \apjs, 172, 29

\bibitem[{{HI4PI Collaboration} {et~al.}(2016){HI4PI Collaboration}, {Ben
  Bekhti}, {Fl{\"o}er}, {Keller}, {Kerp}, {Lenz}, {Winkel}, {Bailin},
  {Calabretta}, {Dedes}, {Ford}, {Gibson}, {Haud}, {Janowiecki}, {Kalberla},
  {Lockman}, {McClure-Griffiths}, {Murphy}, {Nakanishi}, {Pisano}, \&
  {Staveley-Smith}}]{HI4PI2016}
{HI4PI Collaboration}, {Ben Bekhti}, N., {Fl{\"o}er}, L., {et~al.} 2016, \aap,
  594, A116

\bibitem[{{Huchra} {et~al.}(2012){Huchra}, {Macri}, {Masters}, {Jarrett},
  {Berlind}, {Calkins}, {Crook}, {Cutri}, {Erdo{\v{g}}du}, {Falco}, {George},
  {Hutcheson}, {Lahav}, {Mader}, {Mink}, {Martimbeau}, {Schneider},
  {Skrutskie}, {Tokarz}, \& {Westover}}]{Huchra12}
{Huchra}, J.~P., {Macri}, L.~M., {Masters}, K.~L., {et~al.} 2012, \apjs, 199,
  26

\bibitem[{{Ilbert} {et~al.}(2006){Ilbert}, {Arnouts}, {McCracken},
  {Bolzonella}, {Bertin}, {Le F{\`e}vre}, {Mellier}, {Zamorani}, {Pell{\`o}},
  {Iovino}, {Tresse}, {Le Brun}, {Bottini}, {Garilli}, {Maccagni}, {Picat},
  {Scaramella}, {Scodeggio}, {Vettolani}, {Zanichelli}, {Adami}, {Bardelli},
  {Cappi}, {Charlot}, {Ciliegi}, {Contini}, {Cucciati}, {Foucaud}, {Franzetti},
  {Gavignaud}, {Guzzo}, {Marano}, {Marinoni}, {Mazure}, {Meneux}, {Merighi},
  {Paltani}, {Pollo}, {Pozzetti}, {Radovich}, {Zucca}, {Bondi}, {Bongiorno},
  {Busarello}, {de La Torre}, {Gregorini}, {Lamareille}, {Mathez}, {Merluzzi},
  {Ripepi}, {Rizzo}, \& {Vergani}}]{Ilbert2006}
{Ilbert}, O., {Arnouts}, S., {McCracken}, H.~J., {et~al.} 2006, \aap, 457, 841

\bibitem[{{Itoh} {et~al.}(2020){Itoh}, {Utsumi}, {Inoue}, {Ohta}, {Doi},
  {Morokuma}, {Kawabata}, \& {Tanaka}}]{Itoh2020}
{Itoh}, R., {Utsumi}, Y., {Inoue}, Y., {et~al.} 2020, \apj, 901, 3

\bibitem[{{Jones} {et~al.}(2009){Jones}, {Read}, {Saunders}, {Colless},
  {Jarrett}, {Parker}, {Fairall}, {Mauch}, {Sadler}, {Watson}, {Burton},
  {Campbell}, {Cass}, {Croom}, {Dawe}, {Fiegert}, {Frankcombe}, {Hartley},
  {Huchra}, {James}, {Kirby}, {Lahav}, {Lucey}, {Mamon}, {Moore}, {Peterson},
  {Prior}, {Proust}, {Russell}, {Safouris}, {Wakamatsu}, {Westra}, \&
  {Williams}}]{Jones09}
{Jones}, D.~H., {Read}, M.~A., {Saunders}, W., {et~al.} 2009, \mnras, 399, 683

\bibitem[{{Jones} {et~al.}(2016){Jones}, {Hickox}, {Black}, {Hainline},
  {DiPompeo}, \& {Goulding}}]{Jones2016}
{Jones}, M.~L., {Hickox}, R.~C., {Black}, C.~S., {et~al.} 2016, \apj, 826, 12

\bibitem[{{Kaastra} {et~al.}(2000){Kaastra}, {Mewe}, {Liedahl}, {Komossa}, \&
  {Brinkman}}]{Kaastra2000}
{Kaastra}, J.~S., {Mewe}, R., {Liedahl}, D.~A., {Komossa}, S., \& {Brinkman},
  A.~C. 2000, \aap, 354, L83

\bibitem[{{Kaspi} {et~al.}(2002){Kaspi}, {Brandt}, {George}, {Netzer},
  {Crenshaw}, {Gabel}, {Hamann}, {Kaiser}, {Koratkar}, {Kraemer}, {Kriss},
  {Mathur}, {Mushotzky}, {Nandra}, {Peterson}, {Shields}, {Turner}, \&
  {Zheng}}]{Kaspi2002}
{Kaspi}, S., {Brandt}, W.~N., {George}, I.~M., {et~al.} 2002, \apj, 574, 643

\bibitem[{{Kaspi} {et~al.}(2000){Kaspi}, {Smith}, {Netzer}, {Maoz}, {Jannuzi},
  \& {Giveon}}]{Kaspi2000}
{Kaspi}, S., {Smith}, P.~S., {Netzer}, H., {et~al.} 2000, \apj, 533, 631

\bibitem[{{Kelly} \& {Shen}(2013)}]{KellyShen13}
{Kelly}, B.~C. \& {Shen}, Y. 2013, \apj, 764, 45

\bibitem[{{Kenter} {et~al.}(2005){Kenter}, {Murray}, {Forman}, {Jones},
  {Green}, {Kochanek}, {Vikhlinin}, {Fabricant}, {Fazio}, {Brand}, {Brown},
  {Dey}, {Jannuzi}, {Najita}, {McNamara}, {Shields}, \& {Rieke}}]{Kenter2005}
{Kenter}, A., {Murray}, S.~S., {Forman}, W.~R., {et~al.} 2005, \apjs, 161, 9

\bibitem[{{Kollmeier} {et~al.}(2017){Kollmeier}, {Zasowski}, {Rix}, {Johns},
  {Anderson}, {Drory}, {Johnson}, {Pogge}, {Bird}, {Blanc}, {Brownstein},
  {Crane}, {De Lee}, {Klaene}, {Kreckel}, {MacDonald}, {Merloni}, {Ness},
  {O'Brien}, {Sanchez-Gallego}, {Sayres}, {Shen}, {Thakar}, {Tkachenko},
  {Aerts}, {Blanton}, {Eisenstein}, {Holtzman}, {Maoz}, {Nandra}, {Rockosi},
  {Weinberg}, {Bovy}, {Casey}, {Chaname}, {Clerc}, {Conroy}, {Eracleous},
  {G{\"a}nsicke}, {Hekker}, {Horne}, {Kauffmann}, {McQuinn}, {Pellegrini},
  {Schinnerer}, {Schlafly}, {Schwope}, {Seibert}, {Teske}, \& {van
  Saders}}]{Kollmeier17}
{Kollmeier}, J.~A., {Zasowski}, G., {Rix}, H.-W., {et~al.} 2017, arXiv
  e-prints, arXiv:1711.03234

\bibitem[{{Koss} {et~al.}(2017){Koss}, {Trakhtenbrot}, {Ricci}, {Lamperti},
  {Oh}, {Berney}, {Schawinski}, {Balokovi{\'c}}, {Baronchelli}, {Crenshaw},
  {Fischer}, {Gehrels}, {Harrison}, {Hashimoto}, {Hogg}, {Ichikawa}, {Masetti},
  {Mushotzky}, {Sartori}, {Stern}, {Treister}, {Ueda}, {Veilleux}, \&
  {Winter}}]{Koss2017}
{Koss}, M., {Trakhtenbrot}, B., {Ricci}, C., {et~al.} 2017, \apj, 850, 74

\bibitem[{{Koss} {et~al.}(2022{\natexlab{a}}){Koss}, {Ricci}, {Trakhtenbrot},
  {Oh}, {den Brok}, {Mej{\'\i}a-Restrepo}, {Stern}, {Privon}, {Treister},
  {Powell}, {Mushotzky}, {Bauer}, {Ananna}, {Balokovi{\'c}}, {B{\"a}r},
  {Becker}, {Bessiere}, {Burtscher}, {Caglar}, {Congiu}, {Evans}, {Harrison},
  {Heida}, {Ichikawa}, {Kamraj}, {Lamperti}, {Pacucci}, {Ricci}, {Riffel},
  {Rojas}, {Schawinski}, {Temple}, {Urry}, {Veilleux}, \&
  {Williams}}]{Koss2022b}
{Koss}, M.~J., {Ricci}, C., {Trakhtenbrot}, B., {et~al.} 2022{\natexlab{a}},
  \apjs, 261, 2

\bibitem[{{Koss} {et~al.}(2022{\natexlab{b}}){Koss}, {Trakhtenbrot}, {Ricci},
  {Bauer}, {Treister}, {Mushotzky}, {Urry}, {Ananna}, {Balokovi{\'c}}, {den
  Brok}, {Cenko}, {Harrison}, {Ichikawa}, {Lamperti}, {Lein},
  {Mej{\'\i}a-Restrepo}, {Oh}, {Pacucci}, {Pfeifle}, {Powell}, {Privon},
  {Ricci}, {Salvato}, {Schawinski}, {Shimizu}, {Smith}, \& {Stern}}]{Koss2022a}
{Koss}, M.~J., {Trakhtenbrot}, B., {Ricci}, C., {et~al.} 2022{\natexlab{b}},
  \apjs, 261, 1

\bibitem[{{La Franca} {et~al.}(2005){La Franca}, {Fiore}, {Comastri}, {Perola},
  {Sacchi}, {Brusa}, {Cocchia}, {Feruglio}, {Matt}, {Vignali}, {Carangelo},
  {Ciliegi}, {Lamastra}, {Maiolino}, {Mignoli}, {Molendi}, \&
  {Puccetti}}]{LaFranca2005}
{La Franca}, F., {Fiore}, F., {Comastri}, A., {et~al.} 2005, \apj, 635, 864

\bibitem[{{La Franca} {et~al.}(2002){La Franca}, {Fiore}, {Vignali},
  {Antonelli}, {Comastri}, {Giommi}, {Matt}, {Molendi}, {Perola}, \&
  {Pompilio}}]{LaFranca2002}
{La Franca}, F., {Fiore}, F., {Vignali}, C., {et~al.} 2002, \apj, 570, 100

\bibitem[{{Lang}(2014)}]{Lang2014}
{Lang}, D. 2014, \aj, 147, 108

\bibitem[{{Liu} {et~al.}(2022{\natexlab{a}}){Liu}, {Bulbul}, {Ghirardini},
  {Liu}, {Klein}, {Clerc}, {{\"O}zsoy}, {Ramos-Ceja}, {Pacaud}, {Comparat},
  {Okabe}, {Bahar}, {Biffi}, {Brunner}, {Br{\"u}ggen}, {Buchner}, {Ider
  Chitham}, {Chiu}, {Dolag}, {Gatuzz}, {Gonzalez}, {Hoang}, {Lamer}, {Merloni},
  {Nandra}, {Oguri}, {Ota}, {Predehl}, {Reiprich}, {Salvato}, {Schrabback},
  {Sanders}, {Seppi}, \& {Thibaud}}]{LiuA2022}
{Liu}, A., {Bulbul}, E., {Ghirardini}, V., {et~al.} 2022{\natexlab{a}}, \aap,
  661, A2

\bibitem[{{Liu} {et~al.}(2022{\natexlab{b}}){Liu}, {Buchner}, {Nandra},
  {Merloni}, {Dwelly}, {Sanders}, {Salvato}, {Arcodia}, {Brusa}, {Wolf},
  {Georgakakis}, {Boller}, {Krumpe}, {Lamer}, {Waddell}, {Urrutia}, {Schwope},
  {Robrade}, {Wilms}, {Dauser}, {Comparat}, {Toba}, {Ichikawa}, {Iwasawa},
  {Shen}, \& {Medel}}]{LiuT2022_agn}
{Liu}, T., {Buchner}, J., {Nandra}, K., {et~al.} 2022{\natexlab{b}}, \aap, 661,
  A5

\bibitem[{{Liu} {et~al.}(2022{\natexlab{c}}){Liu}, {Merloni}, {Comparat},
  {Nandra}, {Sanders}, {Lamer}, {Buchner}, {Dwelly}, {Freyberg}, {Malyali},
  {Georgakakis}, {Salvato}, {Brunner}, {Brusa}, {Klein}, {Ghirardini}, {Clerc},
  {Pacaud}, {Bulbul}, {Liu}, {Schwope}, {Robrade}, {Wilms}, {Dauser},
  {Ramos-Ceja}, {Reiprich}, {Boller}, \& {Wolf}}]{LiuT2022_sim}
{Liu}, T., {Merloni}, A., {Comparat}, J., {et~al.} 2022{\natexlab{c}}, \aap,
  661, A27

\bibitem[{{Liu} {et~al.}(2018){Liu}, {Merloni}, {Wang}, {Tozzi}, {Shen},
  {Brusa}, {Salvato}, {Nandra}, {Comparat}, {Liu}, {Ponti}, \&
  {Coffey}}]{TLiu2018}
{Liu}, T., {Merloni}, A., {Wang}, J.-X., {et~al.} 2018, \mnras, 479, 5022

\bibitem[{{Luo} {et~al.}(2017){Luo}, {Brandt}, {Xue}, {Lehmer}, {Alexander},
  {Bauer}, {Vito}, {Yang}, {Basu-Zych}, {Comastri}, {Gilli}, {Gu},
  {Hornschemeier}, {Koekemoer}, {Liu}, {Mainieri}, {Paolillo}, {Ranalli},
  {Rosati}, {Schneider}, {Shemmer}, {Smail}, {Sun}, {Tozzi}, {Vignali}, \&
  {Wang}}]{Luo2017}
{Luo}, B., {Brandt}, W.~N., {Xue}, Y.~Q., {et~al.} 2017, \apjs, 228, 2

\bibitem[{{Maiolino} {et~al.}(2001){Maiolino}, {Marconi}, {Salvati},
  {Risaliti}, {Severgnini}, {Oliva}, {La Franca}, \& {Vanzi}}]{Maiolino2001}
{Maiolino}, R., {Marconi}, A., {Salvati}, M., {et~al.} 2001, \aap, 365, 28

\bibitem[{{Maiolino} {et~al.}(2010){Maiolino}, {Risaliti}, {Salvati},
  {Pietrini}, {Torricelli-Ciamponi}, {Elvis}, {Fabbiano}, {Braito}, \&
  {Reeves}}]{Maiolino2010}
{Maiolino}, R., {Risaliti}, G., {Salvati}, M., {et~al.} 2010, \aap, 517, A47

\bibitem[{{Markowitz} {et~al.}(2014){Markowitz}, {Krumpe}, \&
  {Nikutta}}]{Markowitz2014}
{Markowitz}, A.~G., {Krumpe}, M., \& {Nikutta}, R. 2014, \mnras, 439, 1403

\bibitem[{{Massaro} {et~al.}(2015){Massaro}, {Maselli}, {Leto}, {Marchegiani},
  {Perri}, {Giommi}, \& {Piranomonte}}]{Massaro2015}
{Massaro}, E., {Maselli}, A., {Leto}, C., {et~al.} 2015, \apss, 357, 75

\bibitem[{{Merloni} {et~al.}(2019){Merloni}, {Alexander}, {Banerji}, {Boller},
  {Comparat}, {Dwelly}, {Fotopoulou}, {McMahon}, {Nandra}, {Salvato}, {Croom},
  {Finoguenov}, {Krumpe}, {Lamer}, {Rosario}, {Schwope}, {Shanks}, {Steinmetz},
  {Wisotzki}, \& {Worseck}}]{Merloni2019}
{Merloni}, A., {Alexander}, D.~A., {Banerji}, M., {et~al.} 2019, The Messenger,
  175, 42

\bibitem[{{Merloni} {et~al.}(2014){Merloni}, {Bongiorno}, {Brusa}, {Iwasawa},
  {Mainieri}, {Magnelli}, {Salvato}, {Berta}, {Cappelluti}, {Comastri},
  {Fiore}, {Gilli}, {Koekemoer}, {Le Floc'h}, {Lusso}, {Lutz}, {Miyaji},
  {Pozzi}, {Riguccini}, {Rosario}, {Silverman}, {Symeonidis}, {Treister},
  {Vignali}, \& {Zamorani}}]{Merloni2014}
{Merloni}, A., {Bongiorno}, A., {Brusa}, M., {et~al.} 2014, \mnras, 437, 3550

\bibitem[{{Merloni} {et~al.}(2012){Merloni}, {Predehl}, {Becker},
  {B{\"o}hringer}, {Boller}, {Brunner}, {Brusa}, {Dennerl}, {Freyberg},
  {Friedrich}, {Georgakakis}, {Haberl}, {Hasinger}, {Meidinger}, {Mohr},
  {Nandra}, {Rau}, {Reiprich}, {Robrade}, {Salvato}, {Santangelo}, {Sasaki},
  {Schwope}, {Wilms}, \& {German eROSITA Consortium}}]{Merloni2012}
{Merloni}, A., {Predehl}, P., {Becker}, W., {et~al.} 2012, arXiv e-prints,
  arXiv:1209.3114

\bibitem[{{Murray} {et~al.}(2005){Murray}, {Kenter}, {Forman}, {Jones},
  {Green}, {Kochanek}, {Vikhlinin}, {Fabricant}, {Fazio}, {Brand}, {Brown},
  {Dey}, {Jannuzi}, {Najita}, {McNamara}, {Shields}, \& {Rieke}}]{Murray2005}
{Murray}, S.~S., {Kenter}, A., {Forman}, W.~R., {et~al.} 2005, \apjs, 161, 1

\bibitem[{{Nandra} {et~al.}(2015){Nandra}, {Laird}, {Aird}, {Salvato},
  {Georgakakis}, {Barro}, {Perez-Gonzalez}, {Barmby}, {Chary}, {Coil},
  {Cooper}, {Davis}, {Dickinson}, {Faber}, {Fazio}, {Guhathakurta}, {Gwyn},
  {Hsu}, {Huang}, {Ivison}, {Koo}, {Newman}, {Rangel}, {Yamada}, \&
  {Willmer}}]{Nandra2015}
{Nandra}, K., {Laird}, E.~S., {Aird}, J.~A., {et~al.} 2015, \apjs, 220, 10

\bibitem[{{Nenkova} {et~al.}(2008){Nenkova}, {Sirocky}, {Nikutta},
  {Ivezi{\'c}}, \& {Elitzur}}]{Nenkova2008}
{Nenkova}, M., {Sirocky}, M.~M., {Nikutta}, R., {Ivezi{\'c}}, {\v{Z}}., \&
  {Elitzur}, M. 2008, \apj, 685, 160

\bibitem[{{Nishizawa} {et~al.}(2020){Nishizawa}, {Hsieh}, {Tanaka}, \&
  {Takata}}]{Nishizawa2020}
{Nishizawa}, A.~J., {Hsieh}, B.-C., {Tanaka}, M., \& {Takata}, T. 2020, arXiv
  e-prints, arXiv:2003.01511

\bibitem[{{Oh} {et~al.}(2018){Oh}, {Koss}, {Markwardt}, {Schawinski},
  {Baumgartner}, {Barthelmy}, {Cenko}, {Gehrels}, {Mushotzky}, {Petulante},
  {Ricci}, {Lien}, \& {Trakhtenbrot}}]{Oh2018}
{Oh}, K., {Koss}, M., {Markwardt}, C.~B., {et~al.} 2018, \apjs, 235, 4

\bibitem[{{Page} {et~al.}(2011){Page}, {Carrera}, {Stevens}, {Ebrero}, \&
  {Blustin}}]{Page2011}
{Page}, M.~J., {Carrera}, F.~J., {Stevens}, J.~A., {Ebrero}, J., \& {Blustin},
  A.~J. 2011, \mnras, 416, 2792

\bibitem[{{Page} {et~al.}(2001){Page}, {Mittaz}, \& {Carrera}}]{Page2001}
{Page}, M.~J., {Mittaz}, J.~P.~D., \& {Carrera}, F.~J. 2001, \mnras, 325, 575

\bibitem[{{Pavlinsky} {et~al.}(2021){Pavlinsky}, {Tkachenko}, {Levin},
  {Alexandrovich}, {Arefiev}, {Babyshkin}, {Batanov}, {Bodnar}, {Bogomolov},
  {Bubnov}, {Buntov}, {Burenin}, {Chelovekov}, {Chen}, {Drozdova}, {Ehlert},
  {Filippova}, {Frolov}, {Gamkov}, {Garanin}, {Garin}, {Glushenko}, {Gorelov},
  {Grebenev}, {Grigorovich}, {Gureev}, {Gurova}, {Ilkaev}, {Katasonov},
  {Krivchenko}, {Krivonos}, {Korotkov}, {Kudelin}, {Kuznetsova}, {Lazarchuk},
  {Lomakin}, {Lapshov}, {Lipilin}, {Lutovinov}, {Mereminskiy}, {Molkov},
  {Nazarov}, {Oleinikov}, {Pikalov}, {Ramsey}, {Roiz}, {Rotin}, {Ryadov},
  {Sankin}, {Sazonov}, {Sedov}, {Semena}, {Semena}, {Serbinov}, {Shirshakov},
  {Shtykovsky}, {Shvetsov}, {Sunyaev}, {Swartz}, {Tambov}, {Voron}, \&
  {Yaskovich}}]{Pavlinsky2021}
{Pavlinsky}, M., {Tkachenko}, A., {Levin}, V., {et~al.} 2021, \aap, 650, A42

\bibitem[{{Piccinotti} {et~al.}(1982){Piccinotti}, {Mushotzky}, {Boldt},
  {Holt}, {Marshall}, {Serlemitsos}, \& {Shafer}}]{Piccinotti82}
{Piccinotti}, G., {Mushotzky}, R.~F., {Boldt}, E.~A., {et~al.} 1982, \apj, 253,
  485

\bibitem[{{Pierre} {et~al.}(2016){Pierre}, {Pacaud}, {Adami}, {Alis},
  {Altieri}, {Baran}, {Benoist}, {Birkinshaw}, {Bongiorno}, {Bremer}, {Brusa},
  {Butler}, {Ciliegi}, {Chiappetti}, {Clerc}, {Corasaniti}, {Coupon}, {De
  Breuck}, {Democles}, {Desai}, {Delhaize}, {Devriendt}, {Dubois}, {Eckert},
  {Elyiv}, {Ettori}, {Evrard}, {Faccioli}, {Farahi}, {Ferrari}, {Finet},
  {Fotopoulou}, {Fourmanoit}, {Gandhi}, {Gastaldello}, {Gastaud},
  {Georgantopoulos}, {Giles}, {Guennou}, {Guglielmo}, {Horellou}, {Husband},
  {Huynh}, {Iovino}, {Kilbinger}, {Koulouridis}, {Lavoie}, {Le Brun}, {Le
  Fevre}, {Lidman}, {Lieu}, {Lin}, {Mantz}, {Maughan}, {Maurogordato},
  {McCarthy}, {McGee}, {Melin}, {Melnyk}, {Menanteau}, {Novak}, {Paltani},
  {Plionis}, {Poggianti}, {Pomarede}, {Pompei}, {Ponman}, {Ramos-Ceja},
  {Ranalli}, {Rapetti}, {Raychaudury}, {Reiprich}, {Rottgering}, {Rozo},
  {Rykoff}, {Sadibekova}, {Santos}, {Sauvageot}, {Schimd}, {Sereno}, {Smith},
  {Smol{\v{c}}i{\'c}}, {Snowden}, {Spergel}, {Stanford}, {Surdej}, {Valageas},
  {Valotti}, {Valtchanov}, {Vignali}, {Willis}, \& {Ziparo}}]{Pierre2016}
{Pierre}, M., {Pacaud}, F., {Adami}, C., {et~al.} 2016, \aap, 592, A1

\bibitem[{{Predehl} {et~al.}(2021){Predehl}, {Andritschke}, {Arefiev},
  {Babyshkin}, {Batanov}, {Becker}, {B{\"o}hringer}, {Bogomolov}, {Boller},
  {Borm}, {Bornemann}, {Br{\"a}uninger}, {Br{\"u}ggen}, {Brunner}, {Brusa},
  {Bulbul}, {Buntov}, {Burwitz}, {Burkert}, {Clerc}, {Churazov}, {Coutinho},
  {Dauser}, {Dennerl}, {Doroshenko}, {Eder}, {Emberger}, {Eraerds},
  {Finoguenov}, {Freyberg}, {Friedrich}, {Friedrich}, {F{\"u}rmetz},
  {Georgakakis}, {Gilfanov}, {Granato}, {Grossberger}, {Gueguen}, {Gureev},
  {Haberl}, {H{\"a}lker}, {Hartner}, {Hasinger}, {Huber}, {Ji}, {Kienlin},
  {Kink}, {Korotkov}, {Kreykenbohm}, {Lamer}, {Lomakin}, {Lapshov}, {Liu},
  {Maitra}, {Meidinger}, {Menz}, {Merloni}, {Mernik}, {Mican}, {Mohr},
  {M{\"u}ller}, {Nandra}, {Nazarov}, {Pacaud}, {Pavlinsky}, {Perinati},
  {Pfeffermann}, {Pietschner}, {Ramos-Ceja}, {Rau}, {Reiffers}, {Reiprich},
  {Robrade}, {Salvato}, {Sanders}, {Santangelo}, {Sasaki}, {Scheuerle},
  {Schmid}, {Schmitt}, {Schwope}, {Shirshakov}, {Steinmetz}, {Stewart},
  {Str{\"u}der}, {Sunyaev}, {Tenzer}, {Tiedemann}, {Tr{\"u}mper}, {Voron},
  {Weber}, {Wilms}, \& {Yaroshenko}}]{Predehl2021}
{Predehl}, P., {Andritschke}, R., {Arefiev}, V., {et~al.} 2021, \aap, 647, A1

\bibitem[{{Proga} {et~al.}(2000){Proga}, {Stone}, \& {Kallman}}]{Proga2000}
{Proga}, D., {Stone}, J.~M., \& {Kallman}, T.~R. 2000, \apj, 543, 686

\bibitem[{{Ranalli} {et~al.}(2013){Ranalli}, {Comastri}, {Vignali}, {Carrera},
  {Cappelluti}, {Gilli}, {Puccetti}, {Brandt}, {Brunner}, {Brusa},
  {Georgantopoulos}, {Iwasawa}, \& {Mainieri}}]{Ranalli2013}
{Ranalli}, P., {Comastri}, A., {Vignali}, C., {et~al.} 2013, \aap, 555, A42

\bibitem[{{Revnivtsev} {et~al.}(2004){Revnivtsev}, {Sazonov}, {Jahoda}, \&
  {Gilfanov}}]{Revnivtsev2004}
{Revnivtsev}, M., {Sazonov}, S., {Jahoda}, K., \& {Gilfanov}, M. 2004, \aap,
  418, 927

\bibitem[{{Ricci} {et~al.}(2017){Ricci}, {Trakhtenbrot}, {Koss}, {Ueda}, {Del
  Vecchio}, {Treister}, {Schawinski}, {Paltani}, {Oh}, {Lamperti}, {Berney},
  {Gandhi}, {Ichikawa}, {Bauer}, {Ho}, {Asmus}, {Beckmann}, {Soldi},
  {Balokovi{\'c}}, {Gehrels}, \& {Markwardt}}]{Ricci2017}
{Ricci}, C., {Trakhtenbrot}, B., {Koss}, M.~J., {et~al.} 2017, \apjs, 233, 17

\bibitem[{{Richards} {et~al.}(2006){Richards}, {Lacy}, {Storrie-Lombardi},
  {Hall}, {Gallagher}, {Hines}, {Fan}, {Papovich}, {Vanden Berk}, {Trammell},
  {Schneider}, {Vestergaard}, {York}, {Jester}, {Anderson}, {Budav{\'a}ri}, \&
  {Szalay}}]{Richards06}
{Richards}, G.~T., {Lacy}, M., {Storrie-Lombardi}, L.~J., {et~al.} 2006, \apjs,
  166, 470

\bibitem[{{Rothschild} {et~al.}(1979){Rothschild}, {Boldt}, {Holt},
  {Serlemitsos}, {Garmire}, {Agrawal}, {Riegler}, {Bowyer}, \&
  {Lampton}}]{Rothschild79}
{Rothschild}, R., {Boldt}, E., {Holt}, S., {et~al.} 1979, Space Science
  Instrumentation, 4, 269

\bibitem[{{Salvato} {et~al.}(2018){Salvato}, {Buchner}, {Budav{\'a}ri},
  {Dwelly}, {Merloni}, {Brusa}, {Rau}, {Fotopoulou}, \& {Nand
  ra}}]{Salvato2018}
{Salvato}, M., {Buchner}, J., {Budav{\'a}ri}, T., {et~al.} 2018, \mnras, 473,
  4937

\bibitem[{{Salvato} {et~al.}(2022{\natexlab{a}}){Salvato}, {Wolf}, {Dwelly},
  {Georgakakis}, {Brusa}, {Merloni}, {Liu}, {Toba}, {Nandra}, {Lamer},
  {Buchner}, {Schneider}, {Freund}, {Rau}, {Schwope}, {Nishizawa}, {Klein},
  {Arcodia}, {Comparat}, {Musiimenta}, {Nagao}, {Brunner}, {Malyali},
  {Finoguenov}, {Anderson}, {Shen}, {Ibarra-Medel}, {Trump}, {Brandt}, {Urry},
  {Rivera}, {Krumpe}, {Urrutia}, {Miyaji}, {Ichikawa}, {Schneider}, {Fresco},
  {Boller}, {Haase}, {Brownstein}, {Lane}, {Bizyaev}, \&
  {Nitschelm}}]{Salvato22}
{Salvato}, M., {Wolf}, J., {Dwelly}, T., {et~al.} 2022{\natexlab{a}}, \aap,
  661, A3

\bibitem[{{Salvato} {et~al.}(2022{\natexlab{b}}){Salvato}, {Wolf}, {Dwelly},
  {Georgakakis}, {Brusa}, {Merloni}, {Liu}, {Toba}, {Nandra}, {Lamer},
  {Buchner}, {Schneider}, {Freund}, {Rau}, {Schwope}, {Nishizawa}, {Klein},
  {Arcodia}, {Comparat}, {Musiimenta}, {Nagao}, {Brunner}, {Malyali},
  {Finoguenov}, {Anderson}, {Shen}, {Ibarra-Medel}, {Trump}, {Brandt}, {Urry},
  {Rivera}, {Krumpe}, {Urrutia}, {Miyaji}, {Ichikawa}, {Schneider}, {Fresco},
  {Boller}, {Haase}, {Brownstein}, {Lane}, {Bizyaev}, \&
  {Nitschelm}}]{Salvato2022}
{Salvato}, M., {Wolf}, J., {Dwelly}, T., {et~al.} 2022{\natexlab{b}}, \aap,
  661, A3

\bibitem[{Salviander {et~al.}(2007)Salviander, Shields, Gebhardt, \&
  Bonning}]{Salviander_2007}
Salviander, S., Shields, G.~A., Gebhardt, K., \& Bonning, E.~W. 2007, The
  Astrophysical Journal, 662, 131

\bibitem[{{Saxton} {et~al.}(2008){Saxton}, {Read}, {Esquej}, {Freyberg},
  {Altieri}, \& {Bermejo}}]{Saxton2008}
{Saxton}, R.~D., {Read}, A.~M., {Esquej}, P., {et~al.} 2008, \aap, 480, 611

\bibitem[{{Sazonov} \& {Revnivtsev}(2004)}]{Sazonov2004}
{Sazonov}, S.~Y. \& {Revnivtsev}, M.~G. 2004, \aap, 423, 469

\bibitem[{{Schlegel} {et~al.}(1998){Schlegel}, {Finkbeiner}, \& {Davis}}]{SFD}
{Schlegel}, D.~J., {Finkbeiner}, D.~P., \& {Davis}, M. 1998, \apj, 500, 525

\bibitem[{{Schneider} {et~al.}(2022){Schneider}, {Freund}, {Czesla}, {Robrade},
  {Salvato}, \& {Schmitt}}]{Schneider22}
{Schneider}, P.~C., {Freund}, S., {Czesla}, S., {et~al.} 2022, \aap, 661, A6

\bibitem[{{Schulze} {et~al.}(2015){Schulze}, {Bongiorno}, {Gavignaud},
  {Schramm}, {Silverman}, {Merloni}, {Zamorani}, {Hirschmann}, {Mainieri},
  {Wisotzki}, {Shankar}, {Fiore}, {Koekemoer}, \& {Temporin}}]{Schulze15}
{Schulze}, A., {Bongiorno}, A., {Gavignaud}, I., {et~al.} 2015, \mnras, 447,
  2085

\bibitem[{{Shen}(2013)}]{Shen2013}
{Shen}, Y. 2013, Bulletin of the Astronomical Society of India, 41, 61

\bibitem[{Shen \& Liu(2012)}]{Shen_2012}
Shen, Y. \& Liu, X. 2012, The Astrophysical Journal, 753, 125

\bibitem[{{Simmonds} {et~al.}(2018){Simmonds}, {Buchner}, {Salvato}, {Hsu}, \&
  {Bauer}}]{Simmonds2018}
{Simmonds}, C., {Buchner}, J., {Salvato}, M., {Hsu}, L.~T., \& {Bauer}, F.~E.
  2018, \aap, 618, A66

\bibitem[{{Smee} {et~al.}(2013){Smee}, {Gunn}, {Uomoto}, {Roe}, {Schlegel},
  {Rockosi}, {Carr}, {Leger}, {Dawson}, {Olmstead}, {Brinkmann}, {Owen},
  {Barkhouser}, {Honscheid}, {Harding}, {Long}, {Lupton}, {Loomis}, {Anderson},
  {Annis}, {Bernardi}, {Bhardwaj}, {Bizyaev}, {Bolton}, {Brewington}, {Briggs},
  {Burles}, {Burns}, {Castander}, {Connolly}, {Davenport}, {Ebelke}, {Epps},
  {Feldman}, {Friedman}, {Frieman}, {Heckman}, {Hull}, {Knapp}, {Lawrence},
  {Loveday}, {Mannery}, {Malanushenko}, {Malanushenko}, {Merrelli}, {Muna},
  {Newman}, {Nichol}, {Oravetz}, {Pan}, {Pope}, {Ricketts}, {Shelden},
  {Sandford}, {Siegmund}, {Simmons}, {Smith}, {Snedden}, {Schneider},
  {SubbaRao}, {Tremonti}, {Waddell}, \& {York}}]{Smee13}
{Smee}, S.~A., {Gunn}, J.~E., {Uomoto}, A., {et~al.} 2013, \aj, 146, 32

\bibitem[{{Smith} {et~al.}(2001){Smith}, {Brickhouse}, {Liedahl}, \&
  {Raymond}}]{Smith2001}
{Smith}, R.~K., {Brickhouse}, N.~S., {Liedahl}, D.~A., \& {Raymond}, J.~C.
  2001, \apjl, 556, L91

\bibitem[{{Suh} {et~al.}(2020){Suh}, {Civano}, {Trakhtenbrot}, {Shankar},
  {Hasinger}, {Sanders}, \& {Allevato}}]{Suh2020}
{Suh}, H., {Civano}, F., {Trakhtenbrot}, B., {et~al.} 2020, \apj, 889, 32

\bibitem[{{Sunyaev} {et~al.}(2021){Sunyaev}, {Arefiev}, {Babyshkin},
  {Bogomolov}, {Borisov}, {Buntov}, {Brunner}, {Burenin}, {Churazov},
  {Coutinho}, {Eder}, {Eismont}, {Freyberg}, {Gilfanov}, {Gureyev}, {Hasinger},
  {Khabibullin}, {Kolmykov}, {Komovkin}, {Krivonos}, {Lapshov}, {Levin},
  {Lomakin}, {Lutovinov}, {Medvedev}, {Merloni}, {Mernik}, {Mikhailov},
  {Molodtsov}, {Mzhelsky}, {M{\"u}ller}, {Nandra}, {Nazarov}, {Pavlinsky},
  {Poghodin}, {Predehl}, {Robrade}, {Sazonov}, {Scheuerle}, {Shirshakov},
  {Tkachenko}, \& {Voron}}]{Sunyaev2021}
{Sunyaev}, R., {Arefiev}, V., {Babyshkin}, V., {et~al.} 2021, \aap, 656, A132

\bibitem[{{Sutherland} \& {Saunders}(1992)}]{Sutherland1992}
{Sutherland}, W. \& {Saunders}, W. 1992, \mnras, 259, 413

\bibitem[{{Truemper}(1982)}]{Truemper1982}
{Truemper}, J. 1982, Advances in Space Research, 2, 241

\bibitem[{{Trump} {et~al.}(2015){Trump}, {Sun}, {Zeimann}, {Luck}, {Bridge},
  {Grier}, {Hagen}, {Juneau}, {Montero-Dorta}, {Rosario}, {Brandt},
  {Ciardullo}, \& {Schneider}}]{Trump15}
{Trump}, J.~R., {Sun}, M., {Zeimann}, G.~R., {et~al.} 2015, \apj, 811, 26

\bibitem[{Tsuzuki {et~al.}(2006)Tsuzuki, Kawara, Yoshii, Oyabu, Tanabe, \&
  Matsuoka}]{Tsuzuki_2006}
Tsuzuki, Y., Kawara, K., Yoshii, Y., {et~al.} 2006, The Astrophysical Journal,
  650, 57

\bibitem[{{Turner} {et~al.}(1997){Turner}, {George}, {Nandra}, \&
  {Mushotzky}}]{Turner1997}
{Turner}, T.~J., {George}, I.~M., {Nandra}, K., \& {Mushotzky}, R.~F. 1997,
  \apjs, 113, 23

\bibitem[{{Ueda} {et~al.}(2014){Ueda}, {Akiyama}, {Hasinger}, {Miyaji}, \&
  {Watson}}]{Ueda2014}
{Ueda}, Y., {Akiyama}, M., {Hasinger}, G., {Miyaji}, T., \& {Watson}, M.~G.
  2014, \apj, 786, 104

\bibitem[{{Ueda} {et~al.}(2003){Ueda}, {Akiyama}, {Ohta}, \&
  {Miyaji}}]{Ueda2003}
{Ueda}, Y., {Akiyama}, M., {Ohta}, K., \& {Miyaji}, T. 2003, \apj, 598, 886

\bibitem[{{Ueda} {et~al.}(1999){Ueda}, {Takahashi}, {Inoue}, {Tsuru}, {Sakano},
  {Ishisaki}, {Ogasaka}, {Makishima}, {Yamada}, {Akiyama}, \&
  {Ohta}}]{Ueda1999}
{Ueda}, Y., {Takahashi}, T., {Inoue}, H., {et~al.} 1999, \apj, 518, 656

\bibitem[{Vestergaard \& Peterson(2006)}]{Vestergaard_2006}
Vestergaard, M. \& Peterson, B.~M. 2006, The Astrophysical Journal, 641, 689

\bibitem[{Vestergaard \& Wilkes(2001)}]{Vestergaard_2001}
Vestergaard, M. \& Wilkes, B.~J. 2001, The Astrophysical Journal Supplement
  Series, 134, 1

\bibitem[{{Vulic} {et~al.}(2022){Vulic}, {Hornschemeier}, {Haberl},
  {Basu-Zych}, {Kyritsis}, {Zezas}, {Salvato}, {Ptak}, {Bogdan}, {Kovlakas},
  {Wilms}, {Sasaki}, {Liu}, {Merloni}, {Dwelly}, {Brunner}, {Lamer}, {Maitra},
  {Nandra}, \& {Santangelo}}]{Vulic2022}
{Vulic}, N., {Hornschemeier}, A.~E., {Haberl}, F., {et~al.} 2022, \aap, 661,
  A16

\bibitem[{{Waddell} {et~al.}(2023){Waddell}, {Nandra}, {Buchner}, {Wu}, {Shen},
  {Arcodia}, {Merloni}, {Salvato}, {Dauser}, {Boller}, {Liu}, {Comparat},
  {Wolf}, {Dwelly}, {Ricci}, {Brownstein}, \& {Brusa}}]{Waddell2023}
{Waddell}, S. G.~H., {Nandra}, K., {Buchner}, J., {et~al.} 2023, arXiv
  e-prints, arXiv:2306.00961

\bibitem[{{Webb} {et~al.}(2020){Webb}, {Coriat}, {Traulsen}, {Ballet}, {Motch},
  {Carrera}, {Koliopanos}, {Authier}, {de la Calle}, {Ceballos}, {Colomo},
  {Chuard}, {Freyberg}, {Garcia}, {Kolehmainen}, {Lamer}, {Lin}, {Maggi},
  {Michel}, {Page}, {Page}, {Perea-Calderon}, {Pineau}, {Rodriguez}, {Rosen},
  {Santos Lleo}, {Saxton}, {Schwope}, {Tom{\'a}s}, {Watson}, \&
  {Zakardjian}}]{Webb2020}
{Webb}, N.~A., {Coriat}, M., {Traulsen}, I., {et~al.} 2020, \aap, 641, A136

\bibitem[{{Wenger} {et~al.}(2000){Wenger}, {Ochsenbein}, {Egret}, {Dubois},
  {Bonnarel}, {Borde}, {Genova}, {Jasniewicz}, {Lalo{\"e}}, {Lesteven}, \&
  {Monier}}]{Wenger2000}
{Wenger}, M., {Ochsenbein}, F., {Egret}, D., {et~al.} 2000, \aaps, 143, 9

\bibitem[{{Willingale} {et~al.}(2013){Willingale}, {Starling}, {Beardmore},
  {Tanvir}, \& {O'Brien}}]{Willingale2013}
{Willingale}, R., {Starling}, R.~L.~C., {Beardmore}, A.~P., {Tanvir}, N.~R., \&
  {O'Brien}, P.~T. 2013, \mnras, 431, 394

\bibitem[{{Wright} {et~al.}(2010){Wright}, {Eisenhardt}, {Mainzer}, {Ressler},
  {Cutri}, {Jarrett}, {Kirkpatrick}, {Padgett}, {McMillan}, {Skrutskie},
  {Stanford}, {Cohen}, {Walker}, {Mather}, {Leisawitz}, {Gautier}, {McLean},
  {Benford}, {Lonsdale}, {Blain}, {Mendez}, {Irace}, {Duval}, {Liu}, {Royer},
  {Heinrichsen}, {Howard}, {Shannon}, {Kendall}, {Walsh}, {Larsen}, {Cardon},
  {Schick}, {Schwalm}, {Abid}, {Fabinsky}, {Naes}, \& {Tsai}}]{Wright2010}
{Wright}, E.~L., {Eisenhardt}, P. R.~M., {Mainzer}, A.~K., {et~al.} 2010, \aj,
  140, 1868

\bibitem[{{Wu} \& {Shen}(2022)}]{WuShen2022}
{Wu}, Q. \& {Shen}, Y. 2022, \apjs, 263, 42

\bibitem[{{Yaqoob} {et~al.}(1989){Yaqoob}, {Warwick}, \& {Pounds}}]{Yaqoob89}
{Yaqoob}, T., {Warwick}, R.~S., \& {Pounds}, K.~A. 1989, \mnras, 236, 153

\bibitem[{Yip {et~al.}(2004)Yip, Connolly, Vanden~Berk, Ma, Frieman, SubbaRao,
  Szalay, Richards, Hall, Schneider, \& et~al.}]{Yip_2004}
Yip, C.~W., Connolly, A.~J., Vanden~Berk, D.~E., {et~al.} 2004, The
  Astronomical Journal, 128, 2603

\end{thebibliography}

\begin{appendix} 

\section{The eFEDS hard source catalogue}

Description of catalogue contents are listed in Table~\ref{table:cat}.
In the column names, the {\small band} flags indicate a few energy bands as listed in Table.~\ref{table:energy_bands}.
The {\sl\small suffix} flags in the flux-related columns (FluxObsv, FluxCorr, FluxIntr, LumiIntr) include ``Med'' (posterior median), ``Lo1'', ``Lo2'' (1-$\sigma$/2-$\sigma$ percentile lower limit), ``Up1'', ``Up2'' (1-$\sigma$/2-$\sigma$ percentile upper limit), and ``BF'' (best-fit).   
For other X-ray spectral parameters,  the {\sl\small suffix} flags include ``Med'' (posterior median), ``Mean'' (posterior mean), ``Std'' (standard deviation), ``Lo'' (1-$\sigma$ lower limit), ``Up'' (1-$\sigma$ upper limit).
For a few spectral shape parameters, e.g., $\Gamma$ and N$_\mathrm{H}$, we also provide ``HLo'' (1-$\sigma$ HDI lower limit), ``HUp'' (1-$\sigma$ HDI upper limit), and ``KL'' (KL divergence). 
These X-ray spectral parameters are named and calculated in the same way as described in \citet{LiuT2022_agn}.

\centering
\onecolumn
\begin{longtable}{p{0.2\textwidth} p{0.7\textwidth}}
\caption{Columns in the catalog}
\\
\\
\hline
    Column name & Description\\
\hline
\multicolumn{2}{l}{X-ray source properties from \citet{Brunner2022}}\\
\hline
NAME                        & Source name\\
ID\_SRC                  	& Source ID in the eFEDS hard X-ray catalog  \\
ID\_main                  	& ID of the corresponding source in the eFEDS main X-ray catalog  \\
RA\_CORR                   	&X-ray right ascension (deg; J2000), astrometrically corrected  \\
DEC\_CORR                  	&X-ray declination (deg; J2000), astrometrically corrected  \\
RADEC\_ERROR\_CORR          &combined positional uncertainty (arcsec)\\
DET\_LIKE\_{\small band}             	&source detection likelihood in band {\small band}; {\small band} 1,2,3 for 0.2--0.6, 0.6--2.3, and 2.3--5 keV, {\small band} 0 for combined 3-band likelihood  \\
inArea90                  	&Whether located inside the inner 90\%-area region of eFEDS  \\
ML\_RATE\_{\small band}  & source count rate (cts/s) in {\small band} 1,2,3,u,b1,b2,b3,b4\\
ML\_CTS\_{\small band}  & source net counts in {\small band} 1,2,3,u,b1,b2,b3,b4\\
ML\_FLUX\_{\small band}  & source flux (erg cm$^{-2}$ s$^{-1}$) in {\small band} 1,2,3,u,b1,b2,b3,b4\\
\hline
\multicolumn{2}{l}{Multiband properties, mainly from \citet{Salvato2022}}\\
\hline
CTP\_quality               	&Counterpart quality \\
CTP\_CLASS                 	&Classification of the optical counterpart . For AGN it can be 2: ``likely extraGalactic'' or 3: ``secure extraGalactic''.  \\
Redshift     & redshift \\
RedshiftOrig  &       origin of redshift\\
RedshiftGrade &       Redshift Grade . Grade 4 indicates high-quality photo-z. The highest value 5 indicates spec-z. \\
CTP\_LS8\_UNIQUE\_OBJID & ID of the best LS8 counterpart  \\
CTP\_LS8\_RA               	& Right ascension (deg; J2000) of the best LS8 counterpart  \\
CTP\_LS8\_DEC              	& Declination (deg; J2000) of the best LS8 counterpart  \\
W1      	& LS8 Wise W1 magnitude (AB) \\
W1\_ERR  	& LS8 Wise W1 magnitude error \\
W2      	& LS8 Wise W2 magnitude (AB) \\
W2\_ERR  	& LS8 Wise W2 magnitude error \\
LS8\_g   	& LS8 g-band magnitude (AB) \\
LS8\_g\_ERR	& LS8 g-band magnitude error \\
LS8\_r   	& LS8 r-band magnitude (AB) \\
LS8\_r\_ERR	& LS8 r-band magnitude error \\
LS8\_z   	& LS8 z-band magnitude \\
LS8\_z\_ERR	& LS8 z-band magnitude error \\
in\_KiDS                	&Whether located inside the region of the KiDS survey  \\
galNH                       &Total Galactic absorption column density (cm$^{-2}$)\\
AGN         & Boolean flag for the AGN subsample\\
\hline
\multicolumn{2}{l}{X-ray spectral properties}\\
\hline
LxModel                    	&Index of the selected model for X-ray luminosity measurement. 1: single-powerlaw; 3: powerlaw+blackbody; 4: powerlaw with Gamma fixed at $1.8$. 0 means no luminosity measurement.\\
NHclass                     &Class of measurement of AGN $N_{\rm H}$ based on model 1, which can be 1: \texttt{uninformative}, 2: \texttt{unobscured}, 3: \texttt{mildly-measured}, and 4: \texttt{well-measured}.\\
Exposure                    &Exposure time from the X-ray spectral file (s)\\
SrcCts                      &Source net counts in the 0.2--5~keV band measured from the spectra\\
Rate\_{\sl\small band}      & Net count rate in the 0.2--2.3, 0.2--0.5, 0.5--1, 1--2, 2--4.5, 2.3--5, and 5--8~keV bands (with {\sl\small band} suffixes of d2\_2d3, d2\_d5, d5\_1, 1\_2, 2\_4d5, 2d3\_5, and 5\_8) measured from the spectra \\ 
RateErr\_{\sl\small band}   & Net count rate error in the corresponding energy band\\ 
BkgCts\_{\sl\small band}    & Background counts in the 0.2--0.6, 0.6--2.3, 2.3--5, and 5--8~keV bands (with {\sl\small band} suffixes of d2\_d6, d6\_2d3, 2d3\_5, and 5\_8)\\ 
BkgCtsErr\_{\sl\small band} & Background counts error in the corresponding energy band\\ 
FluxObsv\_{\sl\small suffix}\_{\sl\small band}\_{\small m$n$}&Observed energy flux (erg cm$^2$ s$^{-1}$) in an observed-frame energy band {\sl\small s} (0.5--2~keV)  or {\sl\small t} (2.3--5~keV); for all the models  (m0--m5) \\ 
FluxCorr\_{\sl\small suffix}\_{\sl\small band}\_{\small m$n$}&Absorption corrected energy flux (erg cm$^2$ s$^{-1}$) in an observed-frame energy band {\sl\small s} (0.5--2~keV)  or {\sl\small t} (2.3--5~keV); for m1--m5\\ 
FluxIntr\_{\sl\small suffix}\_{\sl\small band}\_{\small m$n$}&Absorption corrected energy flux (erg cm$^2$ s$^{-1}$)) in a rest-frame energy band {\sl\small s} (0.5--2~keV), {\sl\small h} (2--10~keV), or {\sl\small 2keV} (1.999--2.001~keV); for m1--m5\\ 
LumiIntr\_{\sl\small suffix}\_{\sl\small band}\_{\small m$n$}&Intrinsic (absorption corrected) luminosity (erg/s) in a rest-frame energy band {\sl\small s} (0.5--2~keV), {\sl\small h} (2--10~keV), or {\sl\small 2keV} (1.999--2.001~keV); for m1--m5. Having no model index suffix m$n$ indicates the selected luminosity measurement using model LxModel.\\ 
logZ\_m$n$                   &log Bayesian evidence for each mode (m0-- m5)\\
$\Gamma$\_{\sl\small suffix}\_m$n$&                      Powerlaw slope; for m1, m2, m3, m5 \\ 
logPowNorm\_{\sl\small suffix}\_m$n$&                  Power-law normalization for AGN models (m1$\sim$m5)\\ 
lognH\_{\sl\small suffix}\_m$n$&                       AGN absorption column density N$_\mathrm{H}$ (cm$^{-2}$) for AGN models, or Galactic N$_\mathrm{H}$ for m0 \\ 
logBkgNorm\_{\sl\small suffix}\_m$n$&                  Background normalization for all the models\\ 
logApecNorm\_{\sl\small suffix}&                 APEC normalization; only for m0 \\ 
logBBNorm\_{\sl\small suffix}&                   Blackbody normalization; only for m3 \\ 
logkT\_{\sl\small suffix}\_m$n$&                       Temperature (keV) of blackbody (m3) or APEC (m0)\\ 
logAbundanc\_{\sl\small suffix}&                 Abundance of the APEC model; only for m0\\ 
dGm\_{\sl\small suffix}&                        Slope of the additional soft power law minus slope of the primary power law; only for m2\\ 
logFrac\_{\sl\small suffix}&                     Ratio of the additional power-law to the primary power law at 1~keV; only for m2\\ 
\hline
\multicolumn{2}{l}{SDSS optical spectral properties}\\
\hline
SDSS\_DR                    & SDSS data release version \\
PLATE                       & plate ID \\
MJD                         & MJD \\
FIBERID                     & Fiber ID \\
CATALOGID                   & catalog ID \\
SDSS\_RA                    & SDSS target RA [deg] \\
SDSS\_DEC                   & SDSS target DEC [deg] \\
SDSS\_Z                     & SDSS redshift \\
SNR\_conti                  & continum signal-to-noise ratio \\
LogL1350                    & 1350\AA luminosity [erg/s] \\
LogL1350\_err               & 1350\AA luminosity [erg/s] \\
LogL3000                    & 3000\AA luminosity [erg/s] \\
LogL3000\_err               & 3000\AA luminosity [erg/s] \\
LogL5100                    & 5100\AA luminosity [erg/s] \\
LogL5100\_err               & 5100\AA luminosity [erg/s] \\
logMbh\_Hbeta               & Hbeta black hole mass \\
logMbh\_Hbeta\_err          & Hbeta black hole mass 1-$\sigma$ error \\
logMBH\_MgII                & MgII black hole mass \\
logMBH\_MgII\_err           & MgII black hole mass 1-$\sigma$ error \\
logMBH\_CIV                 & CIV black hole mass \\
logMBH\_CIV\_err            & CIV black hole mass 1-$\sigma$ error \\
Ha\_br\_flux            & H$\alpha$ flux\\ 
Ha\_br\_flux\_err       & H$\alpha$ flux uncertainty\\ 
Ha\_br\_LogL            & H$\alpha$ luminosity\\ 
Ha\_br\_LogL\_err       & H$\alpha$ luminosity uncertainty\\ 
Ha\_br\_FWHM            & H$\alpha$ FWHM\\ 
Ha\_br\_FWHM\_err       & H$\alpha$ FWHM uncertainty\\ 
Hb\_br\_flux            & H$\beta$ flux\\ 
Hb\_br\_flux\_err       & H$\beta$ flux uncertainty\\ 
Hb\_br\_LogL            & H$\beta$ luminosity\\ 
Hb\_br\_LogL\_err       & H$\beta$ luminosity uncertainty\\ 
Hb\_br\_FWHM            & H$\beta$ FWHM\\ 
Hb\_br\_FWHM\_err       & H$\beta$ FWHM uncertainty\\ 
MgII\_br\_flux          & MgII flux\\ 
MgII\_br\_flux\_err     & MgII flux uncertainty\\ 
MgII\_br\_LogL          & MgII luminosity\\ 
MgII\_br\_LogL\_err     & MgII luminosity uncertainty\\ 
MgII\_br\_FWHM          & MgII FWHM\\ 
MgII\_br\_FWHM\_err     & MgII FWHM uncertainty\\ 
CIV\_br\_flux           & CIV flux\\ 
CIV\_br\_flux\_err      & CIV flux uncertainty\\ 
CIV\_br\_LogL           & CIV luminosity\\ 
CIV\_br\_LogL\_err      & CIV luminosity uncertainty\\ 
CIV\_br\_FWHM           & CIV FWHM\\ 
CIV\_br\_FWHM\_err      & CIV FWHM uncertainty\\ 
Lbol             & bolometric luminosity \\ 
Lbol\_err         & bolometric luminosity error \\ 
logMBH           & adopted black hole mass \\ 
logMBH\_err       & adopted black hole mass error   \\ 
Ledd             & Eddington luminosity \\ 
Ledd\_err         & Eddington luminosity error \\ 
GoodMBH          & whether the black hole mass is good \\
\hline

\end{longtable}


\twocolumn
\begin{table}
  \small
  \caption{Dictionary of energy band suffixes in the column names}
  \begin{tabular}{p{0.09\textwidth} p{0.26\textwidth}}
   \hline
Band       & Energy range  \\  
\hline	
1 & 0.2--0.6 keV \\
2 & 0.6--2.3 keV\\
3 & 2.3--5 keV\\
s & 0.5--2 keV\\
t & 2.3--5 keV\\
u & 5--8 keV\\
b1 & 0.2--0.5 keV\\
b2 & 0.5--1 keV\\
b3 & 1--2 keV\\
b4 & 2--4.5 keV\\
2keV & 1.999--2.001 keV\\
\hline
  \end{tabular}
      \label{table:energy_bands}
\end{table}

\end{appendix}

\end{document}